\documentclass[12pt]{article}

\usepackage{natbib}
\usepackage{url}
\usepackage{amsmath}
\usepackage{amsmath}
\usepackage{amsfonts}
\usepackage{amssymb}
\usepackage{enumerate}
\usepackage{color}
\usepackage{array}
\usepackage{bbm}
\usepackage{multirow}
\usepackage{rotating}
\usepackage{setspace}
\usepackage{pdflscape}
\usepackage{afterpage}
\def\argmin{{\rm argmin}}  
\usepackage{thmtools, thm-restate}

\newcommand{\BlackBox}{\rule{1.5ex}{1.5ex}}  % end of proof
\ifdefined\proof
    \renewenvironment{proof}{\par\noindent{\bf Proof\ }}{\hfill\BlackBox\\[2mm]}
\else
    \newenvironment{proof}{\par\noindent{\bf Proof\ }}{\hfill\BlackBox\\[2mm]}
\fi
 
\newtheorem{theorem}{Theorem}

\begin{document}
\thispagestyle{empty}
\baselineskip=28pt
\vskip 5mm
\begin{center} {\Large{\bf  Neural Bayes estimators for censored inference with peaks-over-threshold models}}
\end{center}

\begin{center}
\large
Jordan Richards$^{1*}$, Matthew Sainsbury-Dale$^{2,3}$,\\ Andrew Zammit-Mangion$^3$, Rapha\"el Huser$^2$
\end{center}
\footnotetext[1]{
\baselineskip=10pt School of Mathematics,  University of Edinburgh, Edinburgh, United Kingdom.\\ $^*$E-mail: jordan.richards@ed.ac.uk}
\footnotetext[2]{
\baselineskip=10pt Statistics Program, Computer, Electrical and Mathematical Sciences and Engineering (CEMSE) Division, King Abdullah University of Science and Technology (KAUST), Thuwal, 23955-6900, Saudi Arabia.}
\footnotetext[2]{
\baselineskip=10pt School of Mathematics and Applied Statistics, University of Wollongong, Wollongong, Australia}

\baselineskip=17pt
\vskip 4mm
\centerline{\today}
\vskip 6mm

%%%%%%%%%%%%%%%%%%%%%%%%%%%%%%%%%%%%%%%%%%%%%%%%%%%%%%%%%%%%%%%%%%%%%%%%
\begin{center}
{\large{\bf Abstract}}
\end{center}
Making inference with spatial extremal dependence models can be computationally burdensome since they involve intractable and/or censored likelihoods. Building on recent advances in likelihood-free inference with neural Bayes estimators, that is, neural networks that approximate Bayes estimators, we develop highly efficient estimators for censored peaks-over-threshold models that {use data augmentation techniques} to encode censoring information in the neural network {input}. Our new method provides a paradigm shift that challenges traditional censored likelihood-based inference methods for spatial extremal dependence models. Our simulation studies highlight significant gains in both computational and statistical efficiency, relative to competing likelihood-based approaches, when applying our novel estimators to make inference with popular extremal dependence models, such as max-stable, $r$-Pareto, and random scale mixture process models. We also illustrate that it is possible to train a single neural Bayes estimator for a general censoring level, precluding the need to retrain the network when the censoring level is changed. We illustrate the efficacy of our estimators by making fast inference on hundreds-of-thousands of high-dimensional spatial extremal dependence models to assess extreme particulate matter 2.5 microns or less in diameter (${\rm PM}_{2.5}$) concentration over the whole of Saudi Arabia. 
\baselineskip=16pt

\par\vfill\noindent

\pagenumbering{arabic}
\baselineskip=24pt

%%%%%%%%%%%%%%%%%%%%%%%%%%%%%%%%%
%%%%%%%%%%%%%%%%%%%%%%%%%%%%%%%%%
\section{Introduction}
Due to the flexibility and computational scalability of neural networks, recent years have seen a dramatic uptake in their application in the study of spatial extreme values, with their uses ranging from extreme quantile regression \citep{cannon2010flexible,vasiliades2015nonstationary,galib2022deepextrema, wilson2022deepgpd, pasche2022neural,richards2022insights,cisneros2024deep,richards2022unifying} to extremal dependence characterisation \citep{ahmed2022recognizing}, extreme event simulation \citep{boulaguiem2022modeling,Zhang.etal:2023}, and parameter estimation \citep{mcdonald2022comet,lenzi2021neural,majumder2022deep,walchessen2023neural,majumder2022modeling,rai2023fast,sainsbury2022fast, zammit2024neural}.
A promising new approach to likelihood-free inference for spatial (and space-time) dependence models is neural point estimation. In neural point estimation, one trains a neural network to take in data as input and provide parameter point estimates as output; see, for example, \cite{cremanns2017deep}, \cite{zammit2020deep},  \cite{gerber2021fast}, and also \cite{lenzi2021neural} and \cite{sainsbury2022fast} who consider inference with spatial extremal processes. Neural point estimators are accurate, very fast to evaluate (once trained) and they can be trained to be approximately Bayes; we refer to such estimators as neural Bayes estimators \citep[NBEs,][]{sainsbury2022fast}. They are also ``amortised'' in the sense that a trained estimator may be re-used repeatedly at almost no computational cost; once trained, these estimators generate estimates via a single function evaluation and are very fast. Moreover, they provide estimates without requiring evaluation of a likelihood function. This is particularly helpful for models where the likelihood function is computationally intractable or expensive to evaluate, as is often the case with extreme-value models \citep[e.g.,][]{padoan2010likelihood, castruccio2016high, huser2022advances}.\par
Models for spatial extremal dependence can be broadly classified into two categories: models for maxima and models for peaks-over-threshold. Popular models for the former are max-stable processes (MSPs), which arise as the only possible non-degenerate limits of suitably normalised pointwise maxima of independent and identically distributed (i.i.d.) random fields. Likelihood-based inference for MSPs can be computationally difficult, as the number of likelihood terms for a $d$-dimensional observation vector grows super-exponentially with $d$ \citep{padoan2010likelihood}. Since their popularisation by \cite{de1984spectral}, \cite{schlather2002models}, and \cite{padoan2010likelihood}, many efforts have been made to improve the computational speed of making inference with MSPs, and to increase the feasible dimensionality for which they can be applied; see, for example, \cite{genton2011likelihood}, \cite{huser2013composite}, and recent work by \cite{Hector2023} and \cite{huser2022vecchia}. Despite these efforts, traditional techniques for making inference with MSPs remain limited to relatively low-dimensional settings or make use of approximations that lead to a substantial loss in statistical efficiency. \par 
Popular models for spatial peaks-over-threshold data include $r$-Pareto processes \citep{Ferreira2014, de2018high}, inverted MSPs \citep{wadsworth2012dependence}, random scale/location mixtures \citep{huser2017bridging,krupskii2018factor,huser2019modeling}, and the spatial conditional extremes model \citep{wadsworth2022higher}. It has been shown that likelihood estimators for these models can be highly biased if spatial extreme events include marginally non-extreme observations \citep{huser2016likelihood}, and this sub-asymptotic misspecification problem is often solved by treating non-extreme observations as censored when making inference. Although the spatial conditional extremes model usually suffers less from these bias issues, inference with this model may still require censored likelihood estimation in certain applications \citep[e.g., for rainfall extremes, see][]{richards2022modelling}. Moreover, whilst  MSPs are not intended to be used with peaks-over-threshold data (being limits for maxima, not threshold exceedances), their joint tail structure is in one-to-one correspondence with $r$-Pareto processes. It is therefore possible to fit models designed for maxima to high peaks-over-threshold data by censoring low values as, for example, done by \cite{huser2014space}. In these peaks-over-threshold inference approaches, censoring is key to avoid non-extreme values from adversely affecting the estimation of the extremal dependence structure, but can lead to computationally burdensome inference. While computations can be simplified through the use of a nugget effect \citep[see, e.g.,][]{zhang2022hierarchical}, such approaches are still computationally demanding in high dimensions. Censoring is therefore an important technique when making inference with models for spatial extremes, but comes at a high (often prohibitive) computational cost.\par
A way forward to circumvent the computational issues that arise when censoring data is to use likelihood-free inference methods, where the likelihood function need not be evaluated. NBEs are an attractive choice since, as discussed above, they are likelihood-free and optimal in a Bayes sense. 
% \cite{sainsbury2022fast} illustrate the efficacy of their use when estimating the parameters for both the Schlather class of max-stable processes \citep{schlather2002models} and the spatial conditional extremes model. %However, their framework is not built to handle censored data as input.
\cite{lenzi2021neural} and \cite{sainsbury2022fast} illustrate the efficacy of their use when estimating the parameters for several spatial extremal processes, including the Schlather \citep{schlather2002models} and Brown--Resnick \citep{brown1977extreme} classes of max-stable processes, and the spatial conditional extremes model \citep{wadsworth2022higher}. However, the treatment of censored data in this framework has not yet been addressed. 
 In this paper we develop a class of NBEs that allow for, and efficiently extract information from, censored data. 
 The resulting NBEs can be used to make inference from censored data with a variety of popular peaks-over-threshold models. We observe substantial gains in statistical and computational efficiency (when evaluating a trained NBE) over competing censored likelihood-based approaches. Our proposed framework allows us to make inference with spatial extremal dependence models of a dimension that is much higher than was previously possible when used in an ``operational'' manner, that is, when used to repeat an estimation task thousands of times within a short time frame (e.g., a few minutes). In our application, we fit upwards of 130 million separate extremal dependence models {using five different trained NBEs}; the scale of such a study is unprecedented in literature on models for spatial extreme values.\par
The remainder of this paper is organised as follows. In Section~\ref{sec:method}, we review NBEs and outline our new methodology for handling censored inputs. In Section~\ref{sec:spatextproc}, we summarise some popular spatial extremal dependence models and discuss estimation strategies based on censored likelihood maximisation. In Section~\ref{sec:sim}, we conduct simulation studies to illustrate the efficacy of our inference framework, and in Section~\ref{sec:application}, we apply our estimator to analyse extremal dependence of high-resolution Saudi Arabian particulate matter data. %of 2.5 microns or less in diameter (PM2.5) data. 
The paper concludes with a discussion in Section~\ref{sec:discussion}. %An online supplement contains additional details. 
Reproducible code is available from \url{https://github.com/Jbrich95/CensoredNeuralEstimators}.

%%%%%%%%%%%%%%%%%%%%%%%%%%%%%%%%%
%%%%%%%%%%%%%%%%%%%%%%%%%%%%%%%%%
\section{Methodology}
\label{sec:method}
In Section~\ref{sec:NBE}, we review NBEs. In Section~\ref{sec:NBEs}, we detail our methodology for adapting NBEs to allow them to handle censored data as input. 
%%%%%%%%%%%%%%%%%%%%%%%%%%%%%%%%%
\subsection{Neural Bayes estimators}
\label{sec:NBE}
A parametric statistical model is defined as a set of probability distributions $\mathcal{P}$ on a sample space $\Omega$, parameterised by a parameter vector $\boldsymbol{\theta}\in\mathbb{R}^p$ such that ${\mathcal{P}:= \{F(\cdot\mid\boldsymbol{\theta}):\boldsymbol{\theta}\in\Theta\}}$, where $\Theta$ is the parameter space \citep[see][]{mccullagh2002statistical}. For ease of exposition, we take $\Omega\subseteq\mathbb{R}^d$ and assume that $F(\cdot\mid\boldsymbol{\theta})$ admits a density $f(\cdot \;|\;\boldsymbol{\theta})$ with respect to Lebesgue measure. Consider $m$ mutually independent realisations from $F(\cdot\mid\boldsymbol{\theta})\in\mathcal{P}$; a point estimator $\hat{\boldsymbol{\theta}}(\cdot)$ is any mapping from $\Omega^m$ to $\Theta$. Denote the realisations as $\mathbf{Z}:=(\mathbf{Z}_1',\dots,\mathbf{Z}_m')'$. The output of an estimator $\hat{\boldsymbol{\theta}}(\cdot)$, for a given $\boldsymbol{\theta}$ and $\mathbf{Z}$, can be assessed using a non-negative loss function $L(\boldsymbol{\theta},\hat{\boldsymbol{\theta}}(\mathbf{Z}))$; the estimator's risk at $\boldsymbol{\theta}$, $R(\boldsymbol{\theta},\hat{\boldsymbol{\theta}}(\cdot))$, is defined as the loss averaged over all possible data realisations, that is,
\begin{equation}
\label{eq:risk}
R(\boldsymbol{\theta},\hat{\boldsymbol{\theta}}(\cdot)):= \int_{\Omega^m}L(\boldsymbol{\theta},\hat{\boldsymbol{\theta}}(\mathbf{z}))f(\mathbf{z}\mid\boldsymbol{\theta})\mathrm{d}\mathbf{z}.
\end{equation}
The integrated risk $r_\Pi(\cdot)$ is a weighted average of $\eqref{eq:risk}$ over $\boldsymbol{\theta}\in\Theta$, with respect to some prior measure $\Pi(\cdot)$, and is defined as
\begin{equation}
\label{eq:Bayesrisk}
r_\Pi(\hat{\boldsymbol{\theta}}(\cdot)):=\int_\Theta R(\boldsymbol{\theta},\hat{\boldsymbol{\theta}}(\cdot))\mathrm{d}\Pi(\boldsymbol{\theta}).
\end{equation}
An estimator $\hat{\boldsymbol{\theta}}(\cdot)$ that minimises \eqref{eq:Bayesrisk} is a \textit{Bayes estimator} with respect to $L(\cdot,\cdot)$ and $\Pi(\cdot)$. Bayes estimators have attractive theoretical properties under mild regularity conditions, including asymptotic normality, efficiency, and consistency under the squared-error loss \citep[see, e.g.,][]{lehmann2006theory}. \par
A NBE is a neural network that approximately minimises the integrated risk. Bayes estimators for independent replicates satisfy permutation invariance; that is, $\hat{\boldsymbol{\theta}}(\mathbf{Z})=\hat{\boldsymbol{\theta}}(\mathbf{Z}^\star)$ where $\mathbf{Z}^\star$ denotes any permutation of the independent replicates in $\mathbf{Z}$ \citep{sainsbury2022fast}. To construct parsimonious NBEs that have this property, \cite{sainsbury2022fast} propose using a neural network architecture known as DeepSets \citep{zaheer2017deep}. Consider the functions $\boldsymbol{\psi}:\mathbb{R}^d\mapsto \mathbb{R}^q$ and $\boldsymbol{\phi}:\mathbb{R}^q\mapsto\mathbb{R}^p$, and a permutation-invariant set function $\boldsymbol{\mathfrak{a}}:(\mathbb{R}^q)^m\mapsto\mathbb{R}^q$. Here, we let the $j$-th component of $\boldsymbol{\mathfrak{a}}$, $\mathfrak{a}_j(\cdot),$ return the element-wise average over its input set (of cardinality $m$) for $j=1,\dots,q$. We represent $\boldsymbol{\phi}(\cdot)$ and $\boldsymbol{\psi}(\cdot)$ as neural networks, and collect in $\boldsymbol{\gamma}:= (\boldsymbol{\gamma}_{{\phi}}',\boldsymbol{\gamma}_{{\psi}}')'$ their trainable parameters (the so-called ``weights'' and ``biases''). Then, a permutation invariant NBE is of the form
\begin{equation}
\label{eq:NE}
\hat{\boldsymbol{\theta}}(\mathbf{Z}; \boldsymbol{\gamma})=\boldsymbol{\phi}(\mathbf{T}(\mathbf{Z}; \boldsymbol{\gamma}_{{\psi}}); \boldsymbol{\gamma}_{{\phi}}),\quad\text{ with }\quad\mathbf{T}(\mathbf{Z}; \boldsymbol{\gamma}_{{\psi}})=\boldsymbol{\mathfrak{a}}(\{\boldsymbol{\psi}(\mathbf{Z}_t; \boldsymbol{\gamma}_{{\psi}}):t=1,\dots,m\}).
\end{equation}
Permutation invariance allows the NBE to handle independent replicates parsimoniously, which is key for inference with replicated data and spatial extremal dependence models. As our interest is in estimating dependence only, we hereon assume that $\boldsymbol{\theta}$ determines only the dependence characteristics of $\mathbf{Z}$, not its marginal behaviour. Following \cite{sainsbury2022fast}, we use a densely-connected neural network for $\boldsymbol{\phi}(\cdot)$; the architecture for $\boldsymbol{\psi}(\cdot)$ is application-dependent (see the example in Section~\ref{sec:sim_gen}).\par
\label{sec:NBE_CNN}
\label{sec:NBE_pretrain}
The NBE is trained by finding neural network parameters $\boldsymbol{\gamma}^*$ that minimise the integrated risk in the estimator space spanned by $\hat{\boldsymbol{\theta}}(\cdot,\boldsymbol{\gamma})$; as \eqref{eq:Bayesrisk} cannot be directly evaluated, it is approximated using Monte Carlo methods. Given a set of $K$ parameter vectors, $\{\boldsymbol{\theta}^{(k)}: k = 1,\dots,K\},$ drawn from the prior $\Pi(\cdot)$, we simulate, for each $k$, $m$ mutually independent random samples $\mathbf{Z}^{(k)}$ from $F(\cdot\mid\boldsymbol{\theta}^{(k)})$. Then we approximate \eqref{eq:Bayesrisk} using the estimator parameterised by $\boldsymbol{\gamma}$ as
\begin{equation}
\label{Eq:risk}
\hat{r}_\Pi(\hat{\boldsymbol{\theta}}(\cdot \;;\boldsymbol{\gamma}))=\frac{1}{K}\sum_{k=1}^KL(\boldsymbol{\theta}^{(k)},\hat{\boldsymbol{\theta}}(\mathbf{Z}^{(k)};\boldsymbol{\gamma}))\approx {r}_\Pi(\hat{\boldsymbol{\theta}}(\cdot \;;\boldsymbol{\gamma})).
\end{equation}
Estimates $\boldsymbol{\gamma}^*=\argmin_{\boldsymbol{\gamma}}\hat{r}_\Pi(\hat{\boldsymbol{\theta}}(\cdot \;;\boldsymbol{\gamma}))$ can be obtained using, for example, the package \texttt{NeuralEstimators} \citep{sainsbury2022fast}, which is available in \texttt{Julia} and \texttt{R}. \par
{When training an NBE for a large number of replicates $m$, computational speed-ups can be achieved with the use of pre-training \citep[Ch.~8]{goodfellow2016deep}. Let $\boldsymbol{\gamma}^*_{m}$ denote the optimal NBE weights for datasets containing $m$ replicates. We employ pre-training by first finding $\boldsymbol{\gamma}^*_{\tilde{m}}$ for some smaller sample size $\tilde{m}<m$, and then using these as initial estimates when finding $\boldsymbol{\gamma}^*_m$. In practice, we run a recursive pre-training scheme for a sequence of increasing sample sizes, $\tilde{\boldsymbol{m}}:=(\tilde{m}_1,\tilde{m}_2,\dots,m)'$; for details, see Section~\ref{sec:sim_gen}. Throughout we assume that $m$ equals the number of observations from which inference needs to be made.}

\subsection{Neural Bayes estimators for censored data}
\label{sec:NBEs}
In Section~\ref{sec:NBE_Censored}, we review censored likelihood-based inference and detail a way of encoding relevant censoring information into the input of an NBE. In Section~\ref{sec:NBE_Vary}, we describe how to build NBEs that can be used with a range of censoring levels.
\subsubsection{Encoding censored values}
\label{sec:NBE_Censored}

%Consider a spatial process $\{Z(\mathbf{s}):\mathbf{s}\in\mathcal{S}\}$ indexed by the spatial location $\mathbf{s} \in \mathcal{S}\subset \mathbb{R}^2$. Inference on $Z(\cdot)$ is made using 
Suppose that we have (time-)replicated observations $\mathbf{Z}_t:=(Z_{t,1},\dots,Z_{t,d})'$ for $t=1,\dots,m$, 
%where each $\mathbf{Z}_t$ is an independent realisation of $Z(\cdot)$ at sampling locations $\{\mathbf{s}_1,\dots,\mathbf{s}_d\}\subset \mathcal{S}$ at time $t$. Denote by 
sampled independently from the joint distribution $F(\cdot\mid\boldsymbol{\theta})$, and that we want to estimate the unknown parameter vector $\boldsymbol{\theta}$ by prioritising calibration in the joint upper tail. 
% the joint distribution function of $\mathbf{Z}_t$. 
In order to prevent low values from affecting the estimation of extremal dependence, the standard approach in statistics of extremes is to treat all observations falling below some high marginal quantile as \emph{censored}. That is, while we know which observations are treated as censored or not, the exact values of the censored observations are assumed to be unknown. Mathematically, we consider a vector of censoring thresholds $\mathbf{c}:= (c_1,\dots,c_d)'$, where here $c_j=F_j^{-1}(\tau\mid\boldsymbol{\theta})$ is set to be the marginal $\tau$-quantile of $Z_{t,j}$, $j=1,\dots,d$, for some probability $\tau\in[0,1)$ close to one. 
% Let $\mathbf{I}_t\in\{0,1\}^d$ denote a vector of indicator variables with $j$-th component {$\mathbf{I}_{t,j}=\mathbbm{1}\{ Z_{t,j}\leq c_j\}$} indicating which of the variables from the vector $\mathbf{Z}_t:=(Z_{t,1},\dots,Z_{t,d})'$ are being censored (those with $\mathbf{I}_{t,j}=1$), and {where ${d_{t-}=\sum^d_{j=1}\mathbf{I}_{t,j}}\leq d$ is the number of censored components}. 
Let $\mathbf{I}_t := (I_{t,j} : j = 1, \dots, d)'$, with $I_{t,j}=\mathbbm{1}\{ Z_{t,j}\leq c_j\}$, denote the vector of indicator variables with $j$-th component equal to one if the corresponding element of $\mathbf{Z}_t$ is censored and zero otherwise, and {let ${d_{t-} := \sum^d_{j=1}{I}_{t,j}}\leq d$ denote the number of censored components}. 
 We rearrange the observation vector as $\mathbf{Z}_t=({\mathbf{Z}'_{t-}},{\mathbf{Z}'_{t+}})'$, where $\mathbf{Z}_{t-}:=(Z_{t,j}: Z_{t,j} \leq c_j,j=1,\dots,d)'$ is the subvector of censored observations, and $\mathbf{Z}_{t+}:=(Z_{t,j}: Z_{t,j} > c_j,j=1,\dots,d)'$ is the subvector of uncensored observations. We do the same for the vector of censoring thresholds $\mathbf{c}_t:=({\mathbf{c}'_{t-}},{\mathbf{c}'_{t+}})',$ where $\mathbf{c}_{t-}:=(c_{j}: Z_{t,j} \leq c_j,j=1,\dots,d)'$ and $\mathbf{c}_{t+}:=(c_{j}: Z_{t,j} > c_j,j=1,\dots,d)'$. Note that $\mathbf{c}_t$ is a permutation of $\mathbf{c}$, with the ordering time-dependent. Traditional censoring in statistics of extremes assumes that $\mathbf{Z}^{\rm cens}_t:=(\mathbf{I}'_t,{\mathbf{Z}'_{t+}})'$ is the only information available for inference at time $t$. The corresponding likelihood contribution is given in the following proposition.
 \begin{restatable}{proposition}{Firstprop}
\label{prop} 
Let $\mathbf{Z}_t=({Z}_{t,1},\dots,Z_{t,d})',\; d\in\mathbb{N}, t =1,\dots,m,$ be random vectors defined by a statistical model %which is 
 parameterised by $\boldsymbol{\theta}$ with joint distribution $F(\cdot\mid \boldsymbol{\theta})$ and density \mbox{$f(\cdot\mid \boldsymbol{\theta})$}. %, and let $\mathbf{z}_t=({z}_{t,1},\dots,z_{t,d})'$ be the realisation of $\mathbf{Z}_t$. 
 Let ${\mathbf{Z}_{t+}:=(Z_{t,j}: Z_{t,j} > c_j,j=1,\dots,d)'}$ denote the subvector of uncensored values of $\mathbf{Z}_t$, where {$\mathbf{c}:= (c_1,\dots,c_d)'$} is a vector of censoring levels with time-dependent subvector \mbox{$\mathbf{c}_{t-}=(c_j : Z_{t,j} \leq c_j,j=1,\dots,d)'$}. 
% Given indicator variables \mbox{$\mathbf{I}_t:=({I}_{t,1},\ldots,{I}_{t,d})'$} %\in\{0,1\}^d$} 
  % with ${I}_{t,j}=\mathbbm{1}\{ Z_{t,j}\leq c_j\}$, ${j=1,\ldots,d}$, 
% Given indicator variables \mbox{$\mathbf{I}_t:=(\mathbbm{1}\{ Z_{t,j}\leq c_j\}: j = 1, \dots, d)'$},  
Given indicators \mbox{$\mathbf{I}_t:=(\mathbbm{1}\{ Z_{t,j}\leq c_j\}: j = 1, \dots, d)'$},  
the likelihood contribution of %the censored vector 
$\mathbf{Z}_t^{\rm cens}:= (\mathbf{I}'_t,{\mathbf{Z}'_{t+}})'$ is %equal to
\begin{equation}
\label{eq:horrid_integral}
\int_{-\infty}^{\mathbf{c}_{t-}}f(\mathbf{z}_t\mid\boldsymbol{\theta})\mathrm{d}\mathbf{z}_{t-}={\partial^{|\mathcal{J}_t|}\over {\prod_{j\in \mathcal{J}_t}}\partial z_j} F(\mathbf{z}\mid \boldsymbol{\theta})\bigg|_{\mathbf{z}={\mathbf{b}}_t},\; t=1,\dots,m, 
\end{equation}
where $\mathbf{z}_t=({z}_{t,1},\dots,z_{t,d})'$ is the realisation of $\mathbf{Z}_t$, $\mathcal{J}_t := \{j: {z}_{t,j}>c_j \}$ denotes the indices of the uncensored components of $\mathbf{z}_t$ with subvector $\mathbf{z}_{t-}:=({z}_{t,j} : z_{t,j} \leq c_{j},j=1,\dots,d)'$, 
% and $\mathbf{b}_t:=({b}_{t,1},\dots{b}_{t,d})'$ has components ${b}_{t,j}=\max\{z_{t,j},c_j\}, j=1,\dots,d$.
and $\mathbf{b}_t:=(\max\{z_{t,j},c_j\}: j = 1, \dots, d)'$. 
\end{restatable}
For completeness, we provide a formal proof of \eqref{eq:horrid_integral} in Appendix~\ref{supsec:proof}. Note that \eqref{eq:horrid_integral} is a $d_{t-}$-variate integral of the joint density; if $d_{t-}$ is large, \eqref{eq:horrid_integral} quickly becomes computationally intractable and this often prevents censored likelihood inference in moderate-to-large dimensions $d$. While a different censoring framework for extremal processes has been used by \cite{wadsworth2012dependence}, \cite{opitz2016modeling}, and \cite{wadsworth2017modelling}, the formulation in \eqref{eq:horrid_integral} is usually preferred since it is more robust against model misspecification in the bulk, and thus less subject to sub-asymptotic estimation bias \citep{huser2016likelihood}. In this paper, we seek to develop novel NBE methodology that is able to handle censored inputs in a way similar to the censored likelihood contribution in \eqref{eq:horrid_integral}.

%To construct novel likelihood-free estimators that exploit censoring, 
%we require that the NBE training data, $\mathbf{Z}_t$ in \eqref{eq:NE}, communicates the censoring information to the estimator. 
A possible approach could be to discard all censored observations, $\mathbf{Z}_{t-}$, and train our neural estimator based on the uncensored data only, namely $\mathbf{Z}_{t+}$. However, this approach inevitably leads to a loss of information, particularly for large $\tau$ close to one, given that the indicator vector $\mathbf{I}_t$ is then ignored: it is impossible to recover the censored likelihood contribution \eqref{eq:horrid_integral} from the marginal density of $\mathbf{Z}_{t+}$ alone. We note that the information carried by $\mathbf{I}_t$ may not be negligible (especially for large $d$); in the spatial setting, for example, larger (respectively smaller) clusters of zeros or ones are indicative of stronger (respectively weaker) spatial dependence. Moreover, this ``data removal'' approach requires a specialised neural network architecture that can efficiently handle partially-observed vectors with varying patterns of missingness, given that the censored observations may change with every time replicate. Such an architecture can be difficult to design and implement.

Here, we propose a different approach that circumvents the aforementioned issues. We first standardise the data to an appropriate scale by marginally transforming the data $\mathbf{Z}_t=({\mathbf{Z}'_{t-}},{\mathbf{Z}'_{t+}})'$ to $\mathbf{Z}_t^*:=(({\mathbf{Z}_{t-}^{*}})',({\mathbf{Z}_{t+}^{*}})')'$ with marginal distribution $H(\cdot)$. %, which is user-specified and not necessarily related to $F(\cdot\mid\boldsymbol{\theta})$. %, and apply the same transformation to the censoring thresholds $\mathbf{c}=({\mathbf{c}_t^{\leq}}',{\mathbf{c}_t^{>}}')'$, which yields $\mathbf{c}^*=({\mathbf{c}_t^{\leq*}}',{\mathbf{c}_t^{>*}}')'$. 
We then replace the censored component $\mathbf{Z}_{t-}^*$ with a constant $\zeta_\tau< H^{-1}(\tau)$ (which may vary with the censoring probability $\tau$). In other words, we construct the vector $\widetilde{\mathbf{Z}}_t^*:=({\boldsymbol{\zeta}'_\tau},({\mathbf{Z}_{t+}^{*}})')'$, {where $\boldsymbol{\zeta}_\tau:=(\zeta_\tau,\dots,\zeta_\tau)'$} denotes a vector of appropriate length, so that $\widetilde{\mathbf{Z}}_t^*$ is of dimension $d$ and of the same format as the original vector $\mathbf{Z}_t$. Motivated by Proposition~\ref{prop}, we then propose to train the NBE using an augmented dataset ${\mathbf{A}}_t:=(\mathbf{I}'_t,({\widetilde{\mathbf{Z}}_t^{*}})')', t=1,\dots,m$. The new input is passed to the NBE by replacing $\mathbf{Z}_t$ in \eqref{eq:NE} with ${\mathbf{A}}_t$ arranged in a multi-dimensional array with two outer dimensions corresponding to $\mathbf{I}_t$ and $\widetilde{\mathbf{Z}}_t^{*}$, respectively. While the choice of marginal distribution $H(\cdot)$ and constant $\zeta_\tau$ are, in principle, arbitrary, a judicious choice can facilitate the training of the NBE.   
 %In preliminary experiments we observed better performance of our estimator when $H(\cdot)$ has a finite lower bound $z_H< \infty$ and when $\zeta_\tau\leq z_H$. 
  In particular, we find that the estimator is more easily trained when $H(\cdot)$ has a finite lower bound $z_H< \infty$ and when $\zeta_\tau\leq z_H$. 
 For example, one could choose $H(\cdot)$ to be the unit exponential distribution function and $\zeta_\tau\leq 0$; in this way, $\zeta_\tau$ is outside the possible set of uncensored values, which helps the neural network distinguish between the two data types, and thereby learn to treat censored data differently to fully-observed data. We further note that while the augmented data set ${\mathbf{A}}_t$ includes additional (artificial) data, when compared to $\mathbf{Z}_t^{{\rm cens}}$ used to define the censored likelihood contribution \eqref{eq:horrid_integral}, these artificial values are fixed and independent of $\boldsymbol{\theta} \in \Theta$. During training, the NBE therefore learns to ignore these censored values; in our simulation study in Section~\ref{sec:sim}, we verify that our censored NBE has indeed a similar performance to censored likelihood estimators in cases where the latter can be computed. A similar point estimation approach (that uses both data and an indicator variable) was proposed by \cite{Pagendam2023} and \cite{Wang2023.01.09.523219} in the context of missing (not censored) data. There is a subtle but fundamental difference between censoring and missingness; while the censoring mechanism is completely determined by the data and the censoring level $\tau$ (and is thus itself dependent on $\boldsymbol{\theta}$), data missingness mechanisms are often independent of the data generating process.

\subsubsection{Varying censoring level}
\label{sec:NBE_Vary}
%The input $\widetilde{\mathbf{Z}}_t^{*{\rm cens}}=((\mathbf{I}_t)',({\widetilde{\mathbf{Z}}^{*}_t})')'$ implicitly encodes some information on $\tau$, which may be treated as unknown when making inference on the parameters $\boldsymbol{\theta}$. When $\tau$ is not assumed known, one could design an estimator for $\boldsymbol{\theta}$ by minimising, with respect to the weights $\boldsymbol{\gamma}$, a Monte Carlo approximation (similar to \eqref{Eq:risk}) of the modified Bayes risk 

%Here we describe how one can make the NBE adaptive to the censoring level. 
To accommodate varying censoring levels, we feed $\tau$ as an input to the outer neural network of the NBE; that is, with a slight abuse of notation, we replace \eqref{eq:NE} with
\begin{align}
\label{eq:NE2}
%\hat{\boldsymbol{\theta}}(\widetilde{\mathbf{Z}}^{*{\rm cens}}, \tau; \boldsymbol{\gamma})
\hat{\boldsymbol{\theta}}({\mathbf{A}}, \tau; \boldsymbol{\gamma})=\boldsymbol{\phi}(\mathbf{T}({\mathbf{A}}; \boldsymbol{\gamma}_{{\psi}}), \tau; \boldsymbol{\gamma}_{{\phi}}),
\end{align}
where $\mathbf{T}(\mathbf{A}; \boldsymbol{\gamma}_\psi)$ is as defined in \eqref{eq:NE} but with $\mathbf{Z}$ replaced by the augmented data set ${\mathbf{A}}:= ({\mathbf{A}}'_1,\dots,{\mathbf{A}}'_m)'$ and with $\mathbf{Z}_t$ replaced with ${\mathbf{A}}_t$. The NBE $\hat{\boldsymbol{\theta}}({\mathbf{A}}, \tau; \boldsymbol{\gamma})$  is approximately Bayes for any ${\mathbf{A}} \in \Omega_{\mathbf{A},\tau}$ and $\tau \in \mathcal{T} \subseteq [0,1)$ if it minimises, with respect to $\boldsymbol{\gamma}$, the integrated risk,

\begin{equation}
\label{eq:Bayesrisk2}
r_{\Pi}^{\textrm{cens}}(\hat{\boldsymbol{\theta}}(\cdot,\cdot\;; {\boldsymbol{\gamma}})):= \int_\Theta \int_\mathcal{T}\int_{\Omega_{\mathbf{A},\tau}}\!\!L(\boldsymbol{\theta},\hat{\boldsymbol{\theta}}(\mathbf{a},\tau;\boldsymbol{\gamma}))f(\mathbf{a}\mid\tau,\boldsymbol{\theta})p(\tau)\mathrm{d}\mathbf{a}\mathrm{d}\mathbf{\tau}\mathrm{d}\Pi(\boldsymbol{\theta}),
\end{equation}
where $p(\tau)$ is a prior density for $\tau$, and $\Omega_{\mathbf{A},\tau}$ is the sample space of the augmented dataset ${\mathbf{A}}$, which is $\tau$-dependent.
%\begin{equation}
%\label{eq:Bayesrisk2}
%r_{\Pi,\tilde{\Pi}}(\hat{\boldsymbol{\theta}}(\cdot; {\boldsymbol{\gamma}})):= \int_0^1\int_\Theta R^{\rm cens}(\boldsymbol{\theta},\hat{\boldsymbol{\theta}}(\cdot; \boldsymbol{\gamma}); \tau)\mathrm{d}\Pi(\boldsymbol{\theta})\mathrm{d}\tilde{\Pi}({\tau}),
%\end{equation}
%where  $R^{\rm cens}(\boldsymbol{\theta},\hat{\boldsymbol{\theta}}(\cdot); \tau)$ is the risk function defined as in \eqref{eq:risk}, but based on censored data (which implicitly depends on the censoring level $\tau$), and $\tilde{\Pi}(\cdot)$ is some prior distribution for $\tau$ with support on $[0,1)$. 
%Such an estimator will be approximately Bayes with respect to $L(\cdot,\cdot)$, $\Pi(\cdot),$ for values of $\tau$ in the support of $\tilde{\Pi}(\cdot)$. 
This adaptivity to $\tau$ may be useful where, for example, data-collection devices only store values that exceed a certain threshold, but where the threshold is variable. In Appendix~\ref{supsec:invariance}, we show that this augmented Bayes estimator is invariant to the choice of $p(\tau)$, provided that $\mathcal{T}$ is a subset of its support and that $\tau$ is independent of $\boldsymbol{\theta}$ (as is always the case in applications of the peaks-over-threshold censoring framework). 

%In the vast majority of cases where censoring is used, $\tau$ is known and fixed. One could train and evaluate the NBE for a specific value for $\tau$ (i.e., \emph{conditional} on $\tau$), so that the NBE is (approximately) Bayes for that value of $\tau$. However, it might be desirable to have an NBE that targets the Bayes estimator for any $\tau \in[0,1)$ (or a sub-interval thereof), so that it can then be used for a wide range of $\tau$ values without the need for retraining. 
To train the NBE, we use a Monte Carlo approximation to \eqref{eq:Bayesrisk2} similar to \eqref{Eq:risk}, obtained by sampling $\boldsymbol{\theta}$ and $\tau$ from their prior distributions, and simulating ${\mathbf{A}}$ by conditioning on these samples.%Note that the NBE \eqref{eq:NE2} is conditional on $\tau$. 

\section{Spatial extremal processes}
\label{sec:spatextproc}
Here we concisely describe the spatial extremal processes that we consider in our simulations and application; we provide details of the processes and their inference using censored likelihoods. Section~\ref{sec:MSP} details max-stable, $r$-Pareto, and inverted max-stable processes. In Appendix~\ref{supsec:model}, we give further details on $r$-Pareto and Mat\'ern Gaussian processes in Sections~\ref{sec:rPareto} and \ref{supsec:GP}, respectively, with the latter used to build the random scale mixture processes described in Section~\ref{sec:scale_mix}. For comprehensive reviews of the considered processes, see \cite{davison2012}, \cite{davison2019spatial} and \cite{huser2022advances}. For the software used for inference and simulation, see Section~\ref{supsec:comp} of Appendix~\ref{supsec:model}.
%%%%%%%%%%%%%%%%%%%%%%%%%%%%%%%%%
\subsection{Max-stable, $r$-Pareto, and inverted max-stable processes}
\label{sec:MSP}
 Suppose that $\{X_t(\mathbf{s}):\mathbf{s} \in \mathcal{S}\}$ are i.i.d.\ stochastic processes with continuous sample paths on $\mathcal{S}\subset\mathbb{R}^2$ and there exists sequences of functions $a_m(\mathbf{s})>0$ and $b_m(\mathbf{s})$, such that 
%\begin{equation}
%\label{eq:MSP_norm}
$\{a_m(\mathbf{s})\}^{-1}\left[\max\limits_{t=1,\dots,m}\{X_t(\mathbf{s})\}-b_m(\mathbf{s})\right]$, 
%\end{equation}
 converges weakly, as $m\rightarrow \infty$, to a process $\{Z(\mathbf{s}):\mathbf{s} \in \mathcal{S}\}$ with non-degenerate margins. Then $Z(\mathbf{s})$ is a max-stable process (MSP) with generalised extreme value margins \citep{de1984spectral}; see, for example, \cite{davison2015statistics} and Section~\ref{supsec:uni} of Appendix~\ref{supsec:model}. Following marginal standardisation to unit Fr\'echet, denoted by $\tilde{Z}(\mathbf{s})$, and under mild regularity conditions \citep{de1984spectral,haan2006extreme}, a MSP can be constructed through the spectral representation 
\begin{equation}
\label{eq:MSP}
\tilde{Z}(\mathbf{s})=\sup_{k \geq 1} R_kW_k(\mathbf{s}),
\end{equation}
where $\{R_k\}_{k\in\mathbb{N}}$ are points of a Poisson process on $(0,\infty)$ with intensity $r^{-2}\mathrm{d}r$ and $\{W_k(\mathbf{s})\}_{k\in\mathbb{N}}$ are i.i.d.\ copies of a non-negative stochastic process $W(\cdot)$ satisfying $\mathbb{E}[W(\mathbf{s})]=1$ for all $\mathbf{s}\in \mathcal{S}$.\par
Specification of the process $W(\mathbf{s})$ in \eqref{eq:MSP} leads to a limited selection of parametric models for MSPs \citep[see e.g.,][]{smith1990max, schlather2002models, kabluchko2009stationary,opitz2013extremal}. Whilst \cite{sainsbury2022fast} consider inference for the Schlather process \citep{schlather2002models}, we instead focus on the more flexible Brown--Resnick model \citep{brown1977extreme,kabluchko2009stationary,engelke2011equivalent}; \cite{lenzi2021neural} have previously considered neural estimators for this process, albeit in the uncensored setting.  The Brown--Resnick model is defined by specifying $W(\mathbf{s})=\exp\{\varepsilon(\mathbf{s})-\sigma^2(\mathbf{s})\}$ for $\varepsilon(\mathbf{s})$ a zero-mean Gaussian process (GP) with variance $\sigma^2(\mathbf{s})>0$ for all $\mathbf{s}\in\mathcal{S}$, and which satisfies $\varepsilon(\mathbf{0})=0$ almost surely. Extremal dependence in $\tilde{Z}$ is controlled by the semivariogram of the underlying process $\varepsilon(\cdot)$, denoted $\gamma(\mathbf{s}_i,\mathbf{s}_j)$. A typical model for $\gamma$ is the stationary and isotropic semivariogram $\gamma(\mathbf{s}_i,\mathbf{s}_j)=(\|\mathbf{s}_i-\mathbf{s}_j\|/\lambda)^\kappa$, with range and smoothness parameters $\lambda>0$ and $\kappa\in(0,2]$, respectively; in this case, $\sigma^2(\mathbf{s})=2\gamma(\mathbf{s},\mathbf{0})=2(\|\mathbf{s}\|/\lambda)^\kappa$. We adopt this model and hence inference for this process requires estimation of the unknown parameter vector $\boldsymbol{\theta}=(\lambda,\kappa)'$. Note that whilst we focus on a single class of MSP, it is trivial to extend our censored NBE framework to other parametric max-stable processes, provided that simulation is feasible.\par
\par As for $r$-Pareto processes, they arise as the only possible limits of properly renormalised threshold exceedances defined in terms of a homogeneous risk functional $r(\cdot)$ that determines the `magnitude' of the process. Their characterization is intrinsically linked to max-stable processes; in particular, $r$-Pareto processes can be sampled by generating the random functions $R_kW_k(\cdot)$ in \eqref{eq:MSP} and retaining only those such that $r\{R_kW_k(\cdot)\}>1$. For more details on risk functionals and $r$-Pareto processes, see Section~\ref{sec:rPareto} of Appendix~\ref{supsec:model}.  \par
We quantify the strength of asymptotic dependence through the measure $\chi(\cdot,\cdot)$, the upper-tail index \citep{joe1997multivariate}. Assuming ${Z(\mathbf{s}_i)\sim F_i}$, $Z(\mathbf{s}_j) \sim F_j$ with continuous $F_i,F_j$, then $\chi(\mathbf{s}_i,\mathbf{s}_j)=\lim_{q\rightarrow 1}\Pr[F_i\{Z(\mathbf{s}_i)\}>q\mid F_j\{Z(\mathbf{s}_j)\}>q]$, where the process $Z(\cdot)$ exhibits asymptotic independence between sites $\mathbf{s}_i$ and $\mathbf{s}_j$ if $\chi(\mathbf{s}_i,\mathbf{s}_j)=0$, and asymptotic dependence otherwise. Max-stable and $r$-Pareto processes are either inherently asymptotically dependent (i.e., $\chi(\mathbf{s}_i,\mathbf{s}_j)>0$ for all $\mathbf{s}_i,\mathbf{s}_j\in \mathcal{S}$) or exhibit exact independence everywhere; this limitation may not be appropriate for some environmental applications where extreme events become increasingly localised with severity \citep{huser2019modeling}. \citet{wadsworth2012dependence} propose an alternative collection of flexible sub-asymptotic models for spatial extremal processes, which exhibit asymptotic independence (but not exact independence). One such model is the inverted max-stable process (IMSP), which is obtained by applying a monotonically-decreasing marginal transformation to a MSP. An IMSP $Y(\mathbf{s})$ with standard exponential margins can be constructed from a MSP $\tilde{Z}(\cdot)$ by setting $Y(\mathbf{s})=1/\tilde{Z}(\mathbf{s})$ for all $\mathbf{s} \in \mathcal{S}$: if $\tilde{Z}(\cdot)$ is a Brown--Resnick process, then we term $Y(\cdot)$ an ``inverted Brown--Resnick process''. Inference for an inverted Brown--Resnick process requires estimation of the same parameter vector, $\boldsymbol{\theta}=(\lambda,\kappa)'$, as its max-stable counterpart. \par
Inference for MSPs is computationally problematic as the full $d$-dimensional density is generally intractable even for moderate $d$ \citep[i.e., for $d>12$, see][]{castruccio2016high}; note that, by construction, IMSPs inherit the intractable likelihood and computational issues of MSPs \citep{huser2022advances}. Computational speed-ups can be exploited in certain cases, see, for example,  \cite{stephenson2005exploiting}, \cite{davison2012geostatistics}, \cite{wadsworth2014efficient} and \cite{dombry2017bayesian}, but their use can lead to biased parameter estimation \citep{wadsworth2015occurrence,huser2016likelihood} and still incur high computational demand in relatively low dimension, that is, $d < 20$ \citep{huser2019full}. Intractability of the full likelihood has motivated the use of composite pairwise likelihood methods, developed by \cite{padoan2010likelihood}, for inference with MSPs; these approaches falsely assume mutual independence between all observation pairs and construct the (surrogate) pairwise likelihood as a (potentially weighted) product over all bivariate densities. Under mild regularity conditions, pairwise likelihood inference leads to valid inference, whereby the corresponding estimators exhibit consistency and asymptotic normality \citep{padoan2010likelihood, varin2011overview}, but they suffer from a reduction in efficiency relative to methods using the full likelihood \citep{huser2013composite,castruccio2016high}. Computational expense for the pairwise likelihood grows with rate $\mathcal{O}(d^2)$, so a common approach is to use only observation pairs with sampling locations with pairwise distance subceeding a fixed cut-off distance, $h_{max}$ say; this has been shown to improve both statistical, and  computational, efficiency of the pairwise likelihood estimator \citep{bevilacqua2012estimating,sang2014tapered}. Due to their popularity, we adopt censored pairwise likelihood estimators as the competing likelihood approach for MSPs, IMSPs, and GPs, in the simulation study detailed in Section~\ref{sec:sim}.\par
The aforementioned issues with likelihood-based inference with MSPs and IMSPs are compound with the requirement of left-censoring for unbiased inference (see Section~\ref{sec:NBE_Censored}); for expressions of the censored pairwise likelihood for the Brown--Resnick process and associated IMSP, see \citet{huser2013composite} and \citet{thibaud2013}, respectively. 

%%%%%%%%%%%%%%%%%%%%%%%%%%%%%%%%%
 \subsection{Random scale mixtures}
\label{sec:scale_mix}
Max-stable and $r$-Pareto processes can be constructed from scale mixture processes of the form $Z(\mathbf{s})=RW(\mathbf{s})$, for an asymptotically independent stochastic process $W(\mathbf{s})$ with light or bounded upper-tails and an independent Pareto-tailed random variable $R$; such mixture-based processes exhibit asymptotic dependence due to the heavy-tailed random variable $R$ \citep{huser2022advances}.  Modification of $R$ and/or $W(\cdot)$ can lead to processes with more flexible extremal dependence than the inherent asymptotic dependence of MSPs and $r$-Pareto processes. Popular examples include Gaussian scale \citep{huser2017bridging} and location \citep{krupskii2018factor,castro2020local} mixtures, but we constrain our focus to the random scale mixture process proposed by \cite{huser2019modeling}, which provides a model that parsimoniously bridges the gap between asymptotic dependence and asymptotic independence in the interior of the parameter space.  \par
Following \cite{huser2019modeling}, we define a process $\{Z(\mathbf{s})\}=R^\delta \{W(\mathbf{s})^{1-\delta}\}$ for $\delta \in [0,1]$, and where $R \geq 1$ is a unit Pareto random variable and $\{W(\mathbf{s}):\mathbf{s} \in \mathcal{S}\}$ is a stochastic process with unit Pareto margins that exhibits asymptotic independence, for example, a marginally transformed GP or IMSP. The parameter $\delta$ controls the extremal dependence class of the process $Z$; if $\delta < 1/2$, then $Z$ inherits the extremal dependence class of $W$ and, hence, exhibits asymptotic independence; if $\delta \geq 1/2$, $R^\delta$ is heavier-tailed than $W(\mathbf{s})^{1-\delta}$, and so $Z$ is asymptotically dependent. As this model interpolates between an asymptotic dependence, and asymptotic independence, regime, it provides a flexible extremal dependence model that does not require choosing the extremal dependence class of data a priori. We follow \cite{huser2019modeling} and take $W(\cdot)$ to be a (marginally transformed) GP; we adopt the Mat\'ern class with range $\lambda>0$ and smoothness $\kappa>0$, described in Section~\ref{supsec:GP} of Appendix~\ref{supsec:model}. Hence, inference for $Z$ follows by estimating $\boldsymbol{\theta}=(\lambda,\kappa, \delta)'$. Note that whilst $R$ and $W(\cdot)$ have unit Pareto margins, this does not hold for $Z(\cdot)$; hereafter we assume that the margins of $Z(\cdot)$ have been standardised to unit Pareto.  \par
\cite{huser2019modeling,huser2022advances} use the full $d$-dimensional likelihood for inference, rather than the pairwise alternative described in Section~\ref{sec:MSP}; this is approximated via a quasi-Monte Carlo algorithm (described in Section~\ref{sec:rPareto} of Appendix~\ref{supsec:model}). We note that \cite{zhang2022hierarchical} adapt the model of \cite{huser2019modeling} to a hierarchical Bayesian setting, where the full $d$-dimensional censored likelihood becomes more easily amenable to high dimensions by adding a tiny noise term; however, as we are considering the \cite{huser2019modeling} model in the original frequentist setting, we use the quasi-Monte Carlo estimator as the competing likelihood approach for this model (see Section~\ref{sec:sim}). Note that the full likelihood does not scale well with $d$, and so previous applications of this model have been limited to moderate dimensions \citep[mostly $d\leq 30$, and $d \leq 200$ for][]{zhang2022hierarchical}; in Section~\ref{sec:application}, we illustrate that much higher dimensional inference $(d>1000$) is possible using our fast NBE-based approach. 

%%%%%%%%%%%%%%%%%%%%%%%%%%%%%%%%%
%%%%%%%%%%%%%%%%%%%%%%%%%%%%%%%%%
\section{Simulation studies}
\label{sec:sim}
We now conduct several simulation studies to highlight the efficacy of our novel NBE against competing censored (pseudo) likelihood approaches. In Section~\ref{sec:sim_gen}, we outline the general setting for the simulation studies. In Section~\ref{sec:sim_fixed}, we consider studies where the censoring threshold $\tau$ is fixed, while in Section~\ref{sec:sim_vary} we consider studies where $\tau$ varies during training.

%%%%%%%%%%%%%%%%%%%%%%%%%%%%%%%%%
\subsection{General setting}
\label{sec:sim_gen}
We illustrate our methodology by application to a variety of spatial extremal processes $\{Z(\mathbf{s}):\mathbf{s}\in\mathcal{S}\}$, described in Section~\ref{sec:spatextproc}, indexed by the spatial location $\mathbf{s} \in \mathcal{S}\subset \mathbb{R}^2$. Inference on $Z(\cdot)$ is made using replicated observations $\mathbf{Z}_t=(Z_{t,1},\dots,Z_{t,d})'$ for $t=1,\dots,m$, where each $\mathbf{Z}_t$ is an independent realisation of $Z(\cdot)$ at gridded sampling locations $\{\mathbf{s}_1,\dots,\mathbf{s}_d\}\subset \mathcal{S}$ at time $t$.
The domain of interest, $\mathcal{S}$, is taken to be a $16 \times 16$ regular grid on $[0,16]\times[0,16]$, which gives $|\mathcal{S}|=16^2=256$ sampling locations. \par
Our NBEs use the architecture given in Table~\ref{tab:CNN_arch}, with $\boldsymbol{\psi}(\cdot)$ and $\boldsymbol{\phi}(\cdot)$ in \eqref{eq:NE} constructed using the first four and last two rows, respectively, of Table~\ref{tab:CNN_arch}. Following \cite{sainsbury2022fast}, we use a convolutional neural network (CNN) to model $\boldsymbol{\psi}(\cdot)$; for details, see \cite{gu2018recent}. Note that CNNs are only applicable if the inputs are completely observed over a regular grid; hereon we assume this to be the case, but discuss alternative approaches for non-gridded data in Section~\ref{sec:discussion}.
We use $m=200$ independent replicates throughout, and the pre-training scheme described in Section~\ref{sec:NBE_pretrain}, with $\tilde{\mathbf{m}}=(10,50,100,200)'$; $L(\cdot,\cdot)$ in \eqref{eq:Bayesrisk} is taken to be the absolute-error loss. NBEs are trained and validated with parameter sets of length $K$ and $K/5$, respectively, where $K$ differs between studies. 
{For each sampled parameter vector $\boldsymbol{\theta}^{(k)}, k=1,\dots,K$, we generate a random censoring level $\tau$ together with $m$ mutually independent random samples of the augmented (censored) data, i.e., pairs $\{{\mathbf{A}}^{(k)},\tau^{(k)}\}$; see Section~\ref{sec:NBE_Vary}. For the general case where the NBE is dependent on the censoring level (as in \eqref{eq:NE2}), the empirical marginal risk for the parameter corresponding to the $i$-th component $\theta_i^{(k)}$ of $\boldsymbol{\theta}^{(k)}$ is $K^{-1}\sum_{k=1}^KL({\theta}^{(k)}_i,[\hat{\boldsymbol{\theta}}({\mathbf{A}}^{(k)},\tau^{(k)};\boldsymbol{\gamma})]_i)$; we also consider the fixed-$\tau$ case, which is a special case where the prior density $p(\tau)$ is degenerate at a fixed value of $\tau$, i.e., all sampled values of $\tau$ are the same. All NBEs and likelihood-based estimators are tested by evaluating this empirical marginal risk (under both the absolute and squared-error loss) on a test set of $1000$ parameter vectors; as the absolute and squared-error losses are additively separable, summing the  empirical marginal risks for $i=1,\dots,p$ gives a Monte Carlo approximation of the risk in \eqref{eq:Bayesrisk2}.} We assume that the parameters are a priori independent with marginal prior distributions $\lambda \sim {\rm Unif}(2,10)$, $\kappa\sim {\rm Unif}(0.5,2)$ and $\delta \sim{\rm Unif}(0,1)$ across all models.  \par
\begin{table*}
\caption{Neural network architecture used in Section~\ref{sec:sim}. Each convolution filter uses zero padding and unit stride. The function ${\rm vec}(\cdot)$ refers to a flattening of an array to a vector. All layers, except the final layer, used rectified linear unit (ReLU) activation functions; the final layer uses the identity function. Note that $p$ is the number of model parameters. }
\vspace{5pt}
 \begin{tabular}{lcccr} 
\hline
\hline
layer type & input dimension & output dimension & filter dimension & parameters\\
\hline
2D convolution & [16,\;16,\;2] & [7,\;7,\;64] & $10\times 10$ & $6400c+64$ \\
2D convolution & [7,\;7,\;64] & [3,\;3,\;128] & $5\times 5$ & 204,928 \\
2D convolution & [3,\;3,\;128] & [1,\;1,\;256] & $3\times 3$ & 295,168 \\
${\rm vec}(\cdot)$ & [1,\;1,\;256] & [256] &  & 0\\
dense & [256] & [500] &  & 128,500 \\
dense & [500] & [$p$] & & $501p$\\
\hline
\multicolumn{3}{l}{total trainable parameters:} & \multicolumn{2}{r}{641,460+501$p$}\\
\hline
  \end{tabular}
 \label{tab:CNN_arch}
  \end{table*}
NBEs are trained on GPUs randomly selected from within KAUST's Ibex cluster, whilst likelihood-based estimators are evaluated on CPUs from KAUST's Shaheen II supercomputer, see \url{https://www.hpc.kaust.edu.sa/ibex/gpu_nodes} and \url{https://www.hpc.kaust.edu.sa/content/shaheen-ii} for details (last accessed 24/03/2023).

%%%%%%%%%%%%%%%%%%%%%%%%%%%%%%%%%
\subsection{Fixed censoring threshold}
\label{sec:sim_fixed}
We here consider a fixed censoring threshold of $\tau=0.9$ that does not change during training. We compare our NBEs to estimators based on both the pairwise and full likelihood.

\subsubsection{Comparison to pairwise likelihood estimators}
\label{sec:sim_fixed_PL}
We compare NBEs to censored pairwise likelihood estimators for fitting Mat\'ern GPs (see Section~\ref{supsec:GP} of Appendix~\ref{supsec:model}), MSPs and IMSPs; note that inference with each requires estimation of a range and smoothness parameter, $\lambda$ and $\kappa$, respectively. For the censored pairwise likelihoods (CPL), we consider two cut-off distances (see Section~\ref{sec:MSP}): $h_{max}=3$ units and $h_{max}=\infty$ units (i.e., where all pairs are used).  \par
Each NBE takes under three hours to train, with $K=100,000$ training samples, and estimation (given a single parameter set) is incredibly fast, taking on average 0.0016 seconds. In comparison, estimation with the pairwise likelihood is much slower; for $h_{max}=\infty$, the GP and MSP/IMSP required approximately 10 and 20 minutes on average, respectively, with these averages lowering to 2.1 and 1.8 minutes when $h_{max}=3$; thus, the NBE provides a speed-up of the order of 67,500--750,000 (post-training) in this case. \par
Results for the three considered processes are given in Tables~\ref{tab:paper_sim_fixed} and \ref{subtab:sim_fixed1}, which present the marginal test risk with respect to the absolute and squared-error loss, respectively. We observe lower marginal test risk (with respect to both loss functions) for both parameters when using an NBE compared to the censored pairwise likelihood, suggesting that the former is more statistically efficient. Overall, our proposed NBE for censored data exhibits substantial increases in both statistical and computational efficiency relative to the competing likelihood-based approach.\par
\begin{figure}[t]
\centering
\begin{minipage}{0.32\linewidth}
\centering
\includegraphics[width=0.95\linewidth]{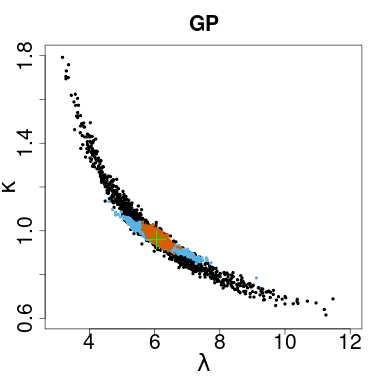} 
\end{minipage}
\begin{minipage}{0.32\linewidth}
\centering
\includegraphics[width=0.95\linewidth]{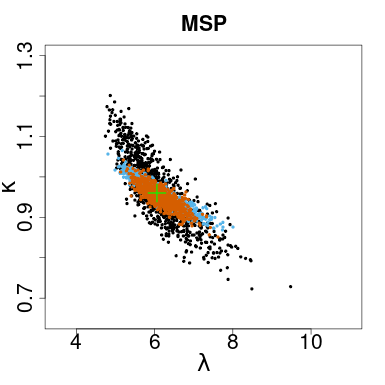} 
\end{minipage}
\begin{minipage}{0.32\linewidth}
\centering
\includegraphics[width=0.95\linewidth]{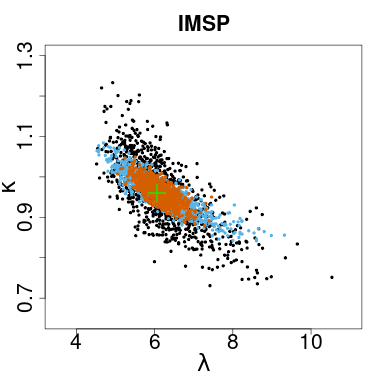} 
\end{minipage}
\caption{Empirical joint distribution from 1000 parameter estimates for (left) GP, (centre) MSP and (right) IMSP, given replicates from a single test parameter set (green cross). Black and blue points denote estimates from the CPL with cut-off distances $h_{max}=\infty$ and $h_{max}=3$, respectively, whilst the brown points correspond to estimates from the NBE.}
\label{fig:joint_dist1}
\end{figure}

Figure~\ref{fig:joint_dist1} presents the joint distribution of parameter estimates for a single test parameter vector with $1000$ sets of replicates. The NBE captures the dependence in estimates of $(\lambda,\kappa)'$ that is also present when utilising the competing pairwise likelihood estimator ($h_{max}=3$), but our NBE clearly has lower sampling variability than the pairwise likelihood estimator.

\begin{table}
\caption{Results for the simulation study detailed in Section~\ref{sec:sim_fixed_PL}. Marginal test risks (s.d.) under the absolute-error loss are provided separately for $\lambda$ ($\times 10^{-1}$) and $\kappa$ ($\times 10^{-2}$). Test risks under the squared-error loss are presented in Table~\ref{subtab:sim_fixed1} of Appendix~\ref{supsec:arch}. The lowest value in each column is given in bold.}
\vspace{5pt}
\centering
\begin{tabular}{lcccccc}
\hline
\hline
 & \multicolumn{2}{c}{GP} & \multicolumn{2}{c}{MSP} & \multicolumn{2}{c}{IMSP} \\
  &  $\lambda$ & $\kappa$&  $\lambda$ & $\kappa$ &  $\lambda$ & $\kappa$ \\
 \hline
NBE &\bf 3.1 (0.1) & \bf 4.5 (0.1) & \bf 2.4 (0.1) & \bf 1.8 (0.1) & \bf 2.6 (0.1) & \bf 2.2 (0.1)\\

CPL ($h_{max}=3$) & 4.4 (0.1) & 5.7 (0.2) & 3.5 (0.1) & 2.2 (0.1) &4.6 (0.2) & 3.2 (0.1)\\

CPL ($h_{max}=\infty$) & 9.3 (0.2) & 16.0 (0.5) & 4.3 (0.1) & 6.4 (0.2) & 5.4 (0.2) & 6.8 (0.2)\\

\hline
\end{tabular}

 \label{tab:paper_sim_fixed}

\end{table}

\subsubsection{Comparison to full likelihood estimators}
\label{sec:sim_fixed_FL}
We now consider inference for the HW and $r$-Pareto processes detailed in Sections~\ref{sec:MSP}--\ref{sec:scale_mix}. Typical likelihood-based inference for both processes uses the full $d$-dimensional likelihood, approximated via quasi-Monte Carlo methods; we compare the efficacy of such an approach against our proposed NBE. Inference with the $r$-Pareto process requires estimation of the Brown--Resnick variogram parameters described in Section~\ref{sec:MSP}.\par
For the HW model, previous attempts at inference have been limited to moderate dimension $d \leq 30$ due to the computational expense of evaluating the likelihood (with the exception of \cite{zhang2022hierarchical}, who pushed it to $d \leq 200$ using specialised Bayesian methods); we find that likelihood inference for replicates on the previously defined $16\times 16$ grid  ($d=256$) is not computationally feasible. Hence, we perform inference for the model simulated on a smaller spatial domain $\mathcal{S}^*$, which is taken to be a regular $6\times 6$ grid on $[0,16]\times[0,16]$ (with $d=6^2=36$). Whilst we compare NBE- and likelihood-based inference on this smaller grid, we further illustrate that inference for the process observed on the larger $16\times 16$ grid is feasible with an NBE; for the $r$-Pareto process, however, we only consider the larger grid. We slightly adjust the architecture in Table~\ref{tab:CNN_arch} to account for the smaller size of $\mathcal{S}^*$ relative to $\mathcal{S}$; this architecture is given in Table~\ref{tab:CNN_arch2}.\par 
We use $K=100,000$ ($K=500,000$) for the $r$-Pareto process and HW process on $\mathcal{S}$ (and $\mathcal{S}^*$). As relatively less information is contained within a single replicate, we found that larger $K$ was required to train the NBEs with the smaller domain $\mathcal{S}^*$; see also, Section~\ref{sec:application}.  NBE estimation time using either architectures  (Table~\ref{tab:CNN_arch} or Table~\ref{tab:CNN_arch2} of Appendix~\ref{supsec:arch}) is comparable with that described in Section~\ref{sec:sim_fixed_PL}. Estimation time with the full likelihood is unpredictable as it is dependent on the number of censored locations within a field, which varies between replicates. For the $r$-Pareto process, the estimation time varies between approximately 2.5 and 12 hours, and for the HW process (on $\mathcal{S}^*$), it varies between approximately 1 and 10 hours, but with some cases requiring upwards of 24 hours. These times clearly show that likelihood-based inference for a HW process observed on $\mathcal{S}$ is not computationally feasible, even when using astute quasi-Monte Carlo approximations.\par
Results for the $r$-Pareto process and HW processes are presented in Tables~\ref{tab:sim_fixed2}--\ref{suptab:sim_fixed_3} of Appendix~\ref{supsec:arch}. For the $r$-Pareto process, our NBE significantly outperforms the likelihood-based estimator; the marginal test risk (with respect to both considered losses) is lower for both parameters, suggesting a gain in statistical efficiency when using the NBE. For the HW process (with domain $\mathcal{S}^*$), the NBE is more computationally efficient than the likelihood-based approach, and it exhibits gains in statistical efficiency, especially for the parameter $\delta$ considered the most difficult parameter to infer \citep{huser2019modeling}. Computational limitations mean that we are prohibited from using the full likelihood-based approach for inference of the process on the larger domain $\mathcal{S}$, but the NBE does not have the same limitation.\par
\begin{figure}[t]
\centering
\begin{minipage}{0.32\linewidth}
\centering
\includegraphics[width=0.95\linewidth]{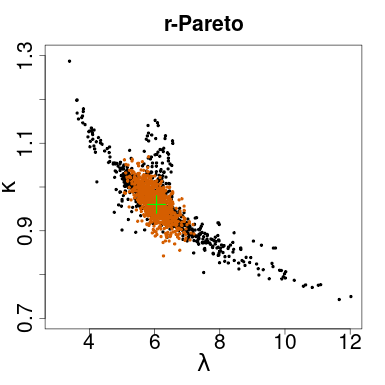} 
\end{minipage}
\begin{minipage}{0.32\linewidth}
\centering
\includegraphics[width=0.95\linewidth]{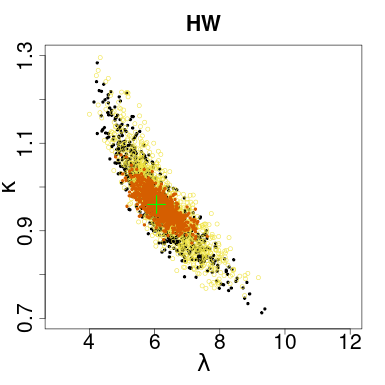} 
\end{minipage}
\begin{minipage}{0.32\linewidth}
\centering
\includegraphics[width=0.95\linewidth]{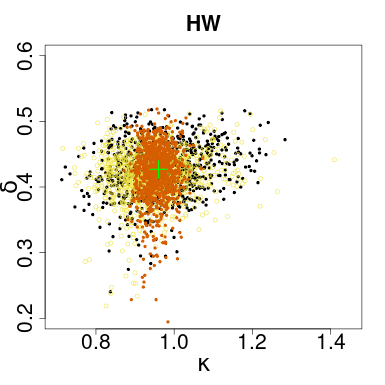} 
\end{minipage}
\caption{Empirical joint distribution with 1000 parameter estimates for (left) $r$-Pareto and (centre/right) HW processes, given replicates from a single test parameter set (green cross); left/centre and right panels plot estimates of $(\lambda,\kappa)'$ and $(\kappa,\delta)'$, respectively. Black points denote estimates from the likelihood inference scheme. Brown points are NBE-based estimates for processes observed on the larger domain $\mathcal{S}$; yellow points are NBE-based estimates for the HW process observed on the small domain $\mathcal{S}^*$.}
\label{fig:joint_dist2}
\end{figure}

Figure~\ref{fig:joint_dist2} presents the joint distribution of parameter estimates, for both the $r$-Pareto, and HW, processes, given a single test parameter vector with $1000$ sets of replicates. For the $r$-Pareto process, we observe that the NBE provides unbiased estimation of $\boldsymbol{\theta}=(\lambda,\kappa)'$, with a reduction in the variance of the estimator relative to the likelihood-based approach. For the HW process, we observe that, when considering $\mathcal{S}^*$, the joint distribution of parameter estimates is similar for NBE- and likelihood-based inference; when the size of $\mathcal{S}$ grows, a substantial reduction in estimator variance is observed. 

%%%%%%%%%%%%%%%%%%%%%%%%%%%%%%%%%
\subsection{Varying censoring threshold}
\label{sec:sim_vary}
We now investigate the amortised nature of NBEs, in particular their efficiency in handling variable censoring levels using a single estimator (trained only once); we constrain our focus to censoring levels $\tau \in [0.85,0.95]$, which is the typical range for such values in applications of spatial peaks-over-threshold models, and set $K=125,000$ training samples throughout. We consider inference for the IMSP and HW processes, defined over the $16\times 16$ grid $\mathcal{S}$, as considered in Section~\ref{sec:sim_fixed_FL} and with the same priors on their respective parameter sets. An estimator, accommodating variable $\tau$, is constructed using the scheme described in Section~\ref{sec:NBE_Vary}: the ``$\tau$-random'' NBE is trained by sampling $\tau$ from a ${\rm Unif}(0.85,0.95)$ prior. For testing, we construct $1000$ parameter sets with a fixed $\tau$ drawn randomly from ${\rm Unif}(0.85,0.95)$ and for comparison, we also train a NBE for this fixed $\tau$ (as in Section~\ref{sec:sim_fixed}), denoted the ``$\tau$-fixed" NBE. The marginal test risk is compared across the two different estimators, with the test value of $\tau$ being $0.893$ and $0.919$ for the IMSP and HW process, respectively. Note that the training parameter sets are the same across both types of NBE. The NBE architecture for $\tau$-fixed is given in Table~\ref{tab:CNN_arch}; the NBE architecture for $\tau$-random differs slightly to Table~\ref{tab:CNN_arch}, with an input dimension of $257$ in the first dense layer. 
\begin{table}[t]
\caption{Results for the variable $\tau$ simulation study detailed in Section~\ref{sec:sim_vary}. Marginal test risks (s.d.) under the absolute-error loss are provided separately for $\lambda$ ($\times 10^{-1}$), $\kappa$ ($\times 10^{-2}$) and $\delta$ ($\times 10^{-2}$). Test risks under the squared-error loss are presented in Table~\ref{suptab:sim_vary} of Appendix~\ref{supsec:arch} The lowest value in each column is given in bold. }
\vspace{5pt}
\centering
\begin{tabular}{lcc|ccc}
\hline
\hline
 & \multicolumn{2}{c|}{IMSP} & \multicolumn{3}{c}{HW}\\
  $\tau$& $\lambda$ & $\kappa$& $\lambda$ & $\kappa$ &$\delta$\\
 \hline
\multirow{1}{*}{\shortstack[l]{random}}  & \bf 2.97 (0.07) & 2.41 (0.06) &  \bf 2.62 (0.07) &  \bf 2.13 (0.05) & \bf 2.98 (0.09)\\

  \multirow{1}{*}{\shortstack[l]{fixed}}  & 3.03 (0.07) & \bf 2.11 (0.05) & 2.75 (0.06) & 2.41 (0.06) & 3.25 (0.10)\\
 \hline
\end{tabular}
\label{tab:sim_vary}
\end{table}
Table~\ref{tab:sim_vary} presents the marginal test risk estimates for all processes and estimators; these are evaluated for $\tau$ at the $\tau$-fixed level (see above). For the IMSP, the marginal tests risk for $\lambda$ and $\kappa$ are minimised by the $\tau$-random and $\tau$-fixed NBE, respectively; for the HW process, the marginal test risks for all parameters are minimised by the NBE trained with $\tau$ varying during training (see Section~\ref{sec:NBE_Vary}). Overall, the results in Table~\ref{tab:sim_vary} suggest that the $\tau$-random NBE often performs slightly better than the $\tau$-fixed NBE, although the differences in the marginal test risk are quite small and not always statistically significant. Therefore, in practice, both NBEs perform almost equally well. This has substantial implications for the generalisability of our estimator and highlights its amortised nature; a single estimator can be trained for a general $\tau$, rather than an estimator for every choice of $\tau$, which greatly reduces the computational expense of training.\par
\begin{figure}[t!]
\centering
\begin{minipage}{0.32\linewidth}
\centering
\includegraphics[width=\linewidth]{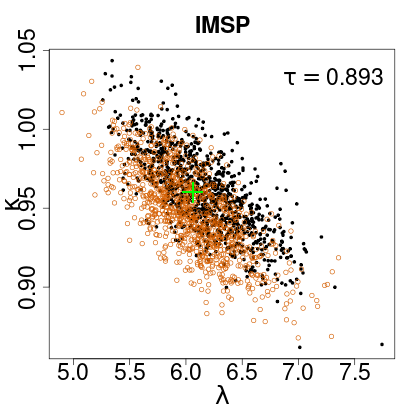} 
\end{minipage}
\vspace{-0.25cm}
\begin{minipage}{0.32\linewidth}
\centering
\includegraphics[width=\linewidth]{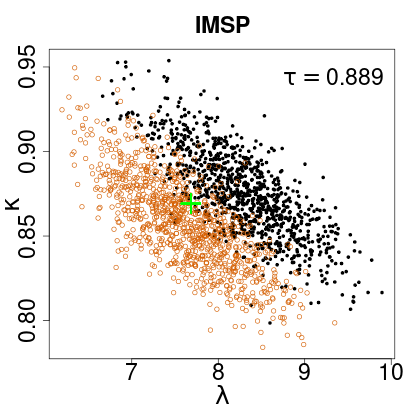} 
\end{minipage}
\begin{minipage}{0.32\linewidth}
\centering
\includegraphics[width=\linewidth]{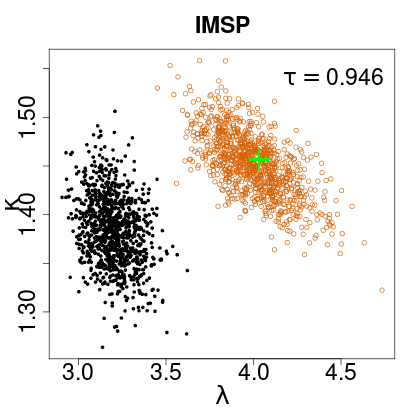} 
\end{minipage}
 \vspace{-0.25cm}
\begin{minipage}{0.32\linewidth}
\centering
\includegraphics[width=\linewidth]{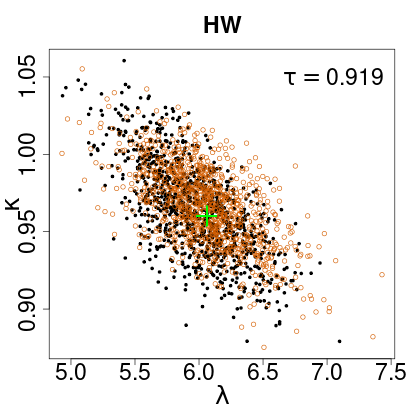} 
\end{minipage}
\begin{minipage}{0.32\linewidth}
\centering
\includegraphics[width=\linewidth]{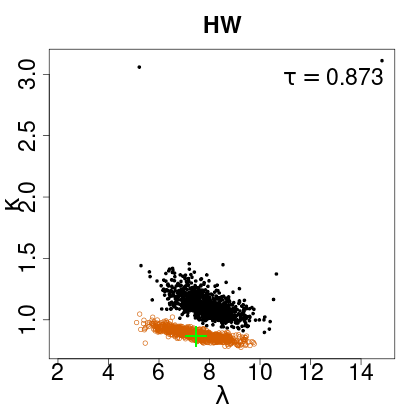} 
\end{minipage}
\begin{minipage}{0.32\linewidth}
\centering
\includegraphics[width=\linewidth]{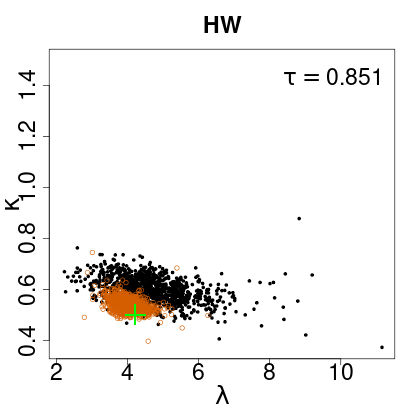} 
\end{minipage}

\begin{minipage}{0.32\linewidth}
\centering
\includegraphics[width=\linewidth]{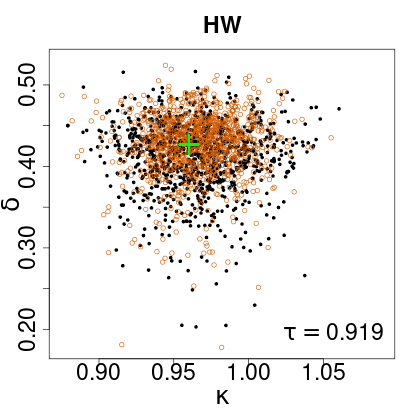} 
\end{minipage}
\begin{minipage}{0.32\linewidth}
\centering
\includegraphics[width=\linewidth]{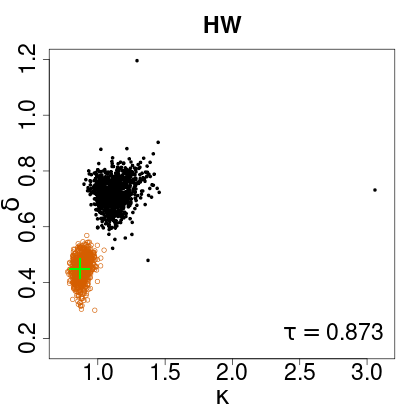} 
\end{minipage}
\begin{minipage}{0.32\linewidth}
\centering
\includegraphics[width=\linewidth]{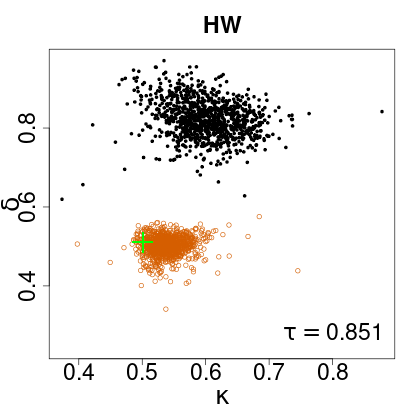} 
\end{minipage}

\caption{Empirical joint distribution with 1000 parameter estimates given $m=200$ replicates from a single test parameter set (green cross); top and centre rows give estimates of $(\lambda,\kappa)'$ for the IMSP and HW processes, respectively, and the bottom row illustrates $(\kappa,\delta)'$ for the HW process. Black and orange points denote estimates using a $\tau$-fixed and $\tau$-random NBE, respectively. Test parameter sets and censoring level $\tau$ differ across panels; note that $\tau$-fixed was trained for the fixed $\tau$ reported in panels in the first column.}
\label{fig:joint_dist3}
\end{figure}
To further evidence the efficacy of the new estimator,  we illustrate in Figure~\ref{fig:joint_dist3} the joint distribution of parameter estimates from the two NBEs, for three different test parameter sets and true censoring levels; results for three additional test sets are provided in Figure~\ref{supfig:joint_dist1} of Appendix~\ref{supsec:figs}. We observe good performance for $\tau$-random, the NBE trained for general $\tau$, across all test parameter sets and censoring levels; the estimates appear generally unbiased as the test parameter estimates are centred on the true parameter set, except in the cases where the true parameter set is close to the boundaries of the prior support (e.g., $\kappa$ close to 0.5); unsurprisingly, the efficacy of the $\tau$-fixed NBE is only comparable with the $\tau$-random NBE when its training $\tau$ is equal (or very close) to the true censoring level (the first column in Figure~\ref{fig:joint_dist3}). The $\tau$-fixed NBE cannot extrapolate to previously unseen censoring levels, and so it should be applied only with the specific $\tau$ used during training.   

%%%%%%%%%%%%%%%%%%%%%%%%%%%%%%%%%
%%%%%%%%%%%%%%%%%%%%%%%%%%%%%%%%%
\section{Application}
\label{sec:application}
Inhalation of ambient fine micro-scale particulate matter (${\rm PM}_{2.5}$) can lead to fatal respiratory and cardiovascular diseases \citep{polichetti2009effects, xing2016impact}; globally, poor air quality is responsible for millions of deaths per annum\  \citep{lelieveld2019cardiovascular}. A country that is particularly affected by poor air quality is Saudi Arabia, due in part to its heavy oil combustion rates, industrial and traffic emissions \citep{khodeir2012source}, and its arid climate, which is conducive to frequent dust storms \citep{alharbi2013march}. In order to mitigate the impacts of extreme air pollution in Saudi Arabia, an understanding of the spatial characteristics of extreme air pollution across the entire region is required; whilst studies of the spatial characteristics of ${\rm PM}_{2.5}$ across the Arabian Peninsula have been conducted by, for example, \cite{munir2017analysing} and \cite{tariq2022spatial}, they often focus on quantifying marginal trends or investigating the spatial profile of events through empirical methods, rather than considering extremal dependence. We provide the first analysis of the spatial extremal dependence of ${\rm PM}_{2.5}$ concentrations across Saudi Arabia using asymptotically-justified spatial extremal dependence models.\par
\begin{figure}[t!]
\centering
\begin{minipage}{0.48\linewidth}
\centering
\includegraphics[width=\linewidth]{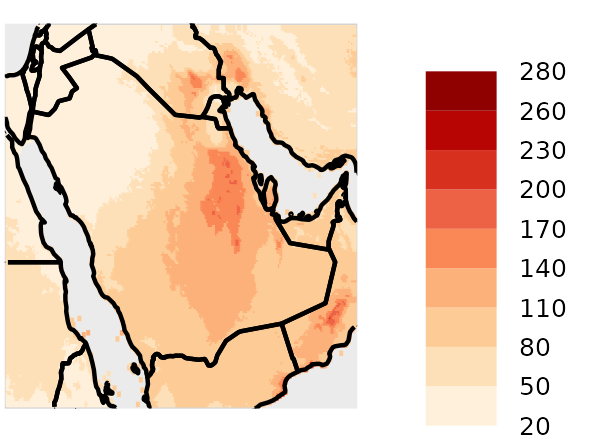} 
\end{minipage}
\hfill
\begin{minipage}{0.48\linewidth}
\centering
\includegraphics[width=\linewidth]{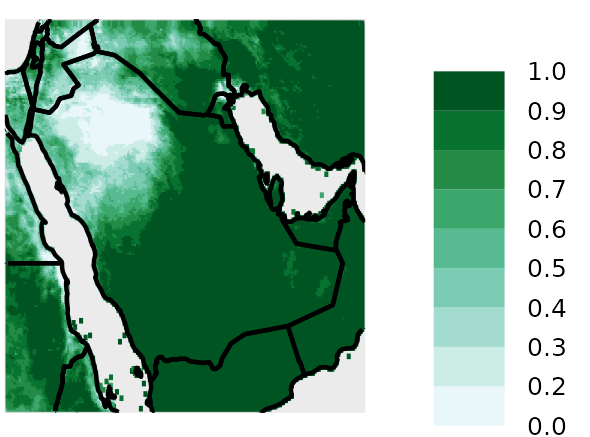} 
\end{minipage}

\caption{Observation of (left) surface average ${\rm PM}_{2.5}$  concentration ($\mu$g/m$^3$) for July 2012 and (right) its corresponding standardised field.}
\label{fig:obs}
\end{figure}
Data are taken from version 5 (V5.GL.02) of the global ground level product developed by \cite{van2021monthly}, which consists of monthly mean surface ${\rm PM}_{2.5}$ concentrations ($\mu$g/m$^3$) from 1998 to 2020 ($m=276$); the product is constructed by combining satellite data from a number of sources with ground level observations, through a geographically-weighted regression, and samples over land only. Our observations are arranged on a regular $242\times 182$ grid encompassing Saudi Arabia; each observation corresponds to the average over a $0.1^{\circ}$ by $0.1^{\circ}$ spatial grid-cell. As our interest is in modelling extremal dependence only, we use a simple rank transform to standardise the data to Unif$(0,1)$ margins; to account for spatial non-stationarity in the margins, we perform the transformation site-wise. The observed field for July 2012, which exhibits the largest pointwise value over the observation period, and its standardised values are presented in Figure~\ref{fig:obs}.\par
We assess extremal dependence in the data through fits of an anisotropic extension of the \cite{huser2019modeling} process. Dependence in our model is represented as a function of the great-circle distance between transformed sites,
\begin{equation}
\mathbf{s}^*=\begin{pmatrix}
1 & 0\\ 
0 & 1/\alpha \end{pmatrix}\begin{pmatrix}
\cos\omega & -\sin\omega \\
\sin\omega & \cos\omega \end{pmatrix}\mathbf{s},
\end{equation}
where $\omega \in [-\pi/2, 0]$ and $\alpha > 0$ control the rotation and coordinate stretching effect, respectively; hence inference requires estimation of $\boldsymbol{\theta}=(\lambda,\kappa,\delta,\alpha,\omega)'$, for $(\lambda,\kappa,\delta)'$ in Section~\ref{sec:scale_mix}. To account for spatial non-stationarity in the data, we perform local extremal dependence model fits. We train a single estimator for a $G \times G$ regular grid of sites, with $G \in \{4,8,16,24,32\}$, and apply it on all subsets of similarly arranged data; the value of $G$ determines the level of smoothing in maps of the parameter estimates. The estimator is only applied to subsets of data that contain no values sampled over water, and so we require between 16,393 ($G=32)$ and 35,589 ($G=4$) model fits for each $G$; {although only five estimators are trained for the analysis, they facilitate estimation, and parameter uncertainty assessment, for over 130,000 local spatial models.} Recall from Section~\ref{sec:scale_mix} that applications of the isotropic \cite{huser2019modeling} model using the full likelihood have been limited to low dimension $d \leq 30$ \citep[or $d\leq 200,$ in the special case of][]{zhang2022hierarchical}; here we consider much higher $d$ up to $32^2=1024$. \par
Each NBE is trained on a reference domain $\mathcal{S}$ taken to be a $G \times G$ regular grid of locations, which is centred within the spatial domain of interest, see Figure~\ref{fig:obs}. We fix $\tau =0.9$ throughout and consequently adopt the $\tau$-fixed estimator, as discussed in Section~\ref{sec:sim_vary}; we choose an NBE optimised for fixed rather than variable $\tau$ as the latter is more computationally expensive to train and the gains in its use are comparably slight (relative to the gains in using an NBE over the competing likelihood approach, see Table~\ref{tab:paper_sim_fixed}). We train each NBE using $\tilde{\mathbf{m}}=(46,138,276)$. To improve model training and reduce computational memory requirements, we adopt ``simulation-on-the-fly'' \citep[for an overview, see][]{chan2018likelihood,gerber2021fast, sainsbury2022fast}. For training, we begin with an initial $|\vartheta_{{\rm train}}|=K$ and $|\vartheta_{{\rm val}}|=K/5$ training and validation parameter sets, respectively. Before every $30$th epoch, the parameters are refreshed and new values are drawn from the prior and, at the end of every fifth epoch, new training data are refreshed using the current parameter sets. We use the maximum, computationally-feasible value of $K$, which, due to increasing computational expense, changes with the dimension $G$; we take $K$ equal to $750,000$, $330,000$, $100,000$ and $38,000$ for dimension $4$ or $8$, 16, 24 and 32, respectively. The architecture of each NBE is also dependent on $G$; these are given in Appendix~\ref{supsec:arch}. Parameter sets are a priori independent with priors $\lambda \sim {\rm Unif}(20,1250)$, ${\kappa\sim {\rm Unif}(0.1,4)}$, $\delta \sim{\rm Unif}(0,1)$, $\alpha\sim{\rm Unif}(0.5,3.5)$ and $\omega\sim{\rm Unif}(-\pi/2,0)$.  Although the various NBEs take up to 72 hours to train, their evaluation on observed datasets is extremely fast: a single evaluation takes approximately $1\times 10^{-3}$ and $4.1 \times 10^{-3}$ seconds for $G=4$ and $G=32$, respectively.  \par
For brevity, we constrain our focus to the case  $G=16$; results for all other values of $G$ are provided in Appendix~\ref{supsec:figs}. As the final activation function in each architecture is the identity (see Tables~\ref{tab:CNN_archG4}--\ref{tab:CNN_archG2432} of Appendix~\ref{supsec:arch}), the parameter estimates are not guaranteed to lie in their respective prior support; we simply truncate such values. Although we could choose an architecture that enforces parameter estimates to lie within their prior support, we found that training of the NBE was more numerically stable with our approach. Figures~\ref{supfig:joint_distG4}--\ref{supfig:joint_distG32} in Appendix~\ref{supsec:figs} provide the joint distribution of parameter estimates for four different test parameter sets and across all values of $G$. We observe good performance for the NBEs across all test parameter sets, except in cases where the test parameter set is on the boundary of the prior parameter space.\par
\begin{figure}[t!]
\centering
\begin{minipage}{0.32\linewidth}
\centering
\includegraphics[width=\linewidth]{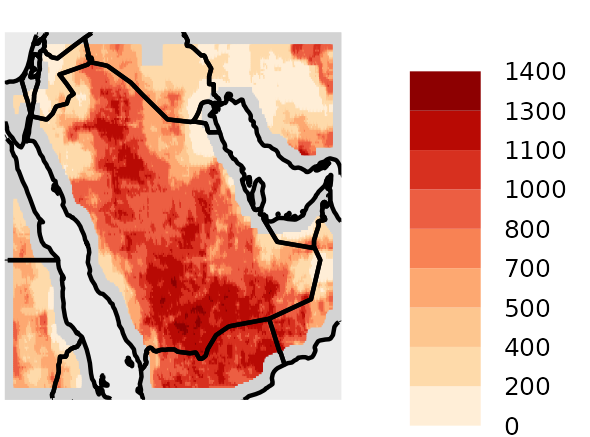} 
\end{minipage}
\begin{minipage}{0.32\linewidth}
\centering
\includegraphics[width=\linewidth]{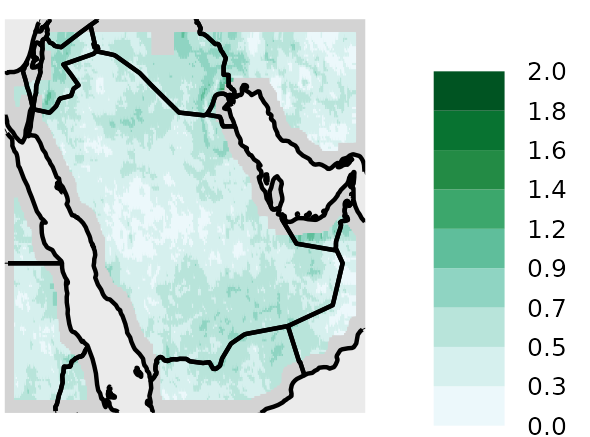} 
\end{minipage}
\begin{minipage}{0.32\linewidth}
\centering
\includegraphics[width=\linewidth]{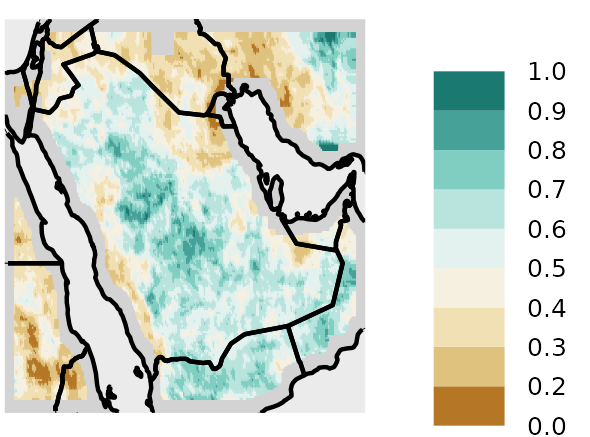} 
\end{minipage}
\begin{minipage}{0.32\linewidth}
\centering
\includegraphics[width=\linewidth]{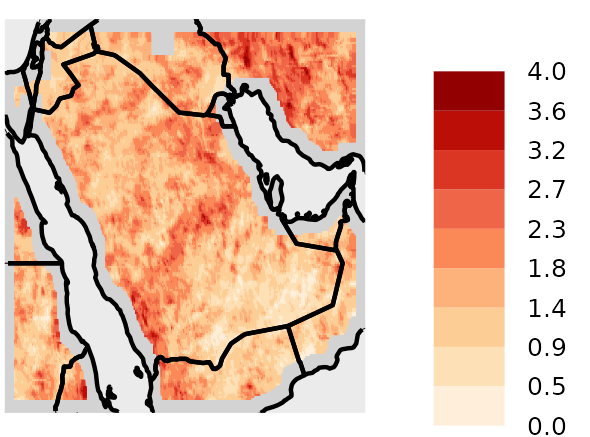} 
\end{minipage}
\begin{minipage}{0.32\linewidth}
\centering
\includegraphics[width=\linewidth]{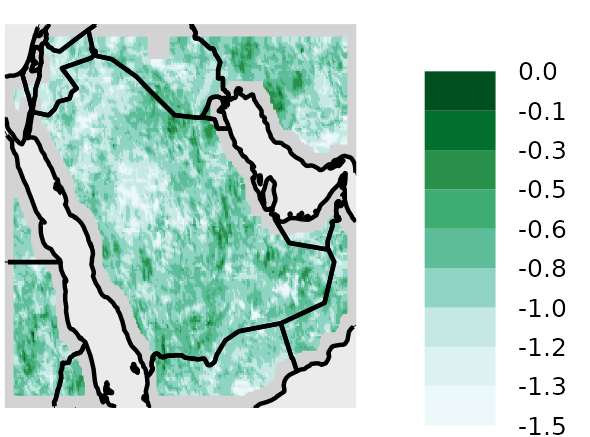} 
\end{minipage}
\caption{Parameter estimates for the HW process model with $G=16$. The top and bottom rows give estimates of dependence and anisotropy parameters, $(\lambda, \kappa, \delta)'$ and $(\alpha,\omega)'$, respectively. {Note that every pixel corresponds to a parameter estimate for a model fitted locally over the surrounding $G\times G$ grid of pixels.}}
\label{fig:results}
\end{figure}

We illustrate estimates of $(\lambda,\kappa,\delta)'$ and $(\alpha,\omega)'$ parameters for $G=16$ in Figure~\ref{fig:results}. As expected, we observe a clear increase in spatial smoothness of parameter estimates as $G$ increases. Estimates of $\lambda$ are high ($>600$km) across most of the Arabian Peninsula, suggesting strong spatial persistence in PM$_{2.5}$ extremes; smaller estimates of $\lambda$ are observed in  parts of Iran, Jordan and Oman, and close to the coastline of the Arabian Gulf, with particularly large estimates in the Empty Quarter to the south of Saudi Arabia and north-east of Yemen. Estimates of $\kappa$ illustrate the smoothness of the extremal process, with larger values in the Empty Quarter suggesting that the process here is smoother. Recall that the parameter $\delta$ determines if the process is asymptotically dependent ($\delta \geq 1/2$) or asymptotically independent ($\delta < 1/2$); Figure~\ref{fig:results} illustrates that Arabian PM$_{2.5}$ is composed of a mixture of locally asymptotically dependent, and asymptotically independent, processes. In central Saudi Arabia and Iran, the estimated process is asymptotically dependent whilst it appears to be increasingly asymptotically independent as we move towards the coast. For other countries such as Iraq, Jordan and Sudan, the estimated process exhibits asymptotic independence. For the anisotropy parameters, $(\alpha,\omega)'$, we observe in Figure~\ref{fig:results} larger estimates of the stretch parameter $\alpha$ along the coastline, and large parts of Iran. 
A directional effect on the extremal dependence structure corresponds to values of $\alpha\neq 1$; we observe strong directionality in the south of Yemen and Iraq, as well as the north of Saudi Arabia. 

Parameter uncertainty assessment is performed using a non-parametric bootstrap. Observed fields are sampled with replacement and used to obtain 1000 bootstrap parameter estimates. These are obtained relatively quickly as the estimators do not require re-training{; all model fits can indeed be performed using the same five trained estimators, each taking only a few millisecond to evaluate post-training}. For all $G$ and each model parameter, marginal $2.5\%$ and $97.5\%$ bootstrap quantile estimates are provided in Figures~\ref{supfig:bootG4}--\ref{supfig:bootG32} of Appendix~\ref{supsec:figs}. Note that upwards of 130 million individual extremal dependence models are estimated to perform our analysis {(using only the five trained estimators)}; to the best of our knowledge, this is a far greater number than any other study utilising the models described in Section~\ref{sec:spatextproc}, and shows that NBEs enable large-scale bootstrap studies with complex models that were impossible to run previously. {While here we use NBEs to fit a large ensemble of models, we still expect to see significant computational gains (compared to a likelihood-based approach) when using NBEs to fit a single model. For $G \geq 8$, fitting the \cite{huser2019modeling} process model using the full likelihood-based approach is expected to take substantially longer than the total training and evaluation time for a single NBE ($>72$ hours; see comparative computation times for $G=6$ in Section~\ref{sec:sim_fixed_FL}). }

%%%%%%%%%%%%%%%%%%%%%%%%%%%%%%%%%
%%%%%%%%%%%%%%%%%%%%%%%%%%%%%%%%%
\section{Discussion}
\label{sec:discussion}
We develop neural Bayes estimators (NBEs) for performing fast likelihood-free inference with popular peaks-over-threshold models, providing a paradigm shift challenging traditional statistical inference with censored data. Our novel approach to handling censored inputs for neural networks facilitates the construction of efficient estimators for a wide range of spatial extremal dependence models, including max-stable, $r$-Pareto, and random scale mixture processes, and we illustrate substantial gains in both the computational, and statistical, efficiency of NBEs relative to competing likelihood-based approaches. We have illustrated our proposed neural methodology with an application to Saudi Arabian ${\rm PM}_{2.5}$ concentrations, where we show that high-dimensional models can be fitted in a matter of milliseconds (post-training). The ability to fit hundreds-of-thousands of models accurately in a reasonable time frame allows us to study our dataset on a scale that is much higher than was previously possible with standard likelihood-based inference techniques, and we reveal new insights into the extremal dependence of air pollution across the Arabian Peninsula.  \par
The efficacy and speed of NBEs opens the door to fitting complex extreme-value processes to data of unprecedented size in an operational and online manner thanks to their amortised nature. While their training times seem long, the training phase needs to occur only once before a NBE can be reapplied, in perpetuity, for a negligible computational cost. Simulation of data during training can be costly, particularly when the size $K$ of the training data or number of sampling locations $d$ are large. Approximate methods for data simulation may alleviate this challenge.

In the simulation study in Section~\ref{sec:sim_vary}, we also demonstrate that it is possible to build a neural estimator for an arbitrary, user-specified censoring level. However, as we take the censoring threshold %$\mathbf{c}$ 
to be a function of a single probability $\tau$, our proposed framework does not cover site-specific or random censoring that may arise, for example, when modelling rainfall \citep{richards2022modelling,richards2023joint}. Estimation under general, user-specified censoring schemes is a future consideration.
 Our framework also has the constraint that $\mathcal{S}$ is a regular grid with fixed dimension. To increase the generality of our framework, this constraint should be removed as it implies that for new data observed over a different domain $\mathcal{S}$, we must retrain a brand new estimator from scratch. For gridded domains, we may be able to build a CNN-based estimator that handles arbitrary grid dimensions by exploiting the scalable CNNs developed by \cite{he2015spatial}, but this would still constrain $\mathcal{S}$ to be a regular grid. \cite{sainsbury2023neural} instead leverage graph neural networks \citep[see the review by][]{wu2020comprehensive} to build neural Bayes estimators which can be applied to irregular data and which are not wedded to a specific number $d$ or a specific configuration of spatial locations. Combining these estimators with our censoring framework is treated as future work. We also note that, whilst here we focus on inference for spatial processes with neural Bayes estimators, our neural estimator can be easily adapted to other application types which utilise left-, right-, or interval-censored data \citep[as, e.g., in survival data analysis,][]{klein2003survival}, or to other types of neural inference methods \citep[see the review by][]{zammit2024neural}, by adapting the architecture and inputs. \par
We did not provide a comparative study to assess the sensitivity of our neural Bayes estimators to the choice of neural network architecture; such a study, and theoretical results pertaining to this question, are left as future work. Our choice of architecture was based on that proposed by \cite{gerber2021fast} for data observed on a $16 \times 16$ grid; we made minor adjustments for different grid dimensions. With this architecture we were able to construct neural estimators that significantly outperform (in terms of both computational speed and estimation accuracy) the gold-standard likelihood-based inference approach. This suggests that this architecture might be suitable for many applications of a similar nature.

We have only considered stationary models. Non-stationary spatial  processes, such as those proposed by \cite{huser2016non}, \cite{zammit2020multi}, \cite{richards2021spatial}, and \cite{shao2022flexible}, provide a more realistic assessment of environmental risk and local impacts for data observed over complex domains. {However, as these models are often characterised by a high-dimensional parameter set, the training of NBEs in this context would require more complex neural network architectures and larger training datasets}; we here limit the parameter set dimension to five and, whilst this is representative for many stationary spatial extremal dependence models, some models, like the spatial conditional extremes model of \cite{wadsworth2022higher}, can have dozens of parameters \citep[see, e.g.,][]{richards2022modelling, richards2023joint}. Whilst \cite{sainsbury2022fast} illustrate inference of models with up to eight parameters, the extension of NBEs to higher dimensions remains the subject of future work. 

{Throughout, we have used uniform priors for model parameters; these are the most commonly used in practice and, by equipping them with a reasonably large support, they allow us to construct NBEs that can be widely applied in various applications. Similarly to traditional Bayes estimators, neural Bayes estimators are optimal with respect to the chosen prior measure, and are thus affected by the choice of prior. Finally, we note that we have here only considered settings where the process model is well-specified. Investigating the efficacy of our trained estimators under model misspecification \citep[see, e.g.,][]{radev2023jana, schmitt2023detecting} remains future work.}

%%%%%%%%%%%%%%%%%%%%%%%%%%%%%%%%%
%%%%%%%%%%%%%%%%%%%%%%%%%%%%%%%%%
\section*{Acknowledgments} This publication is based upon work supported by King Abdullah University of Science and Technology (KAUST) Research Funding under Awards No.\ ORA-2022-5336 and RFS-OFP2023-5550. Additionally, Jordan Richards and Rapha\"el Huser were supported by funding from KAUST under Award No.\ OSR-CRG2020-4394, while Matthew Sainsbury-Dale and Andrew Zammit-Mangion were supported by Australian Research Council (ARC) Discovery Early Career Research Award DE180100203. Matthew Sainsbury-Dale’s research was also supported by an Australian Government Research Training Program Scholarship and a 2022 Statistical Society of Australia (SSA) top-up scholarship. Support from the KAUST Supercomputing Laboratory is gratefully acknowledged.

\renewcommand{\theequation}{A.\arabic{equation}}
\renewcommand{\thefigure}{A\arabic{figure}}
\renewcommand{\thetable}{A\arabic{table}}
\renewcommand{\thesection}{A\arabic{section}}

\setcounter{figure}{0}
\setcounter{table}{0}
\setcounter{equation}{0}
\setcounter{theorem}{0}

\begin{appendix}
\section*{Appendix}
Appendix~\ref{supsec:proof} provides proof of Proposition~\ref{prop}. Appendix~\ref{supsec:invariance} provides a discussion and proof of the Bayes estimators' invariance to the distribution of the marginal censoring level. Appendix~\ref{supsec:model} provides further details for the spatial extremal dependence models that we consider in our simulation studies. Appendices~\ref{supsec:figs} and \ref{supsec:arch} provide supplementary figures and tables, respectively. 
\section{Proof of Proposition~\ref{prop} for the likelihood function of censored data}
\label{supsec:proof}
Here we provide a proof of Proposition~\ref{prop}, which derives the likelihood function of censored data. Recall Proposition~\ref{prop}, as stated in the main paper:
\Firstprop*

\begin{proof}
As in the main paper, we split the observation vector as $\mathbf{Z}_t=({\mathbf{Z}'_{t-}},{\mathbf{Z}'_{t+}})'$, its realised value as $\mathbf{z}_t=({\mathbf{z}'_{t-}},{\mathbf{z}'_{t+}})'$, and the censoring threshold vector as $\mathbf{c}_t=({\mathbf{c}}'_{t-},{\mathbf{c}}'_{t+})'$. Let ${\mathbf{i}_t=(i_{t,1},\dots,i_{t,d})}$ be the realised value of $\mathbf{I}_t$; that is, for $\mathbf{Z}_t$ arranged as $\mathbf{Z}_t=({\mathbf{Z}}'_{t-},{\mathbf{Z}}'_{t+})'$, we have ${\mathbf{i}_t=(\mathbf{1}'_{t-},\mathbf{0}'_{t+})'}$, where $\mathbf{1}_{t-}$ and $\mathbf{0}_{t+}$ are vectors of ones and zeros of dimensions $d_{t-}:= \sum^d_{j=1}i_{t,j}$ and $d_{t+}:= d - d_{t-}$, respectively, which denote the numbers of censored and uncensored values, respectively. Then, since $\mathbf{z}_{t+}>\mathbf{c}_{t+}$ by definition, we note that 
\begin{align}
\Pr\left(\mathbf{I}_t=\mathbf{i}_t,\;\mathbf{Z}_{t+}\leq \mathbf{z}_{t+}\right)&=\Pr\left(\mathbf{Z}_{t-}\leq \mathbf{c_{t-}},\;\mathbf{Z}_{t+}> \mathbf{c}_{t+},\;\mathbf{Z}_{t+}\leq \mathbf{z}_{t+}\right)\nonumber\\
&=\Pr\left(\mathbf{Z}_{t-}\leq \mathbf{c}_{t-},\;\mathbf{c}_{t+}<\mathbf{Z}_{t+}\leq \mathbf{z}_{t+}\right)\nonumber\\
&=\Pr\left(\mathbf{Z}_t\leq {\mathbf{b}}_t\right)-\Pr\left(\mathbf{Z}_t\leq \mathbf{b}^*_t\right),\label{eq:propproof}
\end{align}
where {$\mathbf{b}^*_t=(b^*_{t,1},\dots,b^*_{t,d})'$ and} ${\mathbf{b}}_t=({b}_{t,1},\dots,{b}_{t,d})'$ are vectors with components {$b^*_{t,j}=c_j$ if $z_{t,j} > c_j$ and $b^*_{t,j}=-\infty$, otherwise,} and ${b}_{t,j}=\max\{z_{t,j},c_j\}$ for $j=1,\dots,d$. Therefore, because the rightmost term in \eqref{eq:propproof} does not depend on $\mathbf{z}_{t+}$, the censored density/likelihood contribution (obtained by differentiating \eqref{eq:propproof} with respect to $\mathbf{z}_{t+}$) is 
\begin{align*}
{\partial^{d_{t+}}\over \partial \mathbf{z}_{t+}} \{\Pr\left(\mathbf{Z}_t\leq \mathbf{b}_t\right)-\Pr\left(\mathbf{Z}_t\leq \mathbf{b}^*_t\right)\}&={\partial^{|\mathcal{J}_t|}\over {\prod_{j \in \mathcal{J}_t}}\partial z_j} F(\mathbf{z}\mid \boldsymbol{\theta})\bigg|_{\mathbf{z}={\mathbf{b}}_t},
\end{align*}
which corresponds to the rightmost expression in the proposition. Leveraging the fact that the joint distribution function can be obtained as an integral of the joint density function, this censored likelihood contribution can be rewritten differently as
\begin{align*}
{\partial^{|\mathcal{J}_t|}\over {\prod_{j \in \mathcal{J}_t}}\partial z_j} F(\mathbf{z}\mid \boldsymbol{\theta})\bigg|_{\mathbf{z}={\mathbf{b}}_t}&=\int_{-\infty}^{\mathbf{c}_{t-}}f(\mathbf{z}_t\mid\boldsymbol{\theta})\mathrm{d}\mathbf{z}_{t-},
\end{align*}
which concludes the proof.
\end{proof}

\setcounter{theorem}{0}

\section{Invariance of Bayes estimator to the distribution of the censoring level}
\label{supsec:invariance}

Here we give the conditions for which the Bayes estimator is invariant to the distribution of the censoring level, $p(\tau)$. For ease of exposition, we assume that hereon all distributions admit a density with respect to Lebesgue measure. Notation used in the proof is defined in Section~\ref{sec:method}.

\begin{theorem}
Let $L: \Theta \times \Theta \to \mathbb{R}^{\geq 0}$ denote a strictly convex nonnegative loss function and let $p(\boldsymbol{\theta} \mid \mathbf{a}, \tau)$ denote the posterior density of $\boldsymbol{\theta}$ corresponding to the augmented data $\mathbf{A}={\mathbf{a}}$ and censoring level $\tau$. Assume that the Bayes estimate $\hat{\boldsymbol{\theta}}^\star$ has finite posterior expected loss  $\int_\Theta L(\boldsymbol{\theta}, \hat{\boldsymbol{\theta}}^\star)p(\boldsymbol{\theta} \mid \mathbf{a}, \tau)\mathrm{d}\boldsymbol{\theta}$ for all fixed {$\mathbf{a} \in \Omega_{\mathbf{A},\tau}$} and $\tau \in \mathcal{T}\subseteq [0,1)$. Then the Bayes estimator $\hat{\boldsymbol{\theta}}^\star(\mathbf{a}, \tau)$, $\mathbf{a}\in \Omega_{\mathbf{A},\tau}, \tau \in \mathcal{T}$, is invariant to the distribution of $\tau$, $p(\tau)$, provided that $p(\tau) > 0$ for $\tau \in \mathcal{T}$ and that $\tau$ and $\boldsymbol{\theta}$ are independent.
     \end{theorem}

\begin{proof}
The proof follows a similar argument to that of \citet[][Theorem~1]{sainsbury2023neural}. For a given $\tau \in \mathcal{T}$ and augmented data $\mathbf{a}\in \Omega_{\mathbf{A},\tau}$, the Bayes estimate $\hat{\boldsymbol{\theta}}^\star$ minimises the posterior expected loss, that is, 
\begin{equation}
\hat{\boldsymbol{\theta}}^\star = \argmin_{\hat{\boldsymbol{\theta}}} \int_\Theta L(\boldsymbol{\theta}, \hat{\boldsymbol{\theta}})p(\boldsymbol{\theta} \mid \mathbf{a}, \tau)\mathrm{d}\boldsymbol{\theta},\label{eq:Bayesestimate}
\end{equation}
and, by assumption, $\int_\Theta L(\boldsymbol{\theta}, \hat{\boldsymbol{\theta}}^\star)p(\boldsymbol{\theta} \mid \mathbf{a}, \tau)\mathrm{d}\boldsymbol{\theta} < \infty$. Consider now the Bayes estimator  $\hat{\boldsymbol{\theta}}^\star(\mathbf{a}, \tau)$ that returns the Bayes estimate for each $\mathbf{a} \in \Omega_{\mathbf{A},\tau}$ and censoring level $\tau \in \mathcal{T}$. Since the posterior expected loss is nonnegative, if $p(\tau) > 0$ for $\tau \in \mathcal{T}$ then it is also the case that% and minimised pointwise by the Bayes estimator, 
\begin{equation*}
  \hat{\boldsymbol{\theta}}^\star(\cdot) = 
 \argmin_{\hat{\boldsymbol{\theta}}(\cdot)}
  \int_\mathcal{T}\int_{\Omega_{\mathbf{A},\tau}}\left[\int_\Theta L(\boldsymbol{\theta}, \hat{\boldsymbol{\theta}}(\mathbf{a},\tau))p(\boldsymbol{\theta} \mid \mathbf{a}, \tau)\mathrm{d}\boldsymbol{\theta} \right]p(\mathbf{a} \mid \tau) p(\tau)\mathrm{d}\mathbf{a}\;\mathrm{d}\tau;
  \end{equation*}
see \citet{zammit2024neural} for details. Applying Bayes rule and assuming $\tau$ and $\boldsymbol{\theta}$ are independent yields Equation~\eqref{eq:Bayesrisk2} of the main text, thus completing the proof.
\end{proof}

\section{Spatial extremal dependence model details}
\label{supsec:model}
\subsection{Univariate extreme value distributions}
\label{supsec:uni}
We give here the forms of the generalised extreme value (GEV) and generalised Pareto (GPD) distributions used for modelling univariate extreme values; see \cite{coles2001introduction} for details.\par
The generalised extreme value GEV$(\mu,\sigma,\xi)$ distribution function is
 \begin{equation}
 F_{\rm GEV}(z\mid\mu,\sigma,\xi)= \begin{cases}\exp\left[-\left\{1+\xi\left(\frac{z-\mu}{\sigma}\right)\right\}_+^{-1/\xi}\right], \;\;&\xi\neq 0,\\
 \exp\left\{-\exp\left(-\frac{z-\mu}{\sigma}\right)\right\},\;\;&\xi=0
 ,\end{cases}
 \label{GEVcdf}
 \end{equation}
 with $\{y\}_+=\max\{0,y\}$, location, scale, and shape, parameters $\mu \in \mathbb{R},\sigma >0$, and $\xi \in \mathbb{R}$, respectively, and support $\{z \in \mathbb{R} : 1+\xi(z-\mu)/\sigma > 0\}$. \par
A GPD$(\sigma,\xi)$ random variable has distribution function $F_{ \rm{GPD}}(z)=(1+\xi z/\sigma)^{-1/\xi}$ and support $z\geq 0$ for $\xi \geq 0$ and $0 \leq z \leq -\sigma/\xi$ for $\xi < 0$.

\subsection{Mat\'ern Gaussian processes}
\label{supsec:GP}
Gaussian processes (GPs) have extensive use in classical spatial statistics due to their convenient statistical properties \citep[see][]{gelfand2016spatial}, but are not necessarily appropriate for modelling spatial extremes due to their inability to capture asymptotic dependence \citep{davison2013geostatistics}, that is, a non-zero limit probability of joint threshold exceedances at any two sites.  However, we still consider their inference as they have been previously applied to model spatial extremal dependence and, more importantly, they provide the foundation for many other spatial extremal dependence models \citep{sang2010continuous}, for example, MSPs and IMSPs (see Section~\ref{sec:MSP} of the main text), random scale/location mixtures (see Section~\ref{sec:scale_mix} of the main text) and the spatial conditional extremes model \citep{wadsworth2022higher}.\par
Unlike MSPs and IMSPs , the full $d$-dimensional density for a GP is tractable with computationally inexpensive evaluation, and many methods have been developed to facilitate their fast evaluation for large $d$ \citep[see, e.g.,][]{aune2014parameter}. However, under the censoring regime described in Section~\ref{sec:NBE_Censored} of the main text, evaluation of the full likelihood becomes computability expensive. Hence, for modelling of spatial extremes, inference for GPs usually proceeds with the same censored weighted pairwise likelihood (with cut-off distance $h_{max}$) as described in Section~\ref{sec:MSP} of the main text \citep[see, e.g.,][]{davison2013geostatistics, thibaud2013,richards2022modelling,richards2023joint}. Due to their popularity, we adopt censored pairwise likelihood estimators as the competing likelihood approach for MSPs, IMSPs, and GPs, in the simulation study detailed in Section~\ref{sec:sim} of the main text.\par
We constrain our focus to isotropic and stationary Mat\'ern GPs. Specifically, we consider Gaussian copula models where, after arbitrary marginal transformation, the distribution of the process observed at a finite collection of sites $\{\mathbf{s}_1,\dots,\mathbf{s}_d\}$ is zero mean, unit variance, $d$-variate Gaussian with the correlation between the process observed at sites $\mathbf{s}_i,\mathbf{s}_j \in \mathcal{S}$ determined by the isotropic Mat\'ern correlation function
\[
C(\mathbf{s}_i,\mathbf{s}_j)={2^{1-\kappa}}\{\Gamma(\kappa)\}^{-1}\left({\|\mathbf{s}_i-\mathbf{s}_j\|}/{\lambda}\right)K_\kappa\left({\|\mathbf{s}_i-\mathbf{s}_j\|}/{\lambda}\right),
\]
where $\Gamma(\cdot)$ is the gamma function, $K_\kappa(\cdot)$ is the Bessel function of the second kind of order $\kappa$, and $\lambda>0$ and $\kappa>0$ are range and smoothness parameters, respectively; with the marginal transformations assumed known, inference for these processes involves estimation of $\boldsymbol{\theta}=(\lambda,\kappa)'$. Similar processes were inferred using (uncensored) neural estimators by \cite{gerber2021fast} and \cite{sainsbury2022fast}.
\subsection{$r$-Pareto processes}
\label{sec:rPareto}
Whilst max-stable processes are limits for renormalised maxima, the generalised Pareto process describes the limit of renormalised exceedances, defined as $X\;|\;\{\sup_{\mathbf{s}\in\mathcal{S}}X(\mathbf{s})>u\}$ for large $u$, where the process $X(\cdot)$ has unit Pareto margins; these can be seen as analogues of the univariate generalised extreme value (GEV), and generalised Pareto (GPD), distributions respectively. For details on the univariate extreme value distributions, see Section~\ref{supsec:uni}. Following \cite{Buishand2008} and \cite{Ferreira2014}, consider the process
\begin{equation}
\label{eq:rPAreto}
\left[1+\xi(\mathbf{s})\left\{\frac{X(\mathbf{s})-b_m(\mathbf{s})}{a_m(\mathbf{s})}\right\}\right]^{1/\xi(\mathbf{s})}_+\mid \sup_{\mathbf{s}\in\mathcal{S}}\frac{X(\mathbf{s})-b_m(\mathbf{s})}{a_m(\mathbf{s})}>1,
\end{equation}
where $a_m$ and $b_m$ are as defined in Section~\ref{sec:MSP}, $\xi(\mathbf{s})\in\mathbb{R}$ is the associated GEV or GPD shape parameter (see Section~\ref{supsec:uni}) at site $\mathbf{s}\in\mathcal{S}$, and $z_+=\max\{0,z\}$; then if this process converges weakly, as $m\rightarrow \infty$, to a process with non-degenerate margins, this process is a generalised Pareto process. As with max-stable processes, we typically consider the process on standardised margins; a transformation of $X(\mathbf{s})$ to unit Pareto margins, denoted $\bar{X}(\mathbf{s})$, leads to a standard Pareto process with $\eqref{eq:rPAreto}$ becoming $\bar{X}(\mathbf{s})/m\;|\;\sup_{\mathbf{s}\in\mathcal{S}}\bar{X}(\mathbf{s})>m$; this same standardisation is used to create a max-stable process with unit Fr\'echet margins.\par
\cite{dombry2015functional} extend formulation \eqref{eq:rPAreto} by replacing the conditioning event with an exceedance of some functional of the entire process. They term these processes ``$\ell$-Pareto processes'', whilst later developments by \cite{de2018high,de2022functional} refer to them as ``$r$-Pareto processes''; we adopt the latter terminology. Consider a risk functional $r(\cdot)$, which is homogeneous of order 1 (i.e., $r(\cdot)$ satisfies, for any $c>0$, that $ r(cX)=c\,r(X)$), and takes in as input the entire process $\bar{X}$ and returns some scalar value; examples include $r(\bar{X})=\int_\mathcal{S}\bar{X}(\mathbf{s})\mathrm{d}\mathbf{s}$ or $r(\bar{X})=\sup_{\mathbf{s}\in\mathcal{S}}\bar{X}(\mathbf{s})$, or their finite-dimensional equivalents. An $r$-Pareto process, denoted $Z(\mathbf{s})$ and with standard unit Pareto marginals, is defined by $Z(\mathbf{s})=\lim_{r_0\rightarrow\infty}\bar{X}(\mathbf{s})/r_0\;|\;r(\bar{X})>r_0$. Similarly to \eqref{eq:MSP}, we can construct an $r$-Pareto process through the scale mixture $Z(\mathbf{s})=RW(\mathbf{s})$, but here $R$ is a unit Pareto random variable which is independent of the non-negative stochastic process $W(\mathbf{s})$, which satisfies $r(W)=1$ almost surely; hence we can relate each $r$-Pareto process to a max-stable counterpart. We constrain our focus to inference for $r$-Pareto processes that arise from log-Gaussian random functions, that is, with Brown--Resnick max-stable counterparts, as these are a popular choice \citep[see, e.g,][]{healy2021inference, bewentaore2022space}; however, we note that extremal-$t$ counterparts are an alternative \citep{thibaud2015efficient}.\par
Likelihood-based inference for $r$-Pareto processes follows with specification of a risk functional and use of a censored likelihood. In the case where $r(\bar{X})=\max_{i=1,\dots,d}\bar{X}(\mathbf{s}_i)$, the full $d$-dimensional censored likelihood can be represented as a censored Poisson process likelihood \citep{wadsworth2014efficient,Asadi2015}; hereon we use this risk functional for inference. Evaluation of the full censored likelihood is computationally troublesome as it requires evaluation of $d_{-}$-variate Gaussian distributions for $1\leq d_{-}\leq d-1$. To circumvent this issue, \cite{de2018high} detail a quasi-Monte Carlo algorithm to approximate these integrals and hasten computations, see, also, \cite{genz2009computation}; we take their approach to inference as the competing likelihood-based approach in Section~\ref{sec:sim} of the main text. Inference requires estimation of the Brown--Resnick variogram parameters described in Section~\ref{sec:MSP} of the main text.
%% We note that \cite{de2018high} propose an alternative approach to inference that looks to minimise the gradient score function, rather than a likelihood; this approach is computationally cheaper than the likelihood-based alternative and is applicable for any risk functional $r$, rather than just the maxima. However, it seeks to overcome the bias issues, described in Section~\ref{sec:NBE_Censored}, through application of an appropriate, continuous weighting function to the observations, rather than threshold censoring, and it does not utilise the likelihood; hence it is not truly compatible with the censored likelihood-based inference framework that we wish to emulate with NBEs, and so we do not consider it in our simulation studies.

\subsection{Computational details}
\label{supsec:comp}
For full details on the software used for likelihood-based inference and simulation of the considered spatial extremal processes, see \cite{belzile2022modeler}.\par
To simulate from the Brown--Resnick process, we use the $\texttt{SpatialExtremes}$ package in $\texttt{R}$ \citep{spatialextremes}; realisations from the associated inverted Brown--Resnick process are computed by taking the marginal reciprocal. Simulation of the Brown--Resnick $r$-Pareto processes and their inference with the full $d$-dimensional censored likelihood is conducted using the \texttt{mvPOT} package \citep{de2018mvpot}. For the HW process, both simulation and inference are conducted using the \texttt{R} package \texttt{SpatialADAI} \citep{spatialADAI}, with inference using the quasi-Monte Carlo approximation requiring the  \texttt{mvPOT} package \citep{de2018mvpot}.
\newpage
\section{Supplementary figures}
\label{supsec:figs}
\begin{figure}[h!]
\centering
\begin{minipage}{0.32\linewidth}
\centering
\includegraphics[width=\linewidth]{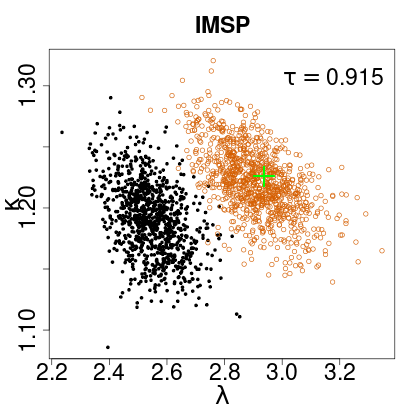} 
\end{minipage}
\begin{minipage}{0.32\linewidth}
\centering
\includegraphics[width=\linewidth]{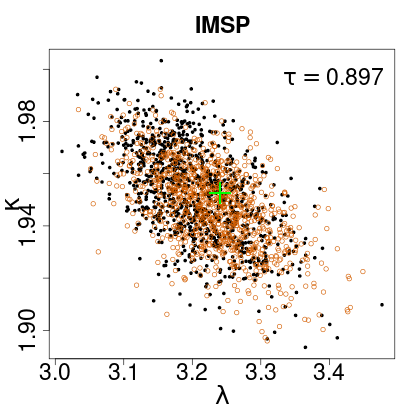} 
\end{minipage}
\begin{minipage}{0.32\linewidth}
\centering
\includegraphics[width=\linewidth]{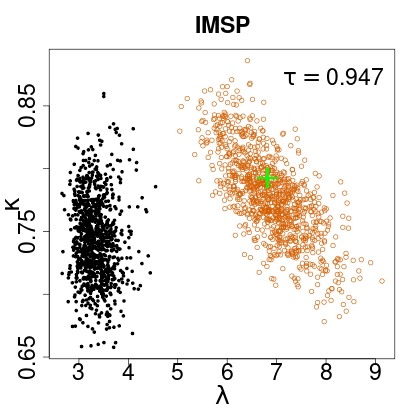} 
\end{minipage}

\begin{minipage}{0.32\linewidth}
\centering
\includegraphics[width=\linewidth]{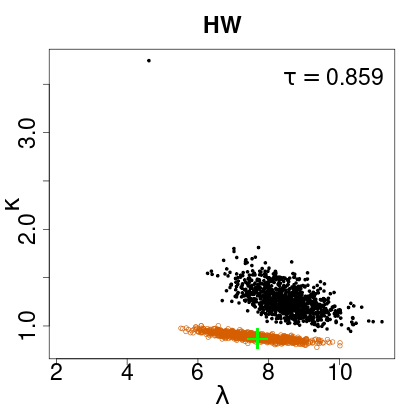} 
\end{minipage}
\begin{minipage}{0.32\linewidth}
\centering
\includegraphics[width=\linewidth]{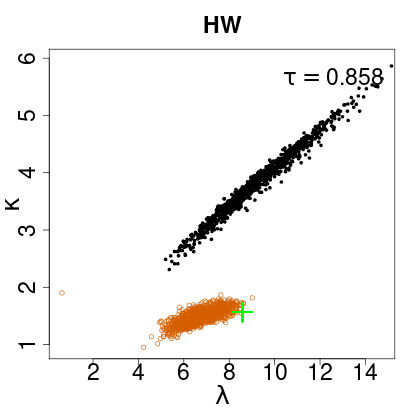} 
\end{minipage}
\begin{minipage}{0.32\linewidth}
\centering
\includegraphics[width=\linewidth]{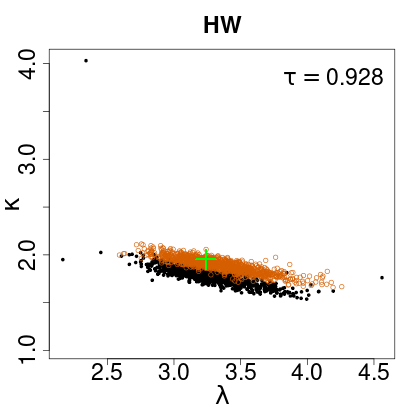} 
\end{minipage}

\begin{minipage}{0.32\linewidth}
\centering
\includegraphics[width=\linewidth]{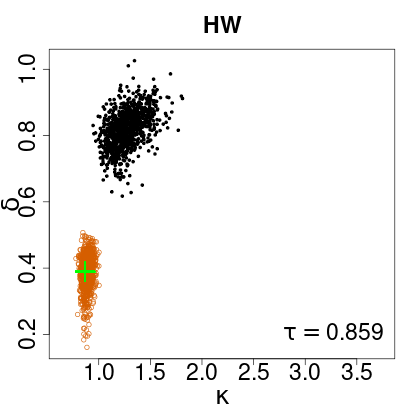} 
\end{minipage}
\begin{minipage}{0.32\linewidth}
\centering
\includegraphics[width=\linewidth]{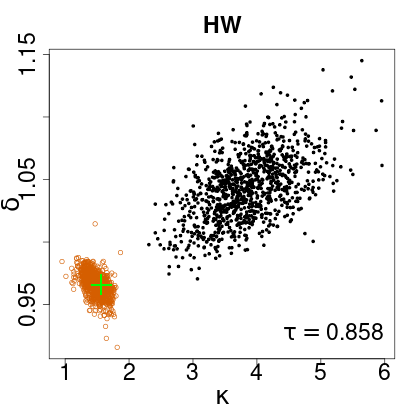} 
\end{minipage}
\begin{minipage}{0.32\linewidth}
\centering
\includegraphics[width=\linewidth]{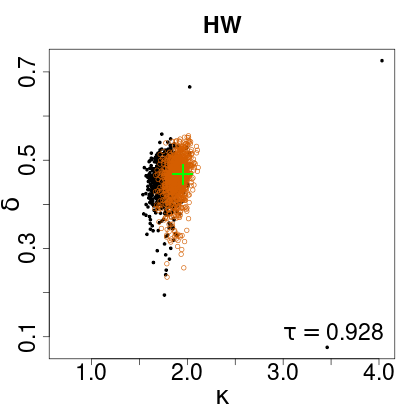} 
\end{minipage}

\caption{Joint distribution of 1000 parameter estimates given $m=200$ replicates from a single test parameter set (green cross); top and centre rows give estimates of $(\lambda,\kappa)'$ for the IMSP and HW processes, respectively, and the bottom row illustrates $(\kappa,\delta)'$ for the HW process. Black, orange and blue points denote estimates using a $\tau$-fixed, $\tau$-random and $\tau$-sequence NBE, respectively (see Section~\ref{sec:sim_vary} of the main text). The test parameter sets and censoring level $\tau$ differ across panels.}
\label{supfig:joint_dist1}
\end{figure}

\begin{figure}
\centering

\begin{minipage}{0.32\linewidth}
\centering
\includegraphics[width=\linewidth]{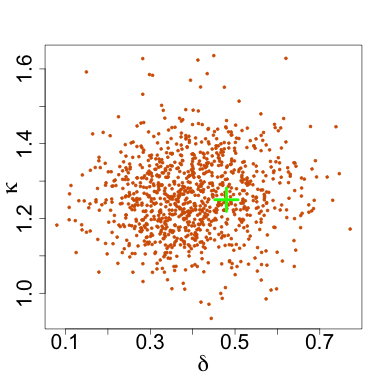} 
\end{minipage}
\begin{minipage}{0.32\linewidth}
\centering
\includegraphics[width=\linewidth]{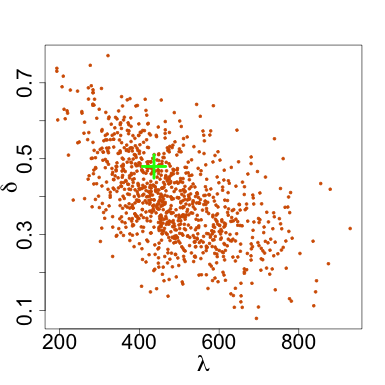} 
\end{minipage}
\begin{minipage}{0.32\linewidth}
\centering
\includegraphics[width=\linewidth]{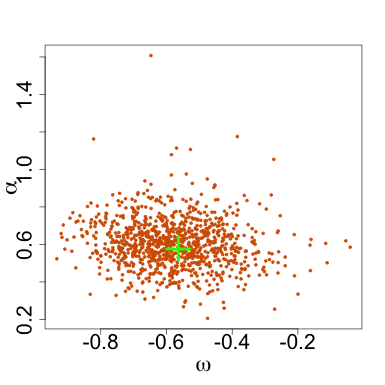} 
\end{minipage}
\begin{minipage}{0.32\linewidth}
\centering
\includegraphics[width=\linewidth]{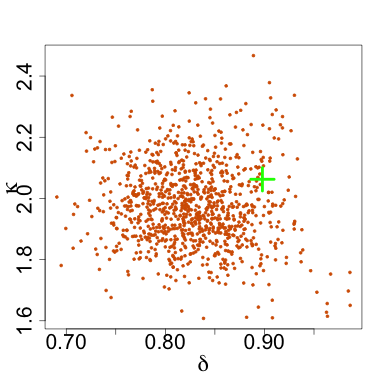} 
\end{minipage}
\begin{minipage}{0.32\linewidth}
\centering
\includegraphics[width=\linewidth]{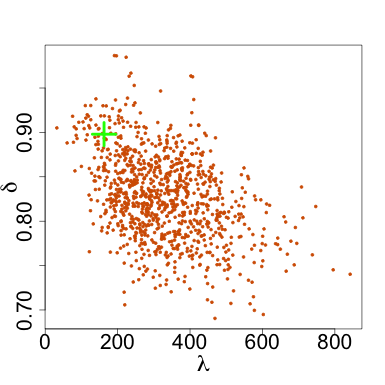} 
\end{minipage}
\begin{minipage}{0.32\linewidth}
\centering
\includegraphics[width=\linewidth]{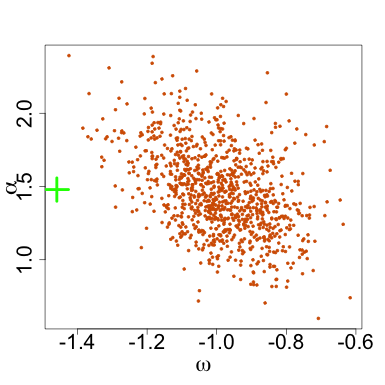} 
\end{minipage}

\begin{minipage}{0.32\linewidth}
\centering
\includegraphics[width=\linewidth]{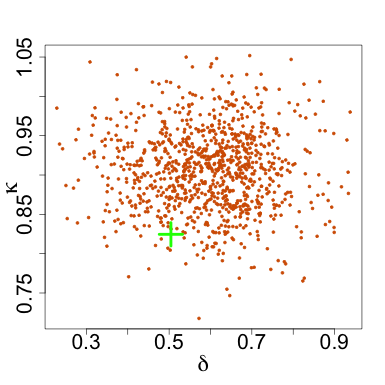} 
\end{minipage}
\begin{minipage}{0.32\linewidth}
\centering
\includegraphics[width=\linewidth]{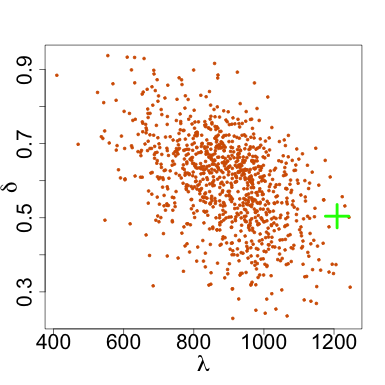} 
\end{minipage}
\begin{minipage}{0.32\linewidth}
\centering
\includegraphics[width=\linewidth]{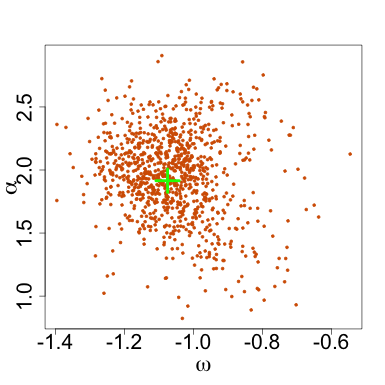} 
\end{minipage}
\begin{minipage}{0.32\linewidth}
\centering
\includegraphics[width=\linewidth]{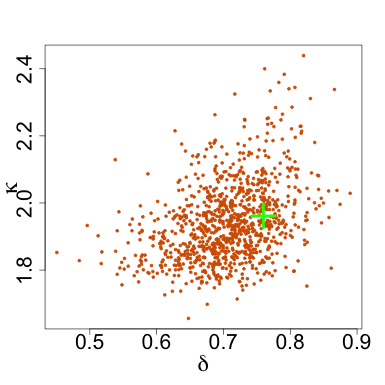} 
\end{minipage}
\begin{minipage}{0.32\linewidth}
\centering
\includegraphics[width=\linewidth]{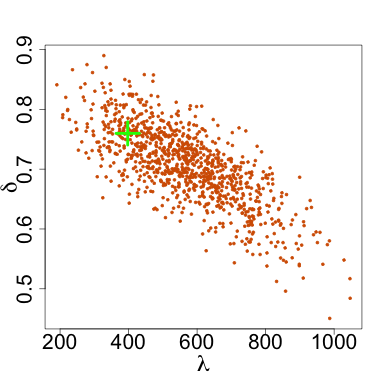} 
\end{minipage}
\begin{minipage}{0.32\linewidth}
\centering
\includegraphics[width=\linewidth]{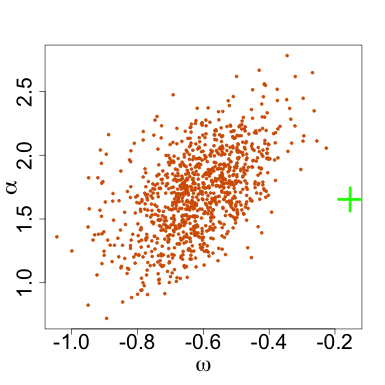} 
\end{minipage}
\caption{Joint distribution of 1000 parameter estimates (orange) when $G=4$ in Section~\ref{sec:application} of the main text. Four test parameter sets (green) are drawn randomly from the prior and differ across rows. The left, centre and right columns give estimates of $(\delta, \kappa)'$, $(\lambda,\delta)'$ and $(\omega,\alpha)'$, respectively. }
\label{supfig:joint_distG4}
\end{figure}

\begin{figure}
\centering
\begin{minipage}{0.32\linewidth}

\includegraphics[width=\linewidth]{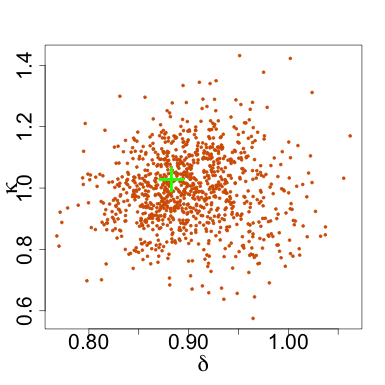} 
\end{minipage}
\begin{minipage}{0.32\linewidth}
\centering
\includegraphics[width=\linewidth]{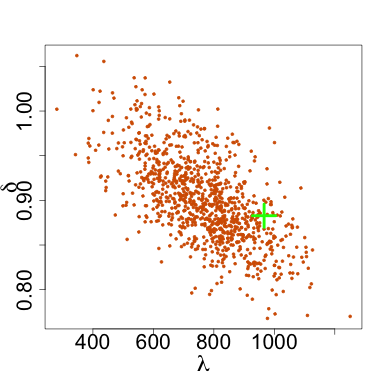} 
\end{minipage}
\begin{minipage}{0.32\linewidth}
\centering
\includegraphics[width=\linewidth]{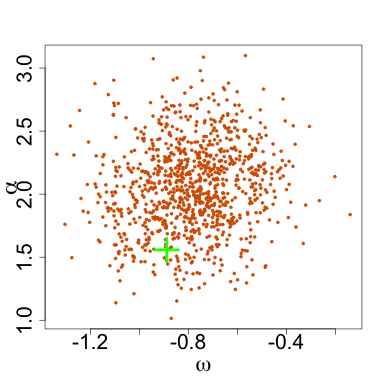} 
\end{minipage}
\begin{minipage}{0.32\linewidth}
\centering
\includegraphics[width=\linewidth]{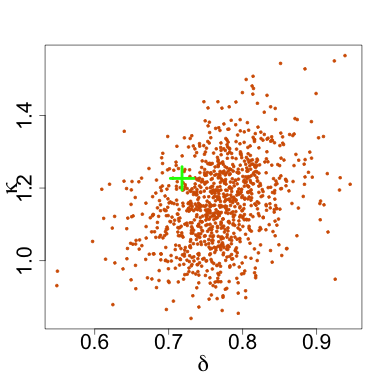} 
\end{minipage}
\begin{minipage}{0.32\linewidth}
\centering
\includegraphics[width=\linewidth]{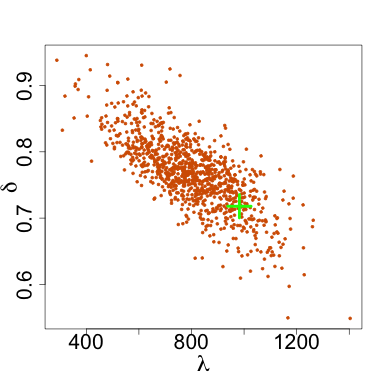} 
\end{minipage}
\begin{minipage}{0.32\linewidth}
\centering
\includegraphics[width=\linewidth]{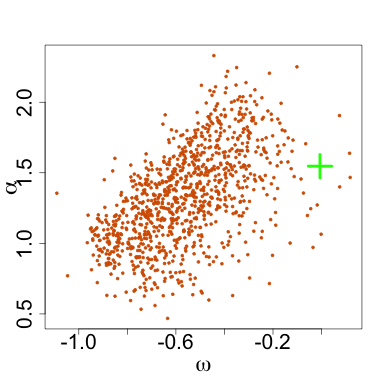} 
\end{minipage}

\begin{minipage}{0.32\linewidth}
\centering
\includegraphics[width=\linewidth]{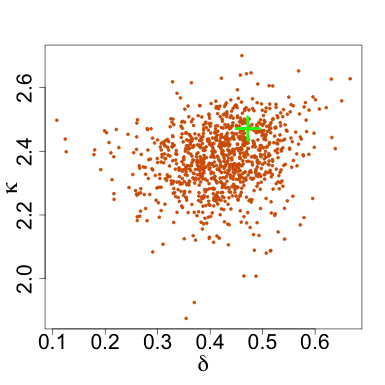} 
\end{minipage}
\begin{minipage}{0.32\linewidth}
\centering
\includegraphics[width=\linewidth]{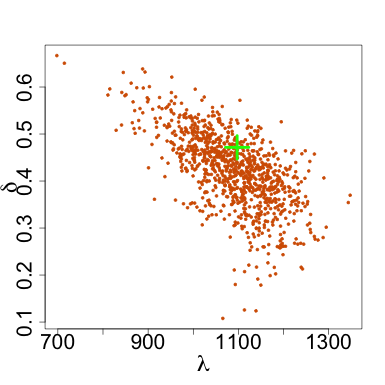} 
\end{minipage}
\begin{minipage}{0.32\linewidth}
\centering
\includegraphics[width=\linewidth]{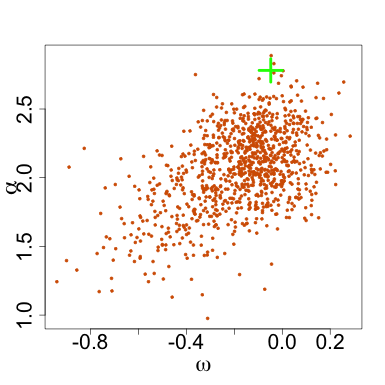} 
\end{minipage}
\begin{minipage}{0.32\linewidth}
\centering
\includegraphics[width=\linewidth]{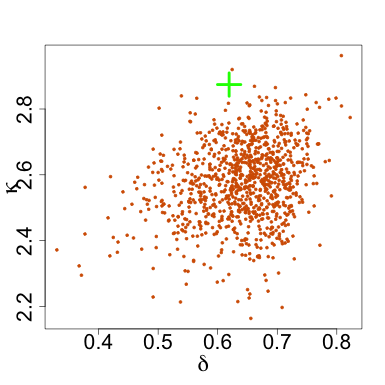} 
\end{minipage}
\begin{minipage}{0.32\linewidth}
\centering
\includegraphics[width=\linewidth]{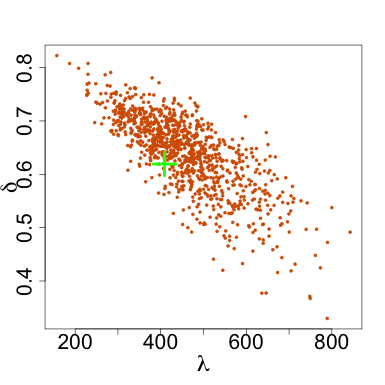} 
\end{minipage}
\begin{minipage}{0.32\linewidth}
\centering
\includegraphics[width=\linewidth]{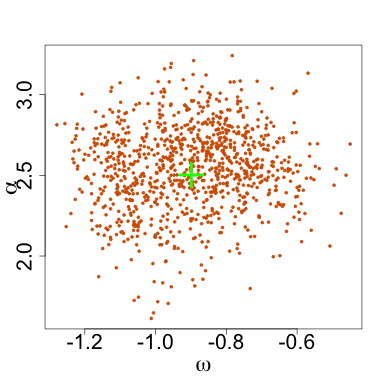} 
\end{minipage}
\caption{Joint distribution of 1000 parameter estimates (orange) when $G=8$ in Section~\ref{sec:application} of the main text. Four test parameter sets (green) are drawn randomly from the prior and differ across rows. The left, centre and right columns give estimates of $(\delta, \kappa)'$, $(\lambda,\delta)'$ and $(\omega,\alpha)'$, respectively. }
\label{supfig:joint_distG8}
\end{figure}

\begin{figure}
\centering

\begin{minipage}{0.32\linewidth}
\centering
\includegraphics[width=\linewidth]{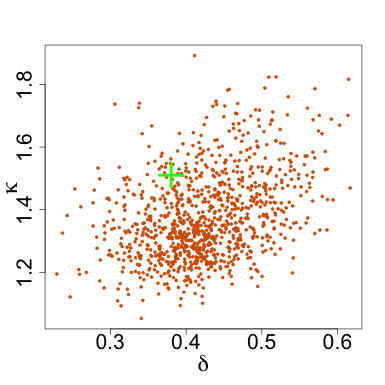} 
\end{minipage}
\begin{minipage}{0.32\linewidth}
\centering
\includegraphics[width=\linewidth]{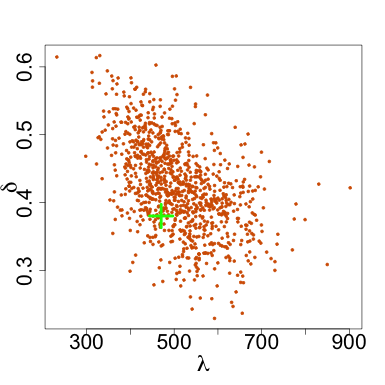} 
\end{minipage}
\begin{minipage}{0.32\linewidth}
\centering
\includegraphics[width=\linewidth]{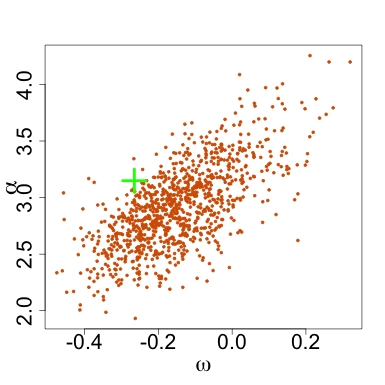} 
\end{minipage}
\begin{minipage}{0.32\linewidth}
\centering
\includegraphics[width=\linewidth]{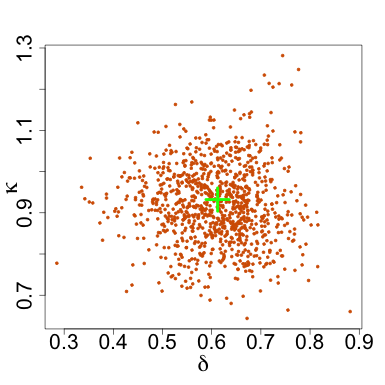} 
\end{minipage}
\begin{minipage}{0.32\linewidth}
\centering
\includegraphics[width=\linewidth]{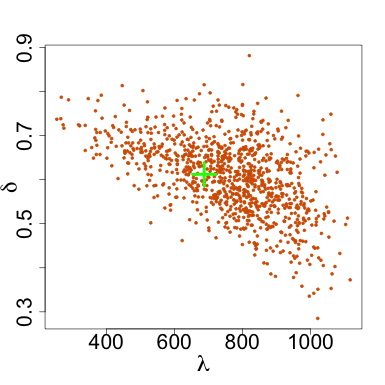} 
\end{minipage}
\begin{minipage}{0.32\linewidth}
\centering
\includegraphics[width=\linewidth]{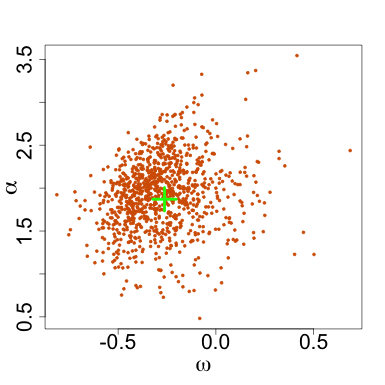} 
\end{minipage}

\begin{minipage}{0.32\linewidth}
\centering
\includegraphics[width=\linewidth]{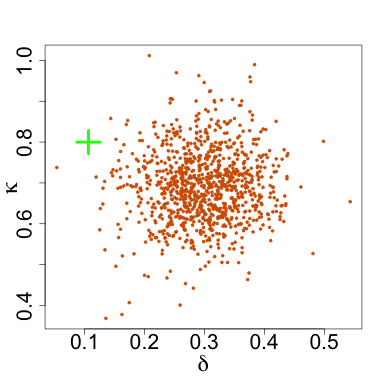} 
\end{minipage}
\begin{minipage}{0.32\linewidth}
\centering
\includegraphics[width=\linewidth]{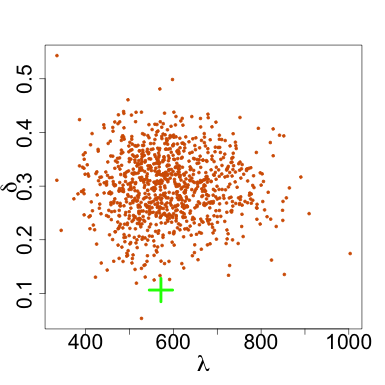} 
\end{minipage}
\begin{minipage}{0.32\linewidth}
\centering
\includegraphics[width=\linewidth]{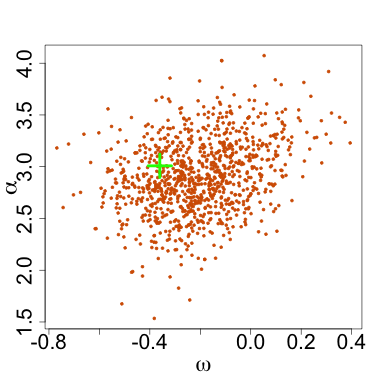} 
\end{minipage}
\begin{minipage}{0.32\linewidth}
\centering
\includegraphics[width=\linewidth]{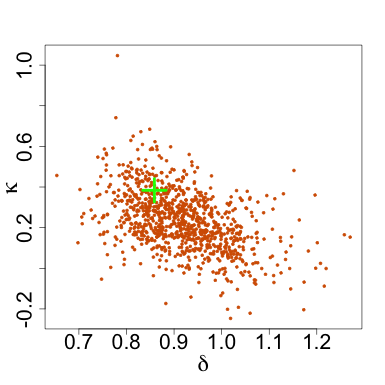} 
\end{minipage}
\begin{minipage}{0.32\linewidth}
\centering
\includegraphics[width=\linewidth]{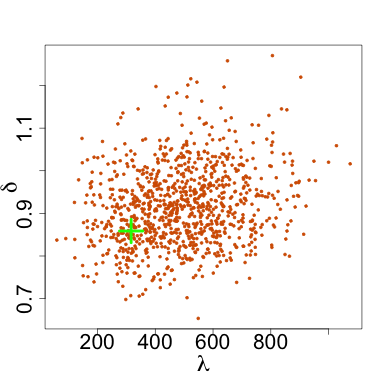} 
\end{minipage}
\begin{minipage}{0.32\linewidth}
\centering
\includegraphics[width=\linewidth]{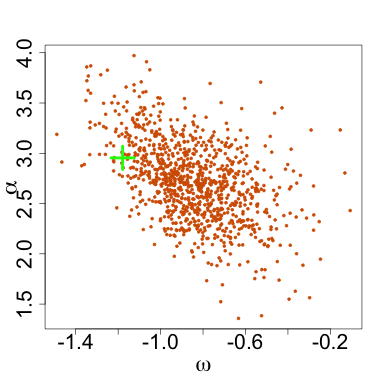} 
\end{minipage}
\caption{Joint distribution of 1000 parameter estimates (orange) when $G=16$ in Section~\ref{sec:application} of the main text. Four test parameter sets (green) are drawn randomly from the prior and differ across rows. The left, centre and right columns give estimates of $(\delta, \kappa)'$, $(\lambda,\delta)'$ and $(\omega,\alpha)'$, respectively. }
\label{supfig:joint_distG16}
\end{figure}

\begin{figure}
\centering

\begin{minipage}{0.32\linewidth}
\centering
\includegraphics[width=\linewidth]{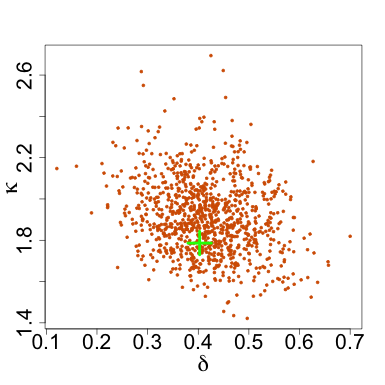} 
\end{minipage}
\begin{minipage}{0.32\linewidth}
\centering
\includegraphics[width=\linewidth]{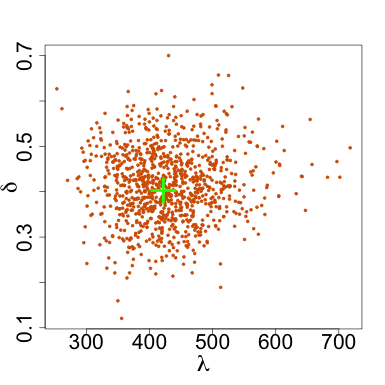} 
\end{minipage}
\begin{minipage}{0.32\linewidth}
\centering
\includegraphics[width=\linewidth]{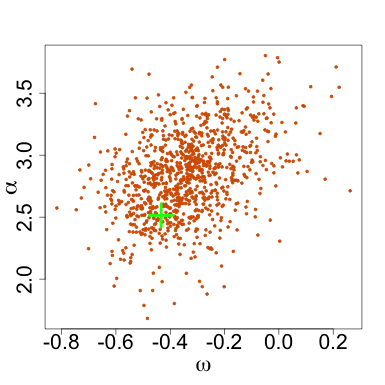} 
\end{minipage}
\begin{minipage}{0.32\linewidth}
\centering
\includegraphics[width=\linewidth]{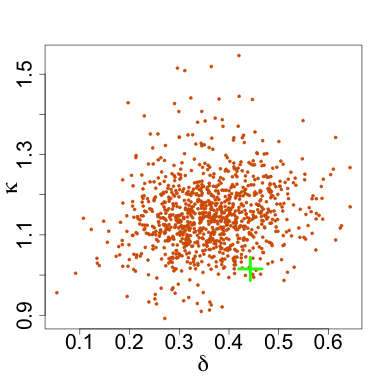} 
\end{minipage}
\begin{minipage}{0.32\linewidth}
\centering
\includegraphics[width=\linewidth]{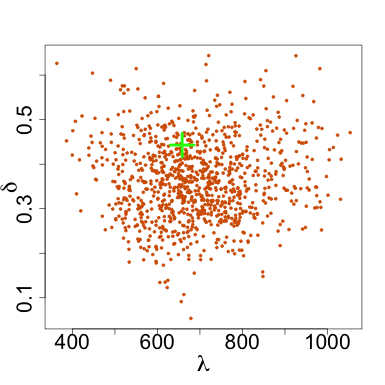} 
\end{minipage}
\begin{minipage}{0.32\linewidth}
\centering
\includegraphics[width=\linewidth]{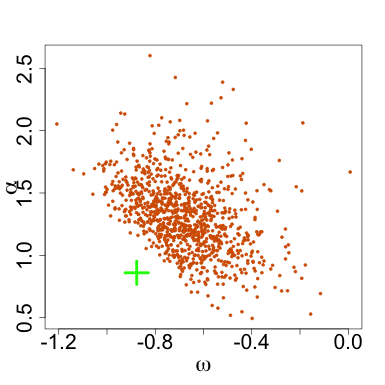} 
\end{minipage}

\begin{minipage}{0.32\linewidth}
\centering
\includegraphics[width=\linewidth]{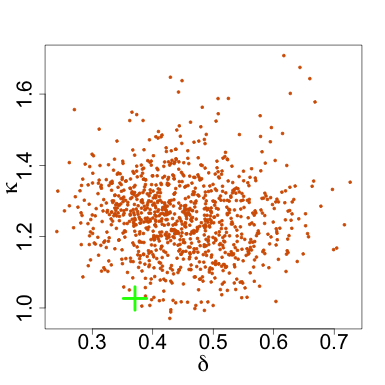} 
\end{minipage}
\begin{minipage}{0.32\linewidth}
\centering
\includegraphics[width=\linewidth]{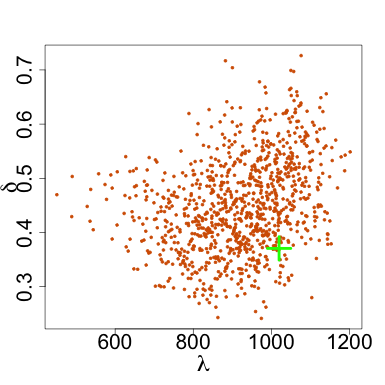} 
\end{minipage}
\begin{minipage}{0.32\linewidth}
\centering
\includegraphics[width=\linewidth]{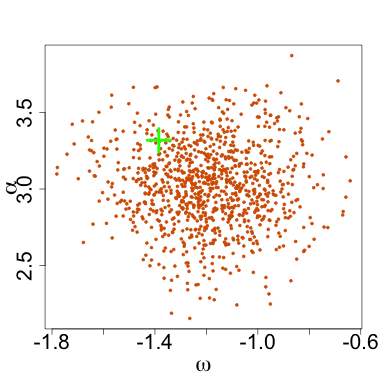} 
\end{minipage}
\begin{minipage}{0.32\linewidth}
\centering
\includegraphics[width=\linewidth]{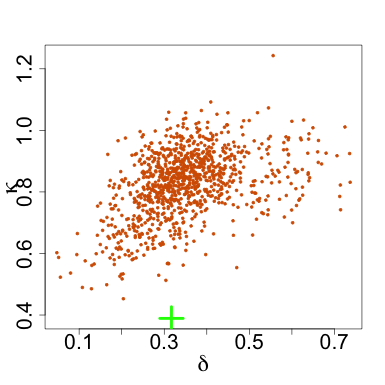} 
\end{minipage}
\begin{minipage}{0.32\linewidth}
\centering
\includegraphics[width=\linewidth]{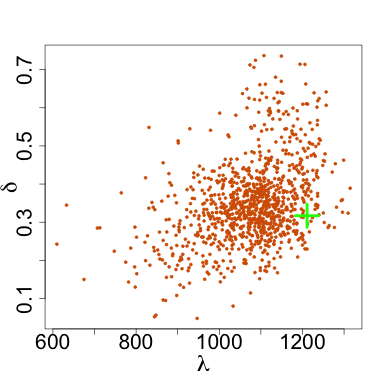} 
\end{minipage}
\begin{minipage}{0.32\linewidth}
\centering
\includegraphics[width=\linewidth]{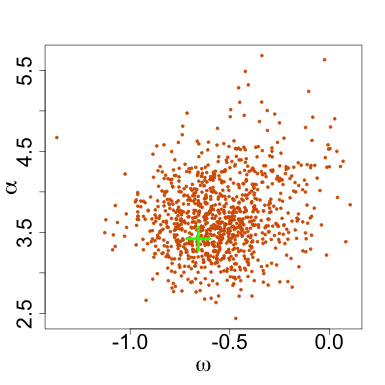} 
\end{minipage}
\caption{ Joint distribution of 1000 parameter estimates (orange) when $G=24$ in Section~\ref{sec:application} of the main text. Four test parameter sets (green) are drawn randomly from the prior and differ across rows. The left, centre and right columns give estimates of $(\delta, \kappa)'$, $(\lambda,\delta)'$ and $(\omega,\alpha)'$, respectively. }
\label{supfig:joint_distG24}
\end{figure}

\begin{figure}
\centering

\begin{minipage}{0.32\linewidth}
\centering
\includegraphics[width=\linewidth]{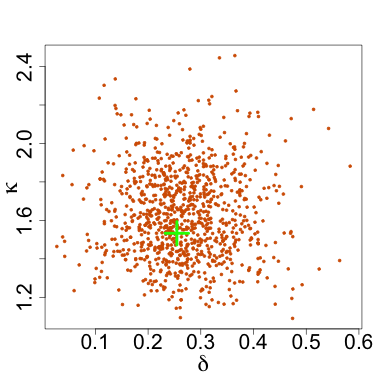} 
\end{minipage}
\begin{minipage}{0.32\linewidth}
\centering
\includegraphics[width=\linewidth]{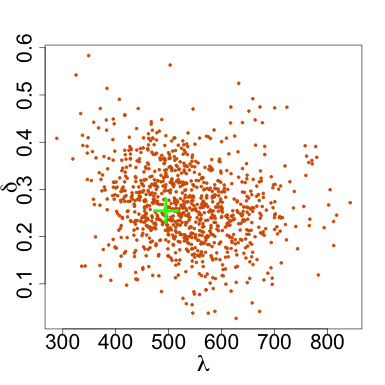} 
\end{minipage}
\begin{minipage}{0.32\linewidth}
\centering
\includegraphics[width=\linewidth]{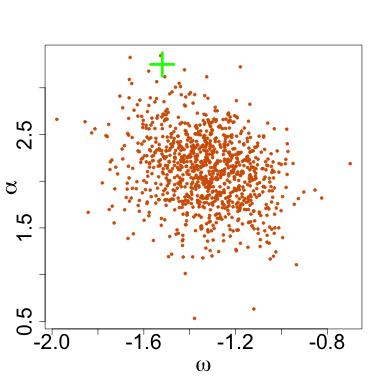} 
\end{minipage}
\begin{minipage}{0.32\linewidth}
\centering
\includegraphics[width=\linewidth]{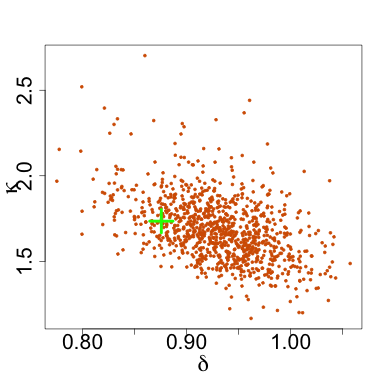} 
\end{minipage}
\begin{minipage}{0.32\linewidth}
\centering
\includegraphics[width=\linewidth]{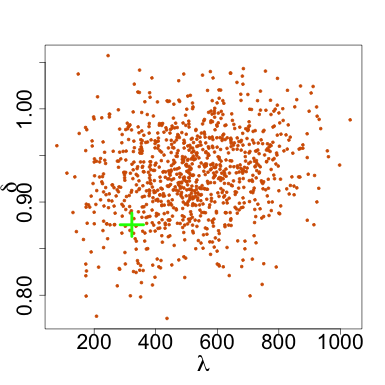} 
\end{minipage}
\begin{minipage}{0.32\linewidth}
\centering
\includegraphics[width=\linewidth]{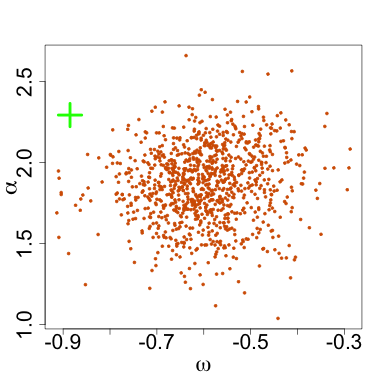} 
\end{minipage}

\begin{minipage}{0.32\linewidth}
\centering
\includegraphics[width=\linewidth]{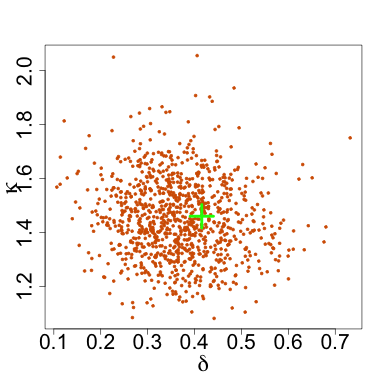} 
\end{minipage}
\begin{minipage}{0.32\linewidth}
\centering
\includegraphics[width=\linewidth]{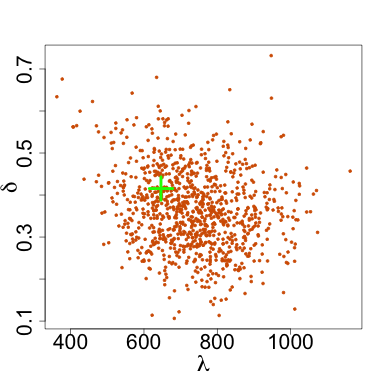} 
\end{minipage}
\begin{minipage}{0.32\linewidth}
\centering
\includegraphics[width=\linewidth]{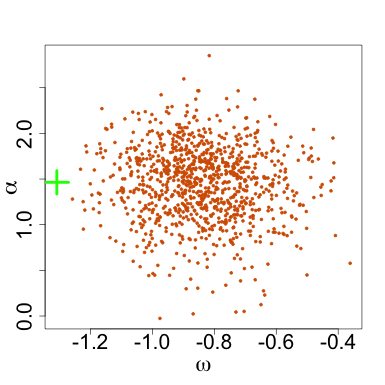} 
\end{minipage}
\begin{minipage}{0.32\linewidth}
\centering
\includegraphics[width=\linewidth]{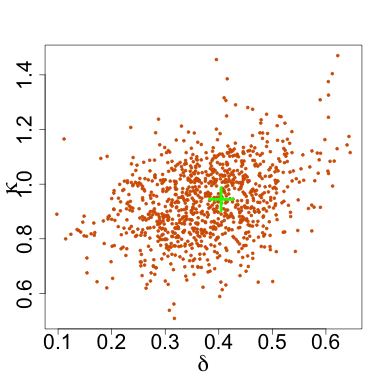} 
\end{minipage}
\begin{minipage}{0.32\linewidth}
\centering
\includegraphics[width=\linewidth]{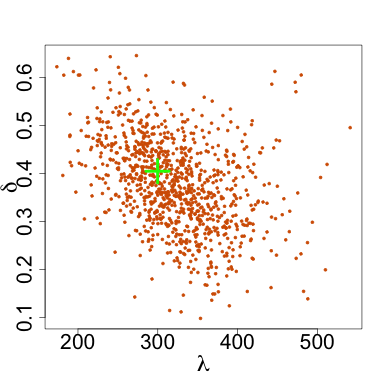} 
\end{minipage}
\begin{minipage}{0.32\linewidth}
\centering
\includegraphics[width=\linewidth]{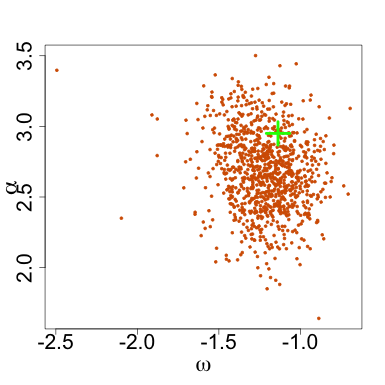} 
\end{minipage}
\caption{ Joint distribution of 1000 parameter estimates (orange) when $G=32$ in Section~\ref{sec:application} of the main text. Four test parameter sets (green) are drawn randomly from the prior and differ across rows. The left, centre and right columns give estimates of $(\delta, \kappa)'$, $(\lambda,\delta)'$ and $(\omega,\alpha)'$, respectively. }
\label{supfig:joint_distG32}
\end{figure}

\begin{figure}[t!]
\centering
\begin{minipage}{0.32\linewidth}
\centering
\includegraphics[width=\linewidth]{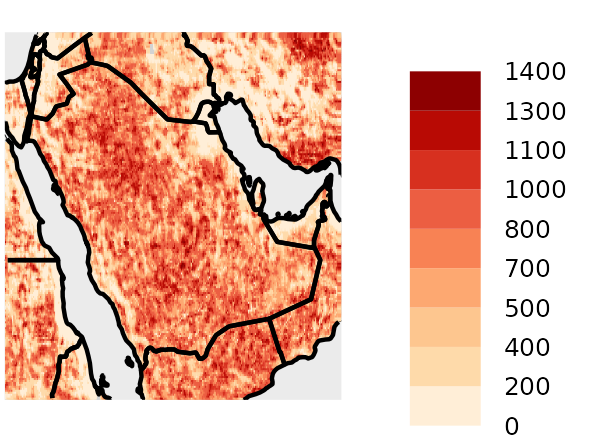} 
\end{minipage}
\hfill
\begin{minipage}{0.32\linewidth}
\centering
\includegraphics[width=\linewidth]{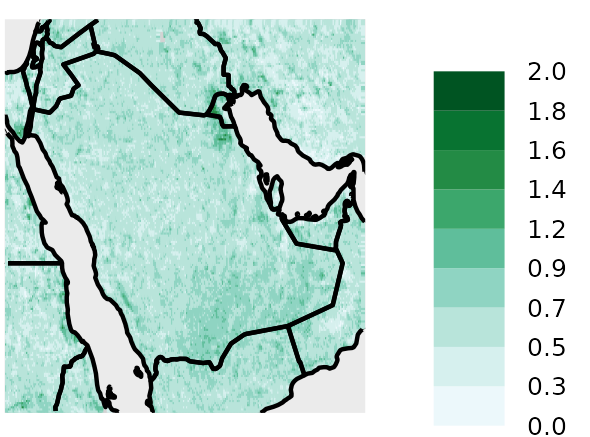} 
\end{minipage}
\hfill
\begin{minipage}{0.32\linewidth}
\centering
\includegraphics[width=\linewidth]{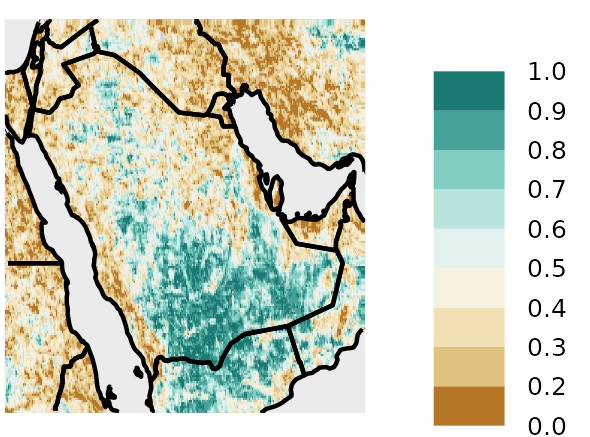} 
\end{minipage}
\begin{minipage}{0.32\linewidth}
\centering
\includegraphics[width=\linewidth]{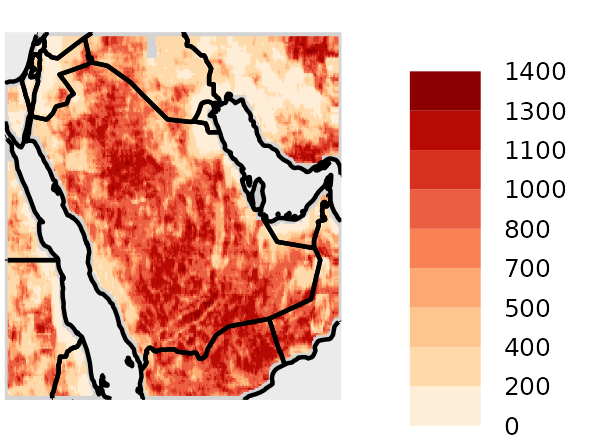} 
\end{minipage}
\hfill
\begin{minipage}{0.32\linewidth}
\centering
\includegraphics[width=\linewidth]{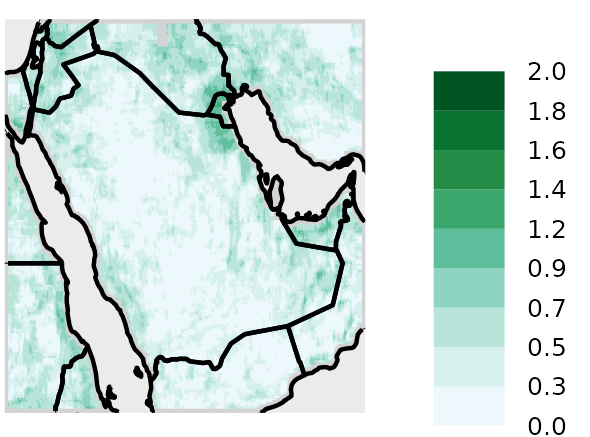} 
\end{minipage}
\hfill
\begin{minipage}{0.32\linewidth}
\centering
\includegraphics[width=\linewidth]{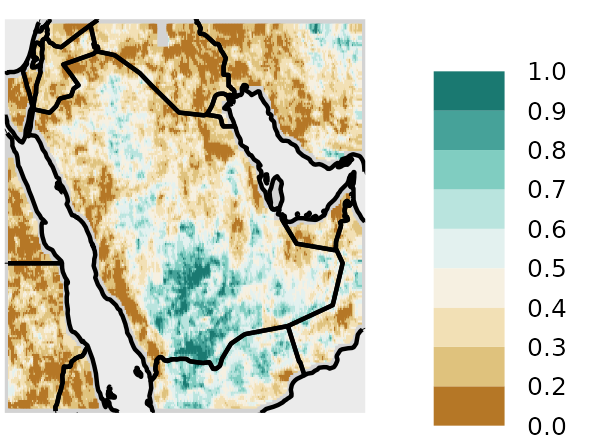} 
\end{minipage}

\begin{minipage}{0.32\linewidth}
\centering
\includegraphics[width=\linewidth]{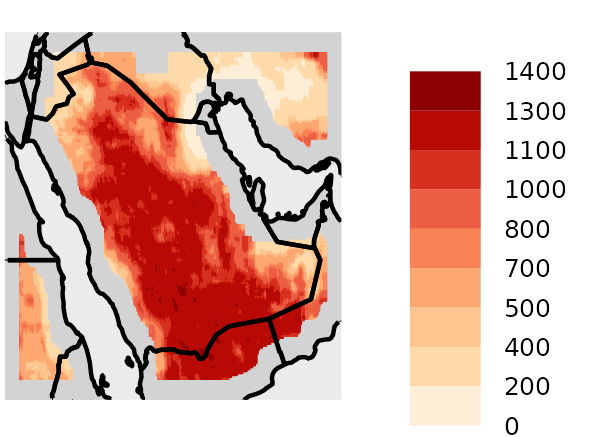} 
\end{minipage}
\hfill
\begin{minipage}{0.32\linewidth}
\centering
\includegraphics[width=\linewidth]{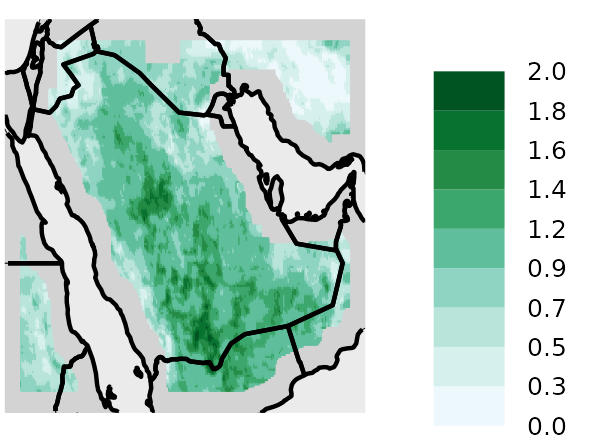} 
\end{minipage}
\hfill
\begin{minipage}{0.32\linewidth}
\centering
\includegraphics[width=\linewidth]{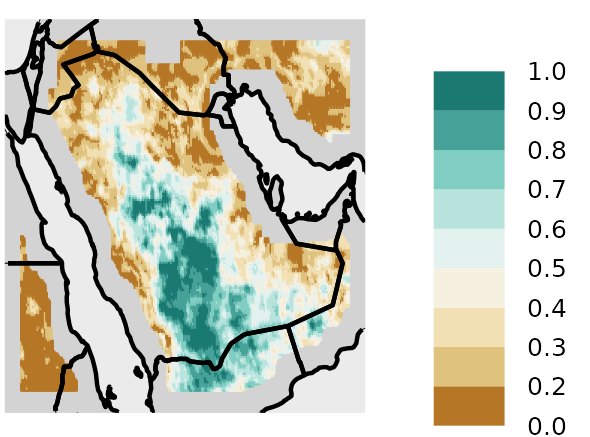} 
\end{minipage}

\begin{minipage}{0.32\linewidth}
\centering
\includegraphics[width=\linewidth]{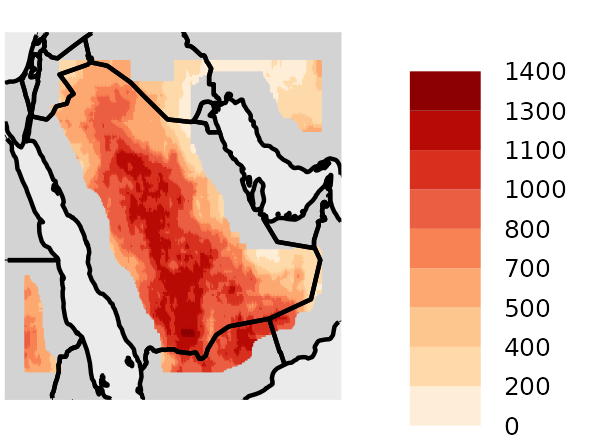} 
\end{minipage}
\hfill
\begin{minipage}{0.32\linewidth}
\centering
\includegraphics[width=\linewidth]{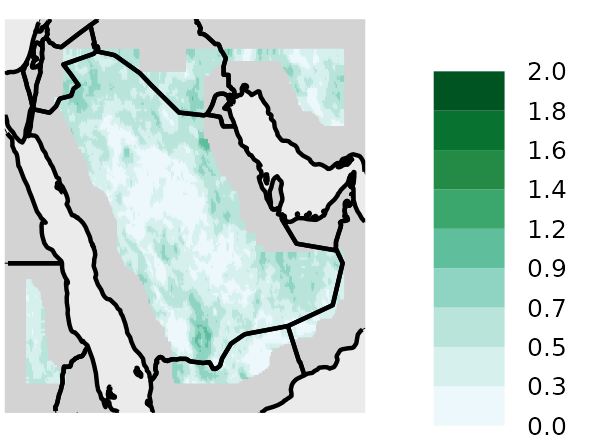} 
\end{minipage}
\hfill
\begin{minipage}{0.32\linewidth}
\centering
\includegraphics[width=\linewidth]{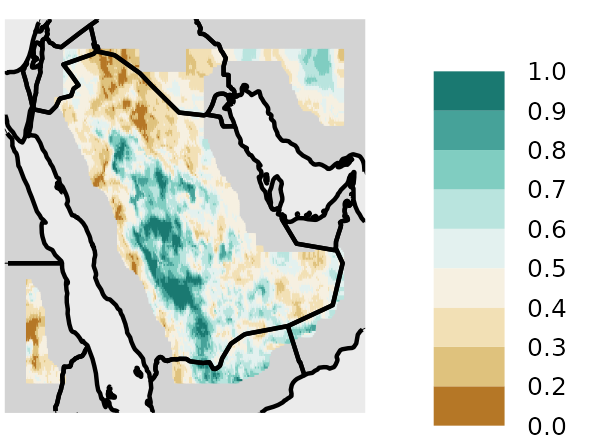} 
\end{minipage}

\caption{Extremal dependence parameter estimates for the HW model used in the application, Section~\ref{sec:application} of the main text. The smoothing level $G$ changes with the row (1st, 4; 2nd, 8; 3rd, 24; 4th, 32). The left, centre and right columns illustrate estimates of $\lambda$, $\kappa$ and $\delta$, respectively.}
\label{supfig:results}
\end{figure}

\begin{figure}[t!]
\centering
\begin{minipage}{0.32\linewidth}
\centering
\includegraphics[width=\linewidth]{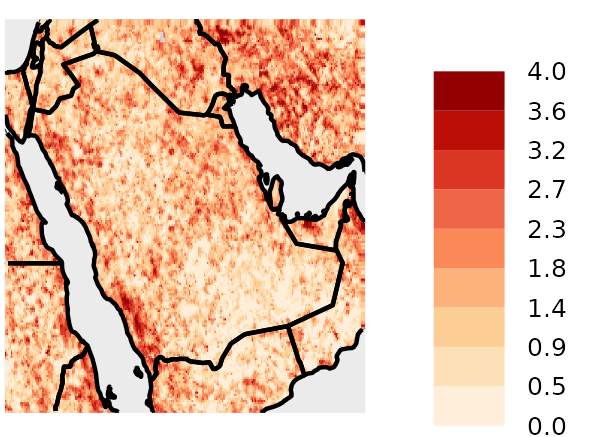} 
\end{minipage}
\hspace{2cm}
\begin{minipage}{0.32\linewidth}
\centering
\includegraphics[width=\linewidth]{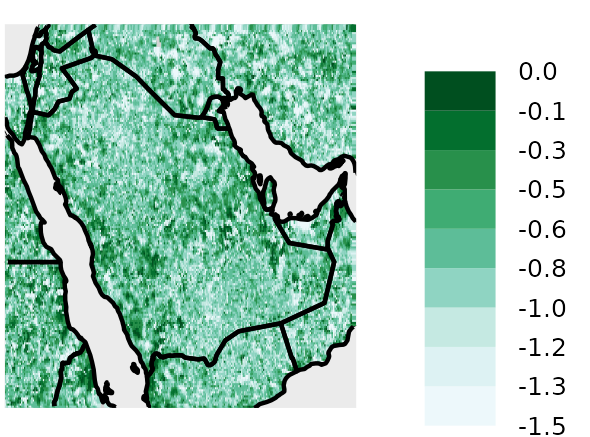} 
\end{minipage}

\begin{minipage}{0.32\linewidth}
\centering
\includegraphics[width=\linewidth]{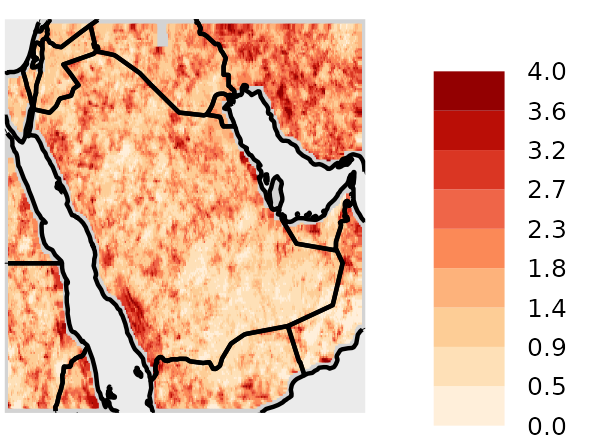} 
\end{minipage}
\hspace{2cm}
\begin{minipage}{0.32\linewidth}
\centering
\includegraphics[width=\linewidth]{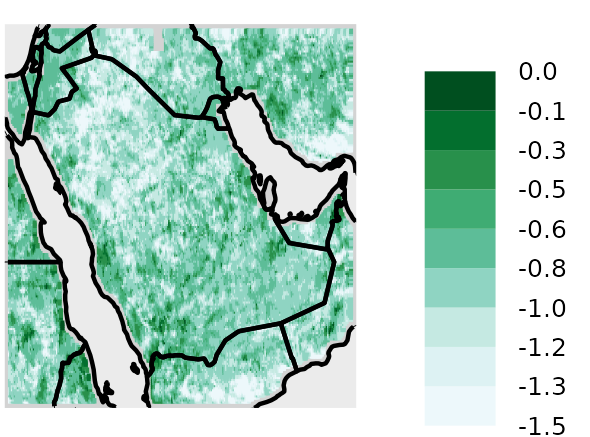} 
\end{minipage}

\begin{minipage}{0.32\linewidth}
\centering
\includegraphics[width=\linewidth]{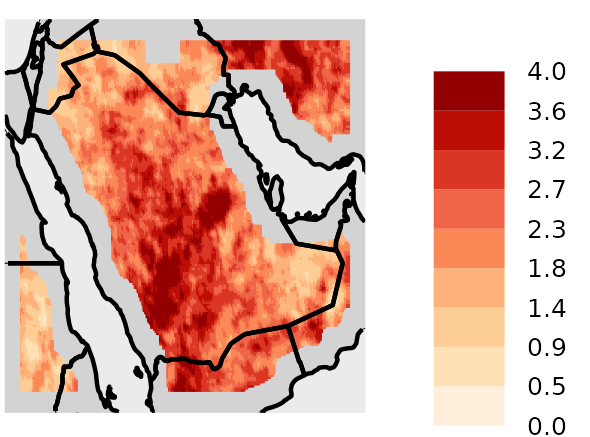} 
\end{minipage}
\hspace{2cm}
\begin{minipage}{0.32\linewidth}
\centering
\includegraphics[width=\linewidth]{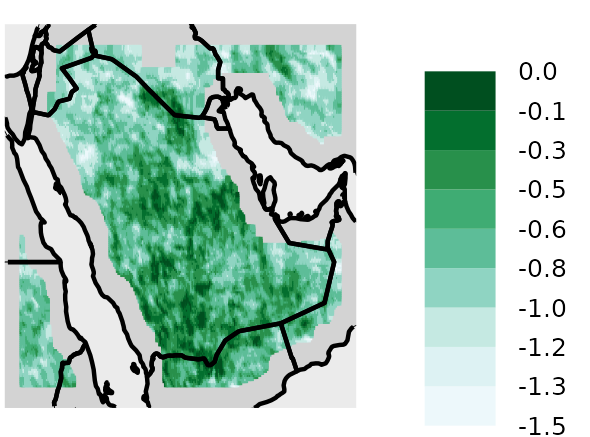} 
\end{minipage}

\begin{minipage}{0.32\linewidth}
\centering
\includegraphics[width=\linewidth]{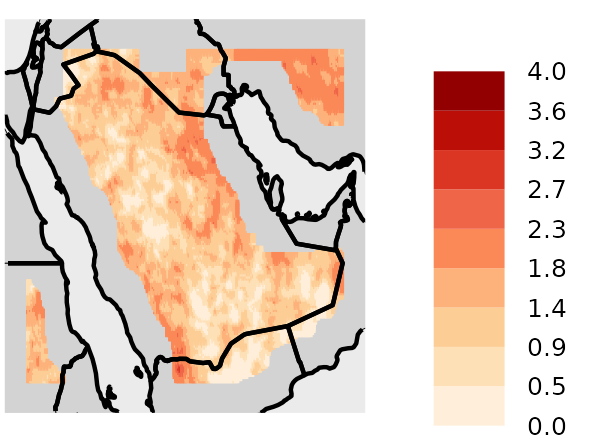} 
\end{minipage}
\hspace{2cm}
\begin{minipage}{0.32\linewidth}
\centering
\includegraphics[width=\linewidth]{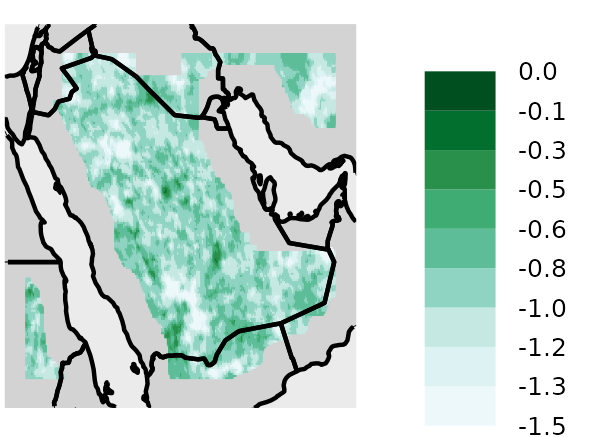} 
\end{minipage}

\caption{Anisotropy parameter estimates for the HW model used in the application, Section~\ref{sec:application}. The smoothing level $G$ changes with the row (1st, 4; 2nd, 8; 3rd, 24; 4th, 32). The left and right columns illustrate estimates of $\alpha$ and $\omega$, respectively.}
\label{supfig:results2}
\end{figure}

\begin{landscape}
\begin{figure}
\centering

\begin{minipage}{0.19\linewidth}
\centering
\includegraphics[width=\linewidth]{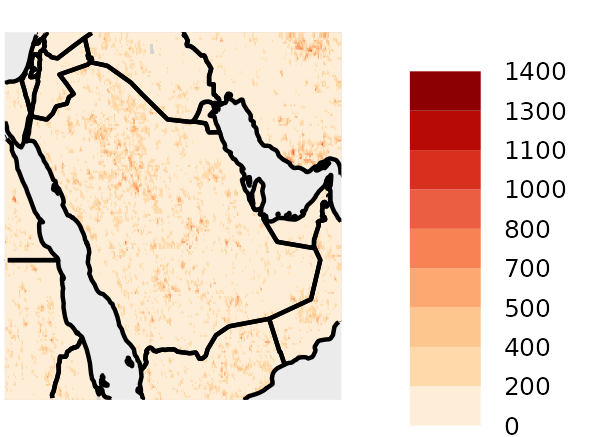} 
\end{minipage}
\begin{minipage}{0.19\linewidth}
\centering
\includegraphics[width=\linewidth]{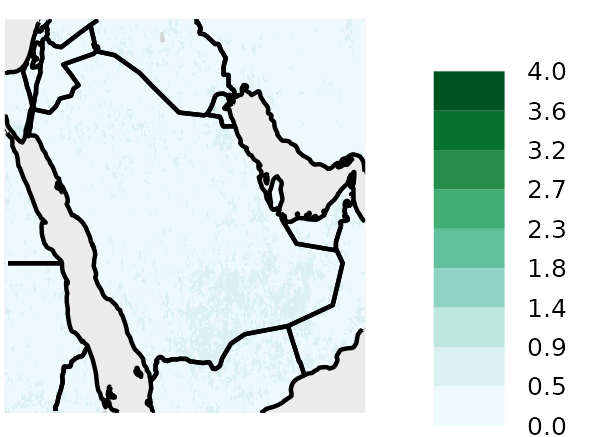} 
\end{minipage}
\begin{minipage}{0.19\linewidth}
\centering
\includegraphics[width=\linewidth]{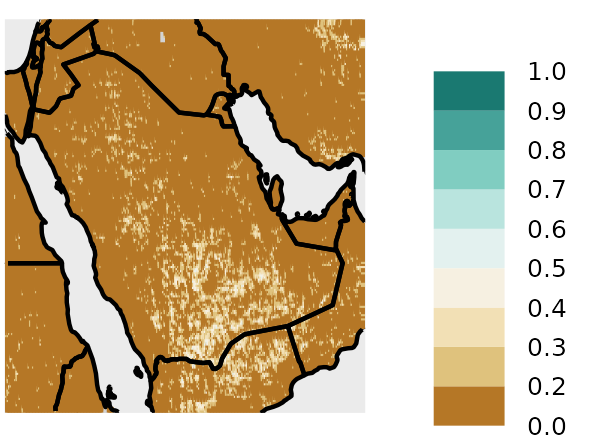} 
\end{minipage}
\begin{minipage}{0.19\linewidth}
\centering
\includegraphics[width=\linewidth]{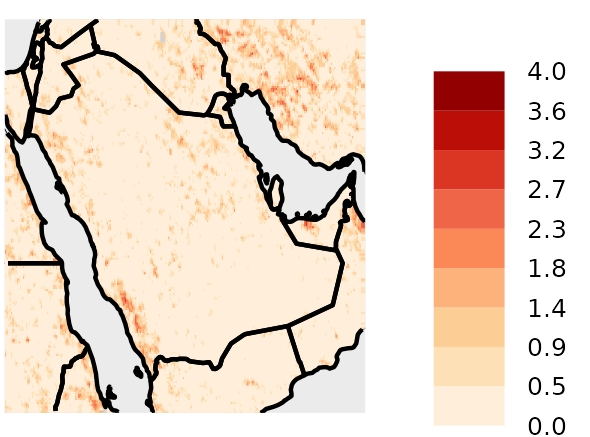} 
\end{minipage}
\begin{minipage}{0.19\linewidth}
\centering
\includegraphics[width=\linewidth]{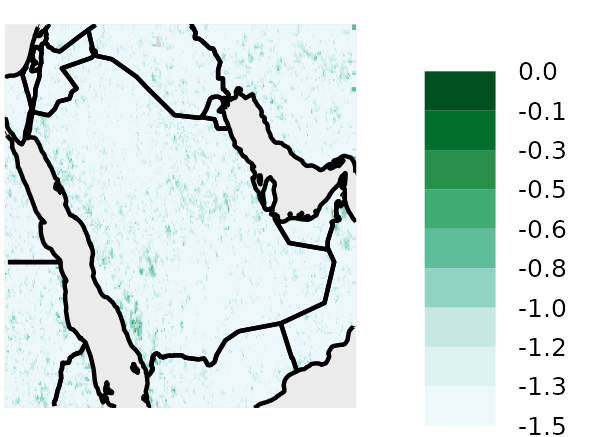} 
\end{minipage}
\begin{minipage}{0.19\linewidth}
\centering
\includegraphics[width=\linewidth]{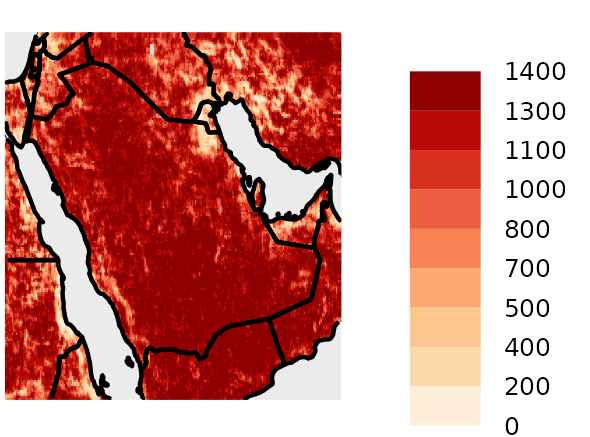} 
\end{minipage}
\begin{minipage}{0.19\linewidth}
\centering
\includegraphics[width=\linewidth]{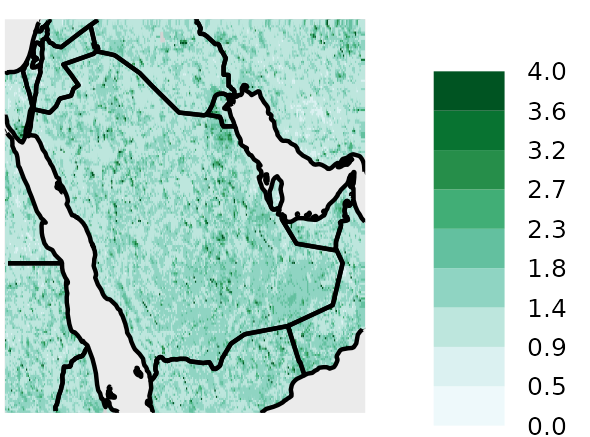} 
\end{minipage}
\begin{minipage}{0.19\linewidth}
\centering
\includegraphics[width=\linewidth]{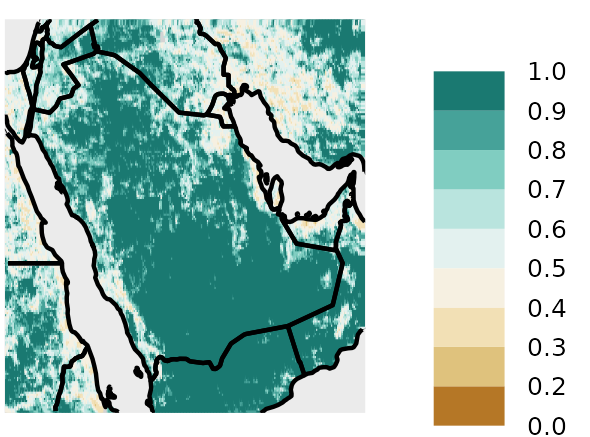} 
\end{minipage}
\begin{minipage}{0.19\linewidth}
\centering
\includegraphics[width=\linewidth]{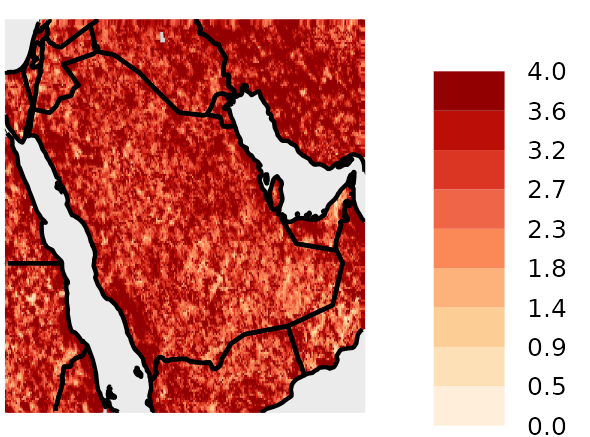} 
\end{minipage}
\begin{minipage}{0.19\linewidth}
\centering
\includegraphics[width=\linewidth]{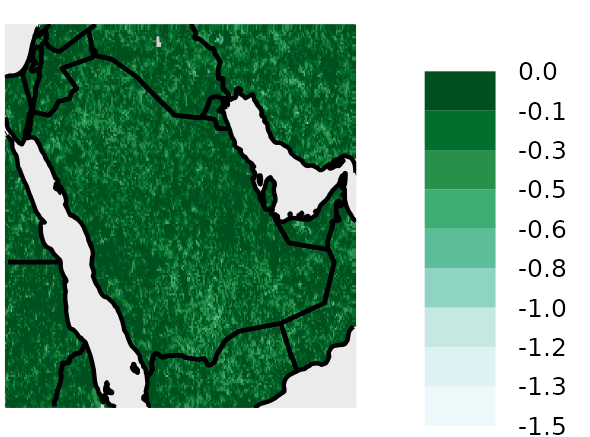} 
\end{minipage}
\begin{minipage}{0.19\linewidth}
\centering
\includegraphics[width=\linewidth]{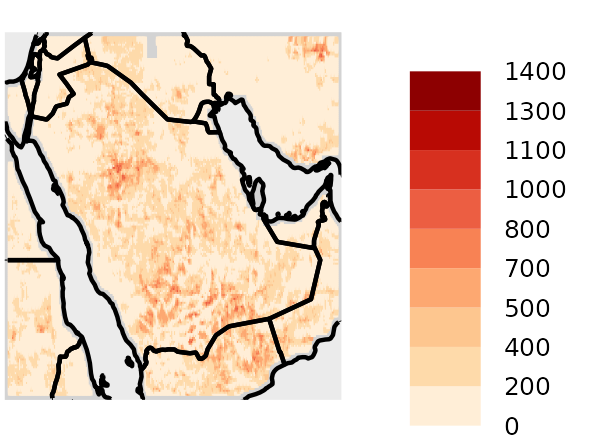} 
\end{minipage}
\begin{minipage}{0.19\linewidth}
\centering
\includegraphics[width=\linewidth]{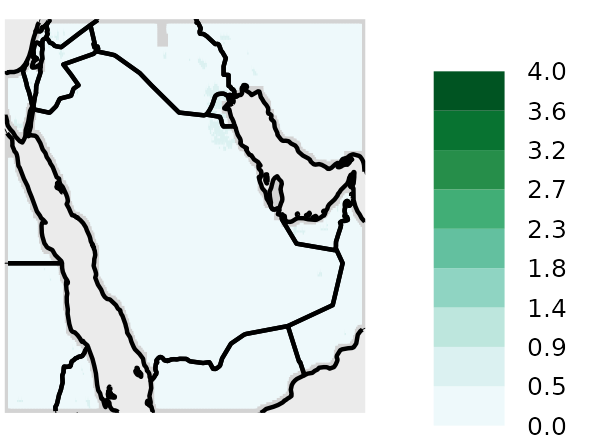} 
\end{minipage}
\begin{minipage}{0.19\linewidth}
\centering
\includegraphics[width=\linewidth]{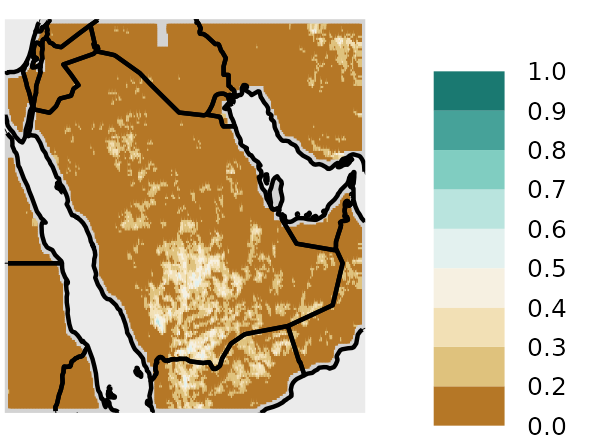} 
\end{minipage}
\begin{minipage}{0.19\linewidth}
\centering
\includegraphics[width=\linewidth]{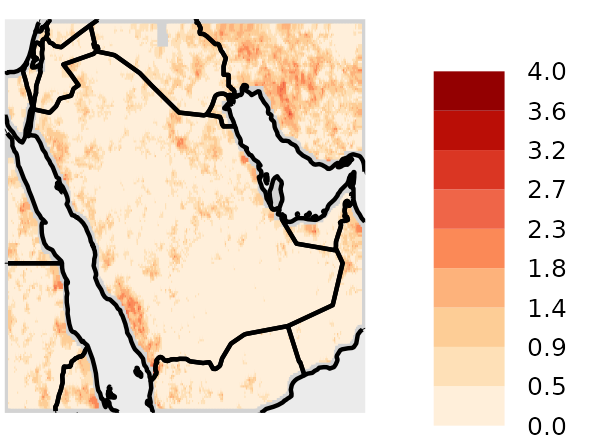} 
\end{minipage}
\begin{minipage}{0.19\linewidth}
\centering
\includegraphics[width=\linewidth]{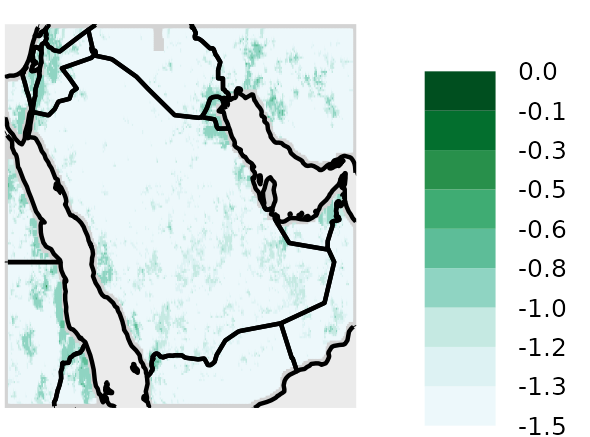} 
\end{minipage}
\begin{minipage}{0.19\linewidth}
\centering
\includegraphics[width=\linewidth]{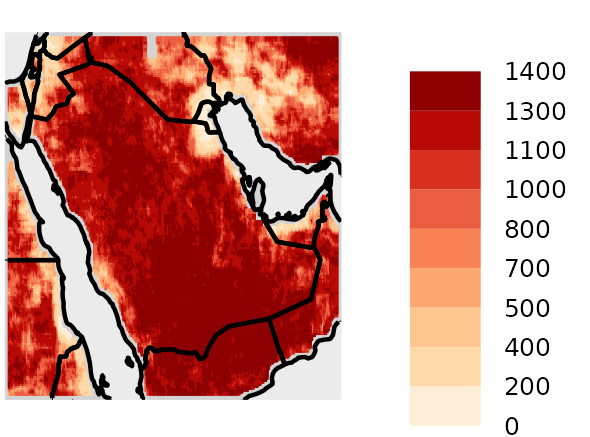} 
\end{minipage}
\begin{minipage}{0.19\linewidth}
\centering
\includegraphics[width=\linewidth]{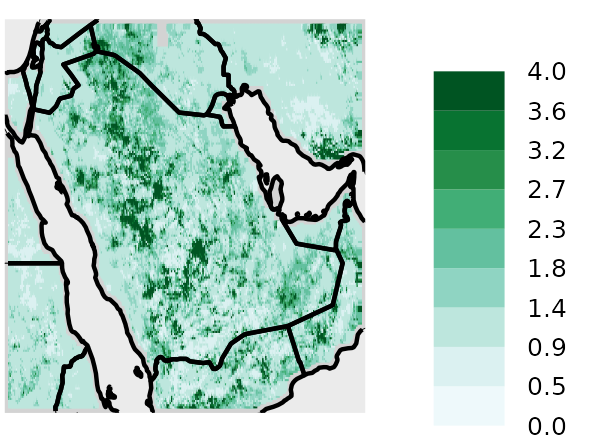} 
\end{minipage}
\begin{minipage}{0.19\linewidth}
\centering
\includegraphics[width=\linewidth]{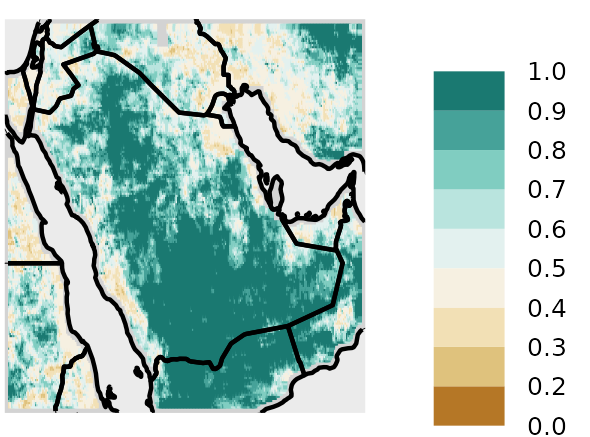} 
\end{minipage}
\begin{minipage}{0.19\linewidth}
\centering
\includegraphics[width=\linewidth]{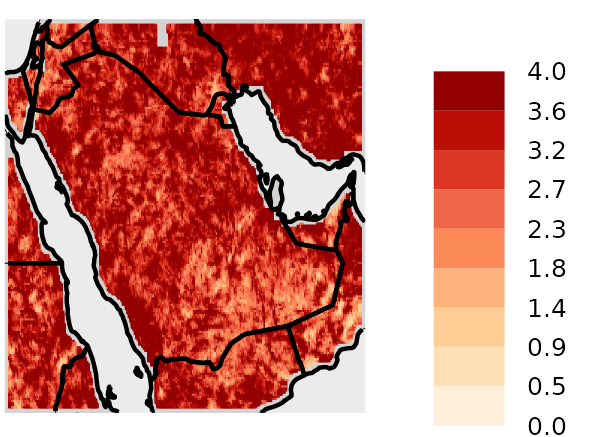} 
\end{minipage}
\begin{minipage}{0.19\linewidth}
\centering
\includegraphics[width=\linewidth]{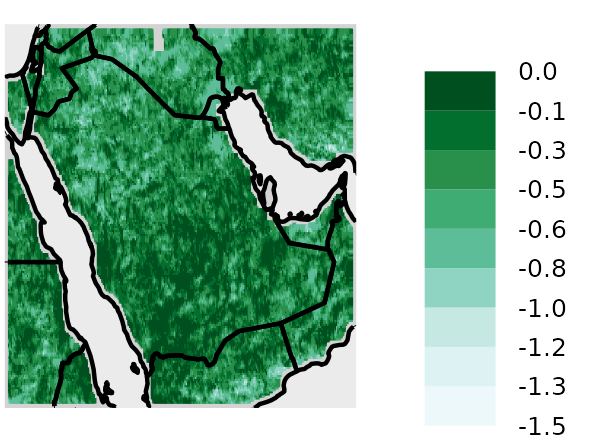} 
\end{minipage}
\caption{Parameter uncertainty assessment for $G=4$ (top two rows) and $G=8$ (last two rows) in the application, Section~\ref{sec:application}. First and second (third and fourth) rows give estimates of the $2.5\%$ and $97.5\%$ bootstrap quantiles, respectively, of (left to right) $\lambda$, $\kappa$, $\delta$, $\alpha$, and $\omega$. }
\label{supfig:bootG4}
\end{figure}
\end{landscape}

\begin{landscape}
\begin{figure}
\centering

\begin{minipage}{0.19\linewidth}
\centering
\includegraphics[width=\linewidth]{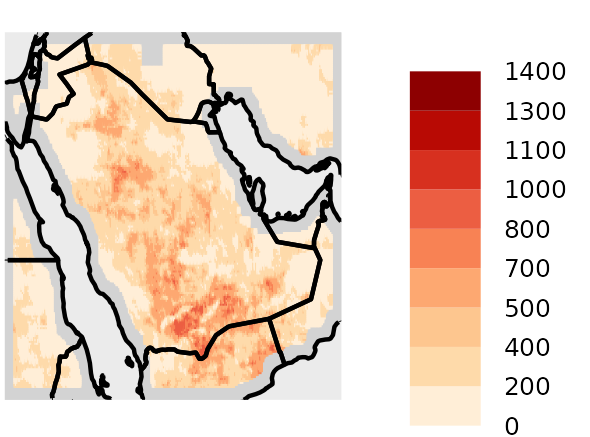} 
\end{minipage}
\begin{minipage}{0.19\linewidth}
\centering
\includegraphics[width=\linewidth]{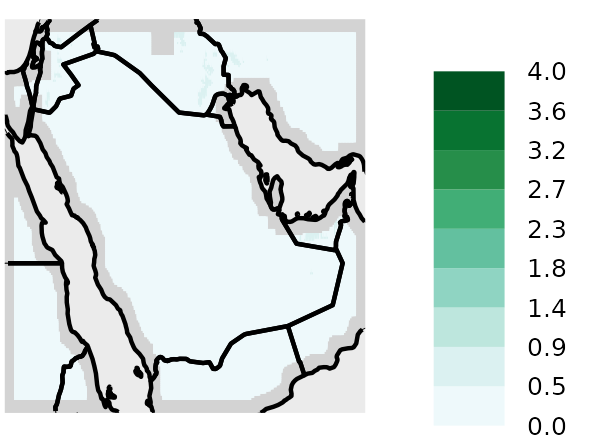} 
\end{minipage}
\begin{minipage}{0.19\linewidth}
\centering
\includegraphics[width=\linewidth]{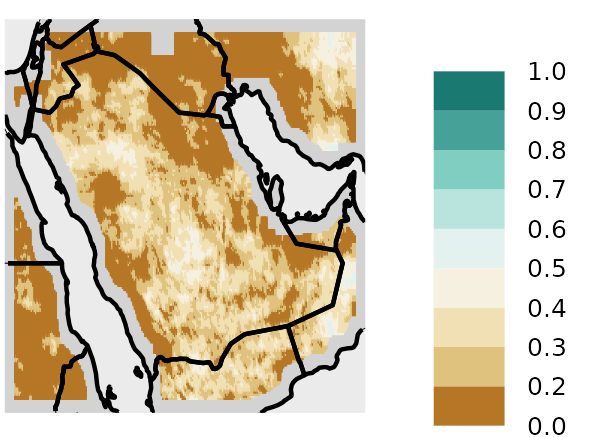} 
\end{minipage}
\begin{minipage}{0.19\linewidth}
\centering
\includegraphics[width=\linewidth]{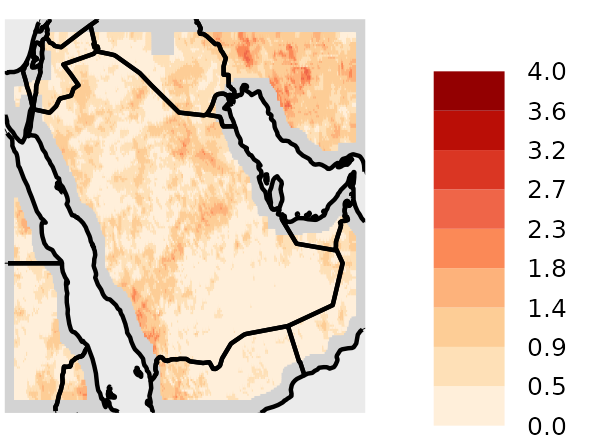} 
\end{minipage}
\begin{minipage}{0.19\linewidth}
\centering
\includegraphics[width=\linewidth]{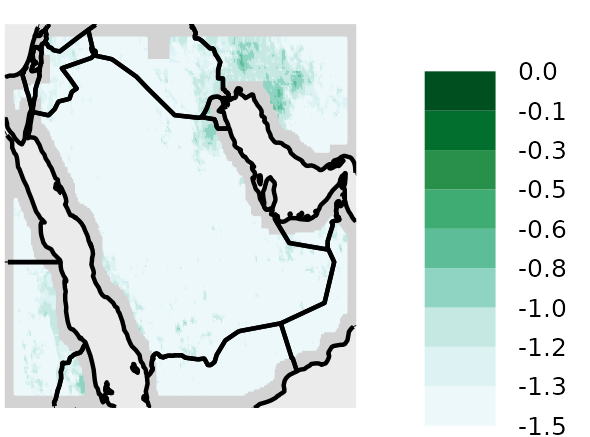} 
\end{minipage}
\begin{minipage}{0.19\linewidth}
\centering
\includegraphics[width=\linewidth]{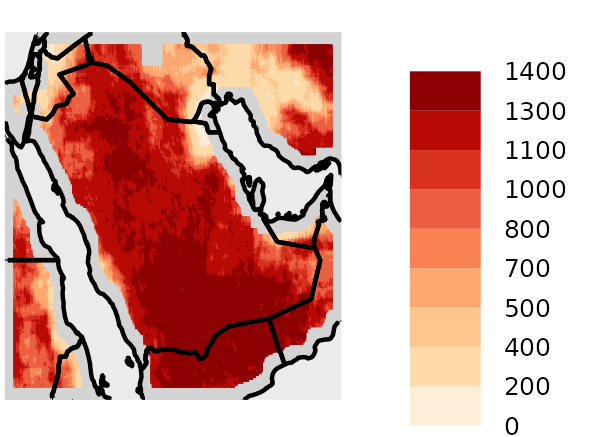} 
\end{minipage}
\begin{minipage}{0.19\linewidth}
\centering
\includegraphics[width=\linewidth]{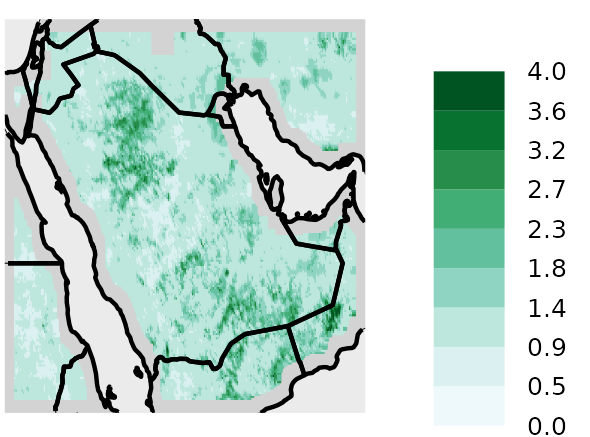} 
\end{minipage}
\begin{minipage}{0.19\linewidth}
\centering
\includegraphics[width=\linewidth]{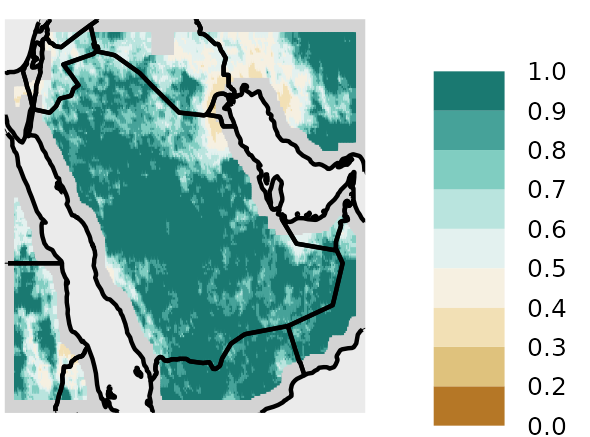} 
\end{minipage}
\begin{minipage}{0.19\linewidth}
\centering
\includegraphics[width=\linewidth]{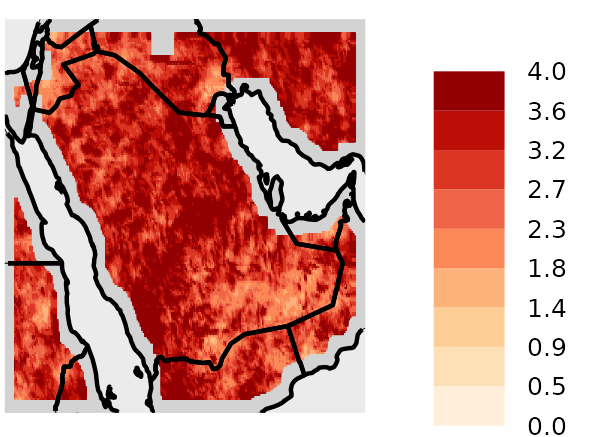} 
\end{minipage}
\begin{minipage}{0.19\linewidth}
\centering
\includegraphics[width=\linewidth]{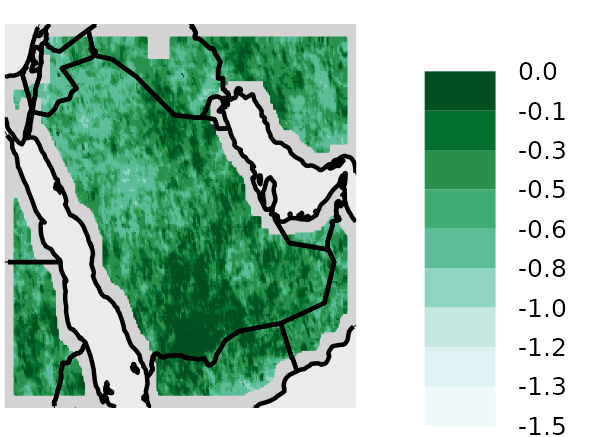} 
\end{minipage}

\caption{Parameter uncertainty assessment for $G=16$ in the application, Section~\ref{sec:application}. Top and bottom rows give estimates of the $2.5\%$ and $97.5\%$ bootstrap quantiles, respectively, of (left to right) $\lambda$, $\kappa$, $\delta$, $\alpha$, and $\omega$. }
\label{supfig:bootG16}
\end{figure}
\end{landscape}

\begin{landscape}
\begin{figure}
\centering
\begin{minipage}{0.19\linewidth}
\centering
\includegraphics[width=\linewidth]{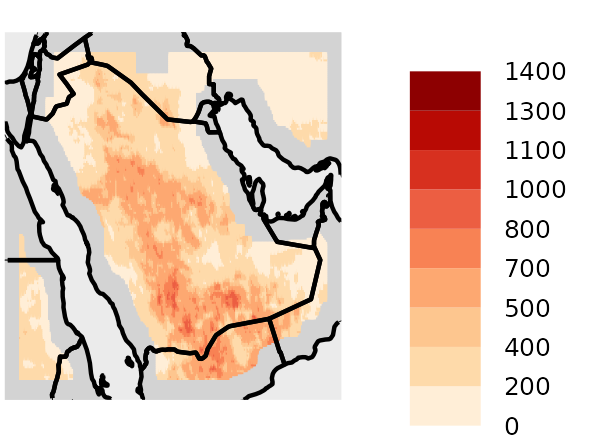} 
\end{minipage}
\begin{minipage}{0.19\linewidth}
\centering
\includegraphics[width=\linewidth]{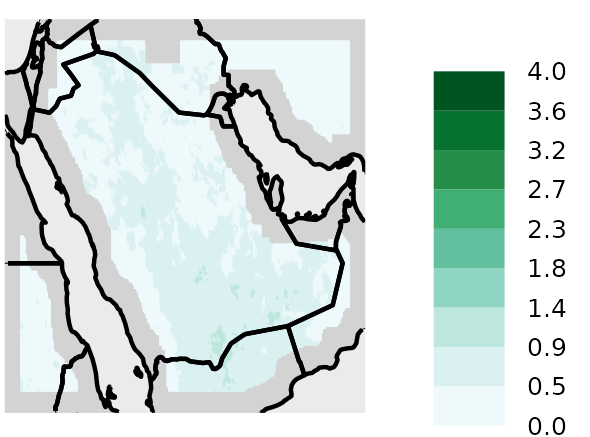} 
\end{minipage}
\begin{minipage}{0.19\linewidth}
\centering
\includegraphics[width=\linewidth]{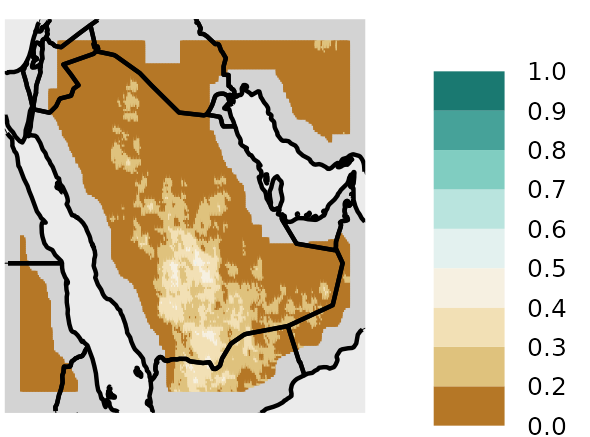} 
\end{minipage}
\begin{minipage}{0.19\linewidth}
\centering
\includegraphics[width=\linewidth]{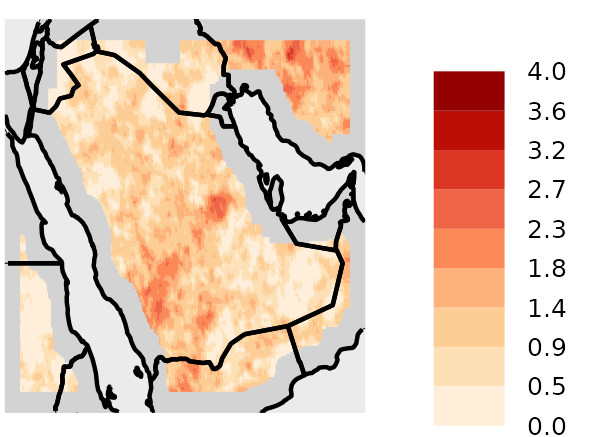} 
\end{minipage}
\begin{minipage}{0.19\linewidth}
\centering
\includegraphics[width=\linewidth]{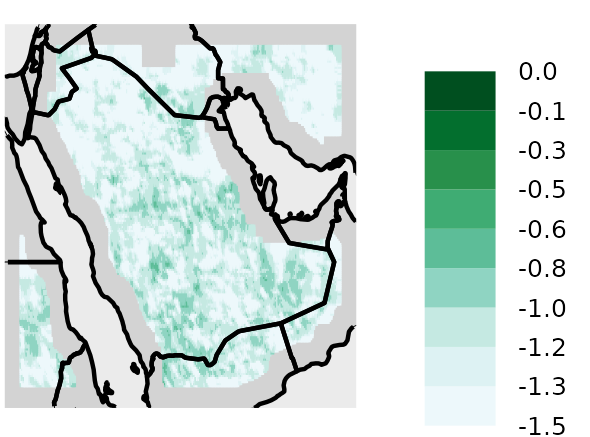} 
\end{minipage}
\begin{minipage}{0.19\linewidth}
\centering
\includegraphics[width=\linewidth]{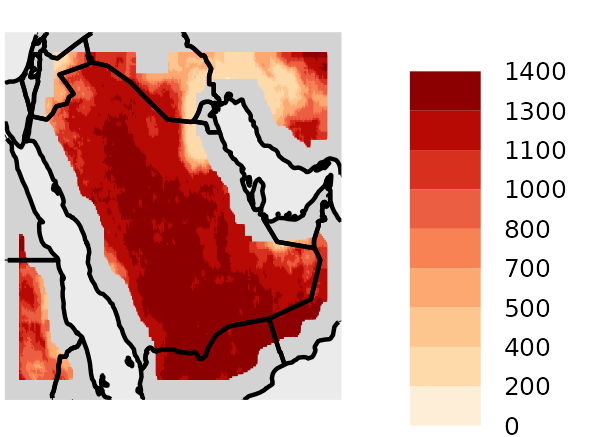} 
\end{minipage}
\begin{minipage}{0.19\linewidth}
\centering
\includegraphics[width=\linewidth]{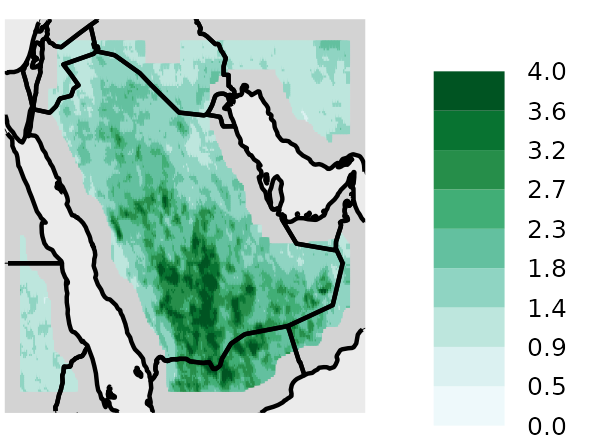} 
\end{minipage}
\begin{minipage}{0.19\linewidth}
\centering
\includegraphics[width=\linewidth]{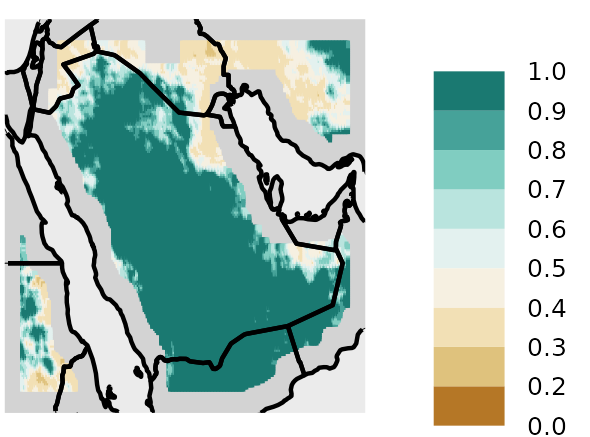} 
\end{minipage}
\begin{minipage}{0.19\linewidth}
\centering
\includegraphics[width=\linewidth]{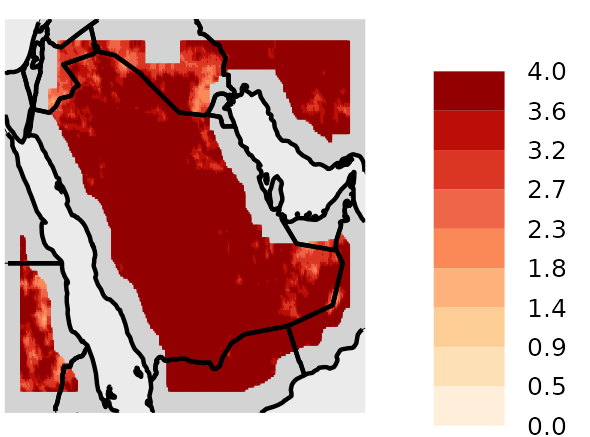} 
\end{minipage}
\begin{minipage}{0.19\linewidth}
\centering
\includegraphics[width=\linewidth]{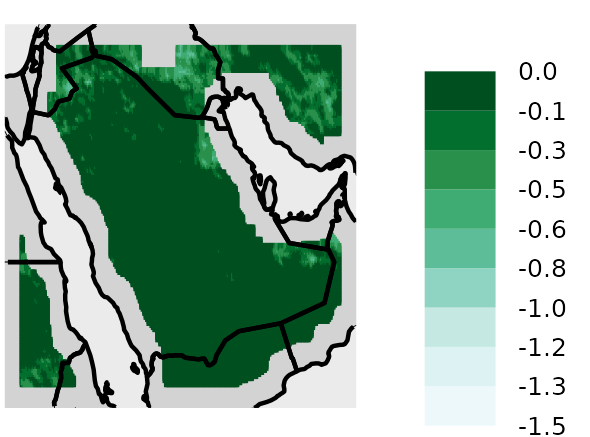} 
\end{minipage}
\begin{minipage}{0.19\linewidth}
\centering
\includegraphics[width=\linewidth]{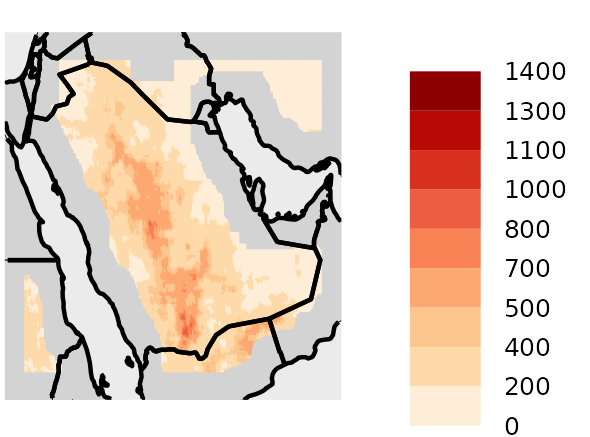} 
\end{minipage}
\begin{minipage}{0.19\linewidth}
\centering
\includegraphics[width=\linewidth]{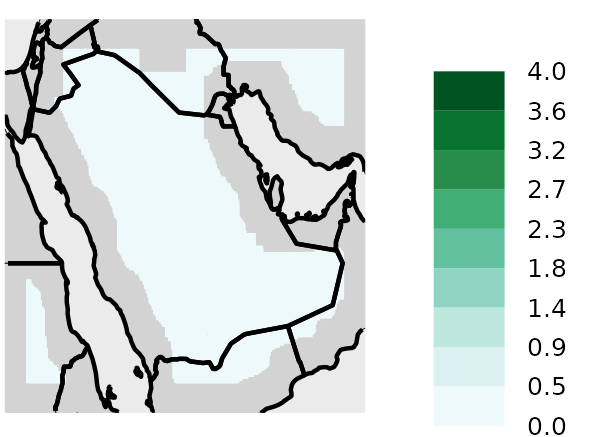} 
\end{minipage}
\begin{minipage}{0.19\linewidth}
\centering
\includegraphics[width=\linewidth]{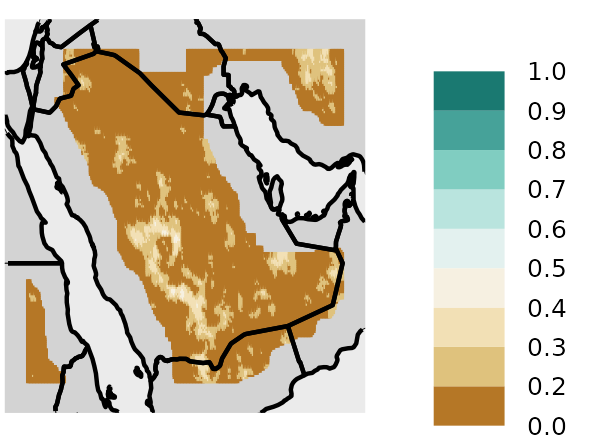} 
\end{minipage}
\begin{minipage}{0.19\linewidth}
\centering
\includegraphics[width=\linewidth]{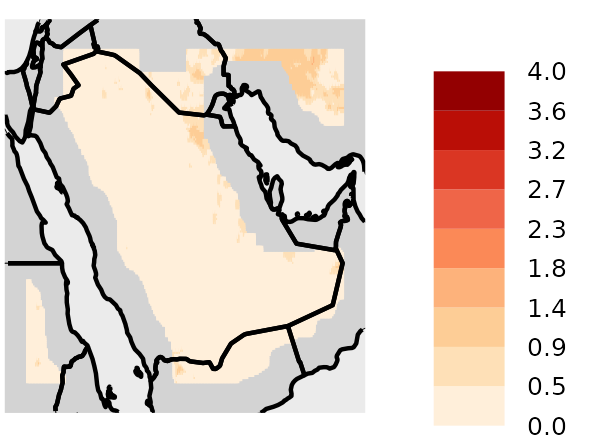} 
\end{minipage}
\begin{minipage}{0.19\linewidth}
\centering
\includegraphics[width=\linewidth]{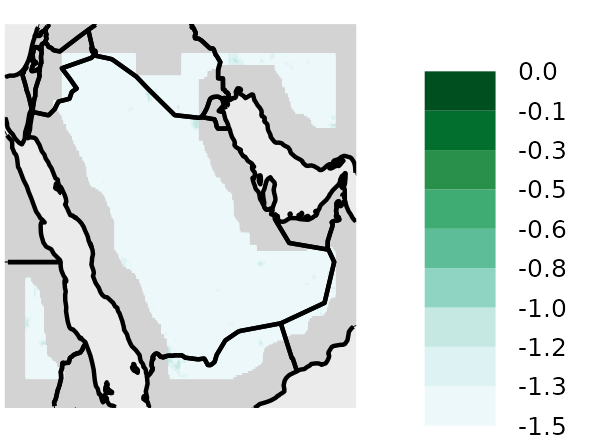} 
\end{minipage}
\begin{minipage}{0.19\linewidth}
\centering
\includegraphics[width=\linewidth]{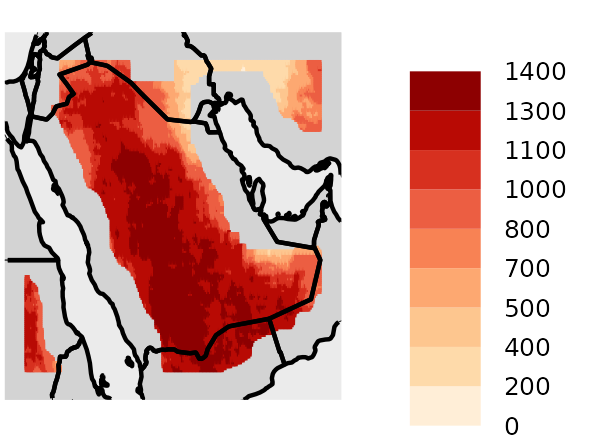} 
\end{minipage}
\begin{minipage}{0.19\linewidth}
\centering
\includegraphics[width=\linewidth]{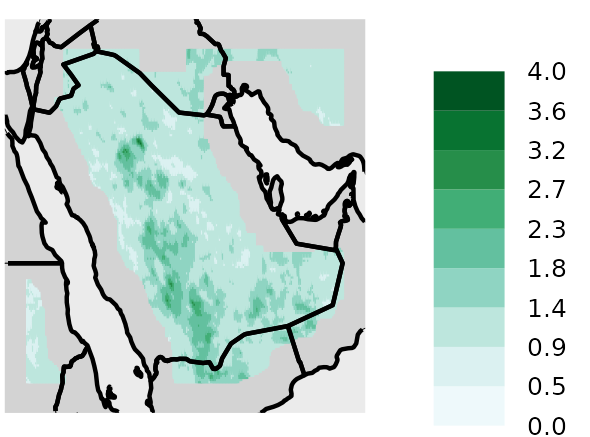} 
\end{minipage}
\begin{minipage}{0.19\linewidth}
\centering
\includegraphics[width=\linewidth]{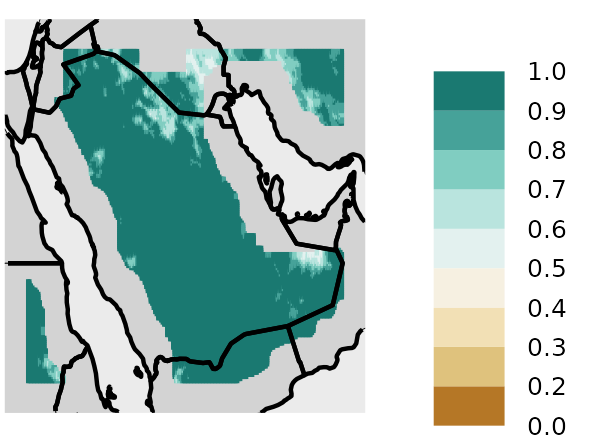} 
\end{minipage}
\begin{minipage}{0.19\linewidth}
\centering
\includegraphics[width=\linewidth]{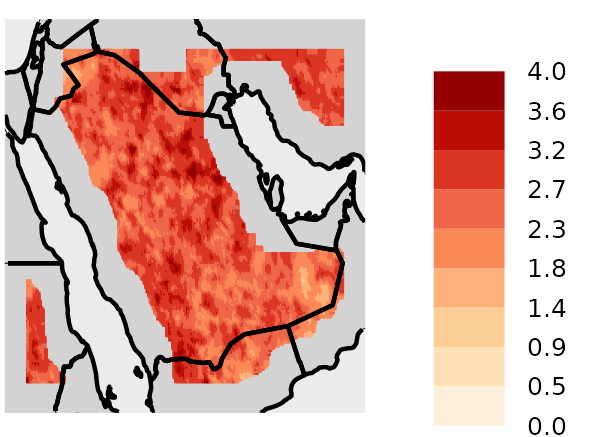} 
\end{minipage}
\begin{minipage}{0.19\linewidth}
\centering
\includegraphics[width=\linewidth]{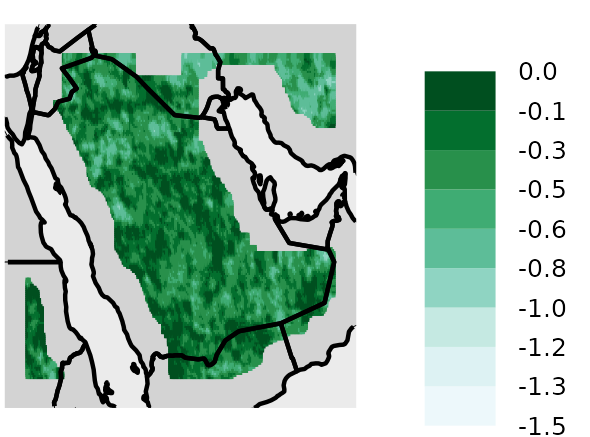} 
\end{minipage}

\caption{ Parameter uncertainty assessment for $G=24$ (first two rows) and $G=32$ (last two rows) in the application, Section~\ref{sec:application}. First and second (third and fourth) rows give estimates of the $2.5\%$ and $97.5\%$ bootstrap quantiles, respectively, of (left to right) $\lambda$, $\kappa$, $\delta$, $\alpha$, and $\omega$. }
\label{supfig:bootG32}
\end{figure}
\end{landscape}
\clearpage

\section{Supplementary tables}
\label{supsec:arch}
\begin{table}[h!]
\caption{Results for the simulation study detailed in Section~\ref{sec:sim_fixed_PL} of the main text. Marginal test risks (s.d.) under the squared-error loss are provided separately for $\lambda$ ($\times 10^{-1}$) and $\kappa$ ($\times 10^{-2}$). The lowest value in each column is given in bold.}
\vspace{5pt}
\centering
\begin{tabular}{l*{2}{c}}
\hline
\hline
& $\lambda$ & $\kappa$\\
 \cline{1-3}
 & \multicolumn{2}{c}{GP}  \\
  \cline{1-3}
NBE &  \bf 1.74 (0.09) &  \bf 0.37 (0.02)\\
CPL ($h_{max}=3$)  &   3.74 (0.24) &  0.71 (0.05) \\
CPL ($h_{max}=\infty$)&   14.62 (0.69) &  4.84 (0.27) \\
 \cline{1-3}
 \cline{1-3}
&  \multicolumn{2}{c}{IMSP} \\
 \cline{1-3}
NBE &  \bf 1.38 (0.10) &   \bf 0.078 (0.004)\\
CPL ($h_{max}=3$)  & 4.82 (0.39) &  0.160 (0.008)\\
CPL ($h_{max}=\infty$) &  6.13 (0.44) &  0.758 (0.036)\\
 \cline{1-3}
  \cline{1-3}
 
 &  \multicolumn{2}{c}{MSP} \\
 \cline{1-3}
NBE & \bf  1.18 (0.10)& \bf 0.055 (0.003)\\
CPL ($h_{max}=3$)  &  2.51 (0.18)&  0.080 (0.003) \\
CPL ($h_{max}=\infty$) &  3.75 (0.25)&  0.686 (0.035) \\
 \cline{1-3}
\end{tabular}
 \label{subtab:sim_fixed1}

\end{table}
\begin{table}[h]
\caption{Results for the $r$-Pareto process simulation study detailed in Section~\ref{sec:sim_fixed_FL} of the main text. Marginal test risks (s.d.) under the absolute-error loss are provided separately for $\lambda$ ($\times 10^{-1}$) and $\kappa$ ($\times 10^{-2}$).  The lowest value in each column is given in bold.}
\vspace{5pt}
\centering

\begin{tabular}{l*{2}{c}}
\hline
\hline
  &  $\lambda$ & $\kappa$\\
 \hline
NBE &\bf 2.29 (0.07) & \bf 1.87 (0.05) \\
 
Quasi-Monte Carlo censored likelihood & 4.92 (0.20) & 4.31 (0.23) \\
\hline
\end{tabular}
\label{tab:sim_fixed2}
\end{table}

\begin{table}[h!]
\caption{Results for the $r$-Pareto process simulation study detailed in Section~\ref{sec:sim_fixed_FL} of the main text. Marginal test risks (s.d.) under the squared-error loss are provided separately for $\lambda$ ($\times 10^{-1}$) and $\kappa$ ($\times 10^{-2}$). The lowest value in each column is given in bold.}
\vspace{5pt}
\centering

\begin{tabular}{l*{2}{c}}
\hline
\hline  & $\lambda$ & $\kappa$\\
 \hline
NBE & \bf 0.99 (0.06) &  \bf 0.060 (0.003) \\

Quasi-Monte Carlo  censored likelihood  &  6.30 (0.64) &  0.731 (0.101) \\
\hline
\end{tabular}
\label{suptab:sim_fixed_2}
\end{table}

\begin{table}[t]
\caption{Results for the HW process simulation study detailed in Section~\ref{sec:sim_fixed_FL} of the main text; the process is observed over either $\mathcal{S}$, a $16\times 16$ regular grid, or $\mathcal{S}^*$, a $6\times 6$ regular grid. Marginal test risks (s.d.) under the \emph{absolute-error} loss are provided separately for $\lambda$ ($\times 10^{-1}$), $\kappa$ ($\times 10^{-2}$) and $\delta$ ($\times 10^{-2}$). The lowest value in each column is given in bold.}
\vspace{5pt}
\centering
\begin{tabular}{l*{3}{c}}
\hline
\hline
  &  $\lambda$ & $\kappa$&  $\delta$\\
 \hline
NBE ($\mathcal{S}$)&  \bf 5.77 (0.19) & \bf  7.42 (0.31) &3.94 (0.17) \\
NBE ($\mathcal{S}^*$)& 6.30 (0.19) &   9.49 (0.26) &   \bf  3.24 (0.11)\\
Quasi-Monte Carlo  censored likelihood& 5.82 (0.18) & 9.53  (0.36) & 4.80 (0.30)\\
\hline
\end{tabular}
 \label{tab:sim_fixed_3}
\end{table}

\begin{table}[h!]
\caption{Results for the HW process simulation study detailed in Section~\ref{sec:sim_fixed_FL} of the main text; the process is observed over either $\mathcal{S}$, a $16\times 16$ regular grid, or $\mathcal{S}^*$, a $6\times 6$ regular grid. Marginal test risks (s.d.) under the \emph{squared-error} loss are provided separately for $\lambda$ ($\times 10^{-1}$), $\kappa$ ($\times 10^{-2}$) and $\delta$ ($\times 10^{-2}$). The lowest value in each column is given in bold.}
\vspace{5pt}
\centering
\begin{tabular}{l*{3}{c}}
\hline
\hline
  &  $\lambda$ & $\kappa$&  $\delta$\\
 \hline
NBE ($\mathcal{S}$)&  \bf 6.87 (0.47)&  \bf 1.54 (0.25) &   0.46 (0.14) \\
NBE ($\mathcal{S}^*$)& 7.42 (0.50) &   1.57 (0.09) &  \bf  0.23 (0.02)\\
Quasi-Monte Carlo  censored likelihood& 6.61 (0.48) &   2.19 (0.28) &   1.13 (0.24)\\
\hline
\end{tabular}
 \label{suptab:sim_fixed_3}
\end{table}

\begin{table*}[h!]
\caption{Neural network architecture for the HW process observed on the smaller domain $\mathcal{S}^*$, see Section~\ref{sec:sim_fixed_FL} of the main text. Each convolution filter uses zero padding and unit stride. The function $\mbox{vec}(\cdot)$ refers to a flattening of an array to a vector. All layers, except the final layer, used rectified linear unit (ReLU) activation functions; the final layer uses the identity function.}
\vspace{5pt}
 \begin{tabular}{lcccr} 
\hline
\hline
layer type & input dimension & output dimension & filter dimension & parameters\\
\hline
2D convolution & [6,\;6,\;2] & [4,\;4,\;128] & $3\times 3$ & 2,432\\
2D convolution & [4,\;4,\;128] & [2,\;2,\;256] & $3\times 3$ & 147,584 \\
2D convolution & [2,\;2,\;256] & [1,\;1,\;256] & $2\times 2$ & 262,400 \\
$\mbox{vec}(\cdot)$ & [1,\;1,\;256] & [256] &  & 0\\
dense & [256] & [500] &  & 128,500 \\
dense & [500] & [$3$] & & $1503$\\
\hline
\multicolumn{4}{l}{total trainable parameters:} &542,419\\
\hline
  \end{tabular}
 \label{tab:CNN_arch2}
  \end{table*}
  \begin{table}[h!]
\caption{Results for the variable $\tau$ simulation study detailed in Section~\ref{sec:sim_vary}. Marginal test risks (s.d.) under the squared-error loss are provided separately for $\lambda$ ($\times 10^{-1}$), $\kappa$ ($\times 10^{-2}$) and $\delta$ ($\times 10^{-2}$). The lowest value in each column is given in bold.}
\vspace{5pt}
\centering

\begin{tabular}{l*{5}{c}}
\hline
\hline
 & \multicolumn{2}{c}{IMSP} & \multicolumn{3}{c}{HW}\\
  $\tau$& $\lambda$ & $\kappa$& $\lambda$ & $\kappa$ &$\delta$\\
 \hline
\multirow{1}{*}{\shortstack[l]{random}}  &  \bf 1.40 (0.07)  &0.089 (0.004) & \bf 1.14 (0.06) & \bf 0.070 (0.004) &  \bf 0.16 (0.01)\\
  \multirow{1}{*}{\shortstack[l]{fixed}}  & 1.43 (0.07)  &\bf 0.068 (0.003) & 1.18 (0.05) &  0.092 (0.004) &  0.21 (0.01)\\
 \hline
\end{tabular}
\label{suptab:sim_vary}
\end{table}

\begin{table*}[h!]
\caption{Neural network architectures used for the application, see Section~\ref{sec:application} of the main text, when $G=4$. Each convolution filter uses zero padding and unit stride. The function $\mbox{vec}(\cdot)$ refers to a flattening of an array to a vector. All layers, except the final layer, used rectified linear unit (ReLU) activation functions; the final layer uses the identity function. }
\vspace{5pt}
 \begin{tabular}{lcccr} 
\hline
\hline
\multicolumn{5}{c}{$G=4$}\\
\hline
layer type & input dimension & output dimension & filter dimension & parameters\\
\hline
2D convolution & [4,\;4,\;2] & [3,\;3,\;64] & $2\times 2$ & 576\\
2D convolution & [3,\;3,\;64] & [2,\;2,\;128] & $2\times 2$ & 32,896 \\
2D convolution & [2,\;2,\;128] & [1,\;1,\;256] & $2\times 2$ & 131,328 \\
$\mbox{vec}(\cdot)$ & [1,\;1,\;256] & [256] &  & 0\\
dense & [256] & [500] &  & 128,500 \\
dense & [500] & [$5$] & & $2505$\\
\hline
\multicolumn{4}{l}{total trainable parameters:} &295,805\\
\hline
  \end{tabular}

 \label{tab:CNN_archG4}
  \end{table*}

\begin{table*}[h!]
\caption{Neural network architectures used for the application, see Section~\ref{sec:application} of the main text, when $G=8$. Each convolution filter uses zero padding and unit stride. The function $\mbox{vec}(\cdot)$ refers to a flattening of an array to a vector. All layers, except the final layer, used rectified linear unit (ReLU) activation functions; the final layer uses the identity function. }
\vspace{5pt}
 \begin{tabular}{lcccr} 
\hline
\hline
\multicolumn{5}{c}{$G=8$}\\
\hline
layer type & input dimension & output dimension & filter dimension & parameters\\
\hline
2D convolution & [8,\;8,\;2] & [4,\;4,\;64] & $5\times 5$ & 3,264\\
2D convolution & [4,\;4,\;64] & [2,\;2,\;128] & $3\times 3$ & 73,856 \\
2D convolution & [2,\;2,\;128] & [1,\;1,\;256] & $2\times 2$ & 131,328 \\
$\mbox{vec}(\cdot)$ & [1,\;1,\;256] & [256] &  & 0\\
dense & [256] & [500] &  & 128,500 \\
dense & [500] & [$5$] & & $2505$\\
\hline
\multicolumn{4}{l}{total trainable parameters:} &339,453\\
\hline
  \end{tabular}

 \label{tab:CNN_archG8}
  \end{table*}
  
  \begin{table*}[h!]
\caption{Neural network architectures used for the application, see Section~\ref{sec:application} of the main text, when $G=16$. Each convolution filter uses zero padding and unit stride. The function $\mbox{vec}(\cdot)$ refers to a flattening of an array to a vector. All layers, except the final layer, used rectified linear unit (ReLU) activation functions; the final layer uses the identity function. Note that the layer denoted by the asterisk ($*$) features in Table~\ref{tab:CNN_archG2432}.}
\vspace{5pt}
 \begin{tabular}{lcccr} 
\hline
\hline
\multicolumn{5}{c}{$G=16$}\\
\hline
layer type & input dimension & output dimension & filter dimension & parameters\\
\hline
2D convolution & [16,\;16,\;2] & [7,\;7,\;64] & $10\times 10$ & 12,864\\
2D convolution$^*$ & [7,\;7,\;64] & [3,\;3,\;128] & $5\times 5$ & 204,928 \\
2D convolution & [3,\;3,\;128] & [1,\;1,\;256] & $3\times 3$ & 295,168 \\
$\mbox{vec}(\cdot)$ & [1,\;1,\;256] & [256] &  & 0\\
dense & [256] & [500] &  & 128,500 \\
dense & [500] & [$5$] & & $2505$\\
\hline
\multicolumn{4}{l}{total trainable parameters:} &643,965\\
\hline
  \end{tabular}

 \label{tab:CNN_archG16}
  \end{table*}
  
    \begin{table*}[h!]
\caption{Neural network architectures used for the application, see Section~\ref{sec:application} of the main text, when $G=24$ and $G=32$. Each convolution filter uses zero padding and unit stride. The function $\mbox{vec}(\cdot)$ refers to a flattening of an array to a vector. All layers, except the final layer, used rectified linear unit (ReLU) activation functions; the final layer uses the identity function.}
\vspace{5pt}
 \begin{tabular}{lcccr} 
\hline
\hline
\multicolumn{5}{c}{$G=24$}\\
\hline
layer type & input dimension & output dimension & filter dimension & parameters\\
\hline
2D convolution & [24,\;24,\;2] & [16,\;16,\;32] & $9\times 9$ & 5,216\\
2D convolution & [16,\;16,\;32] & [7,\;7,\;64] & $10\times 10$ & 204,864\\
\multicolumn{5}{c}{As in $*$ and subsequent layers of Table~\ref{tab:CNN_archG16}.}\\

\hline
\multicolumn{4}{l}{total trainable parameters:} &841,181\\
\hline
  \end{tabular}
 \begin{tabular}{lcccr} 
\hline
\multicolumn{5}{c}{$G=32$}\\
\hline
layer type & input dimension & output dimension & filter dimension & parameters\\
\hline
2D convolution & [32,\;32,\;2] & [16,\;16,\;32] & $17\times 17$ & 18,528\\
2D convolution & [16,\;16,\;32] & [7,\;7,\;64] & $10\times 10$ & 204,864\\
\multicolumn{5}{c}{As in $*$ and subsequent layers of Table~\ref{tab:CNN_archG16}.}\\

\hline
\multicolumn{4}{l}{total trainable parameters:} &854,493\\
\hline
  \end{tabular}
 \label{tab:CNN_archG2432}
  \end{table*}
%%%%%%%%%%%%%%%%%%%%%%%%%%%%%%%%%
%%%%%%%%%%%%%%%%%%%%%%%%%%%%%%%%%
%%%%%%%%%%%%%%%%%%%%%%%%%%%%%%%%% 

\end{appendix}

\clearpage

\baselineskip=14pt
\begingroup
\setstretch{0.75}
\bibliographystyle{apalike}
\bibliography{ref}

\begin{thebibliography}{}

\bibitem[Ahmed et~al., 2022]{ahmed2022recognizing}
Ahmed, M., Maume-Deschamps, V., and Ribereau, P. (2022).
\newblock Recognizing a spatial extreme dependence structure: A deep learning approach.
\newblock {\em Environmetrics}, 33(4):e2714.

\bibitem[Alharbi et~al., 2013]{alharbi2013march}
Alharbi, B., Maghrabi, A., and Tapper, N. (2013).
\newblock {The March 2009 dust event in Saudi Arabia: Precursor and supportive environment}.
\newblock {\em Bulletin of the American Meteorological Society}, 94(4):515--528.

\bibitem[Asadi et~al., 2015]{Asadi2015}
Asadi, P., Davison, A.~C., and Engelke, S. (2015).
\newblock {Extremes on river networks}.
\newblock {\em The Annals of Applied Statistics}, 9(4):2023 -- 2050.

\bibitem[Aune et~al., 2014]{aune2014parameter}
Aune, E., Simpson, D.~P., and Eidsvik, J. (2014).
\newblock Parameter estimation in high dimensional {G}aussian distributions.
\newblock {\em Statistics and Computing}, 24:247--263.

\bibitem[Belzile et~al., 2023]{belzile2022modeler}
Belzile, L.~R., Dutang, C., Northrop, P.~J., and Opitz, T. (2023).
\newblock A modeler’s guide to extreme value software.
\newblock {\em Extremes}, 26(4):595--638.

\bibitem[Bevilacqua et~al., 2012]{bevilacqua2012estimating}
Bevilacqua, M., Gaetan, C., Mateu, J., and Porcu, E. (2012).
\newblock Estimating space and space-time covariance functions for large data sets: a weighted composite likelihood approach.
\newblock {\em Journal of the American Statistical Association}, 107(497):268--280.

\bibitem[B{\'e}wentaor{\'e} and Barro, 2022]{bewentaore2022space}
B{\'e}wentaor{\'e}, S. and Barro, D. (2022).
\newblock {Space-time trend detection and dependence modeling in extreme event approaches by functional peaks-over-thresholds: application to precipitation in Burkina Faso}.
\newblock {\em International Journal of Mathematics and Mathematical Sciences}, 2022.

\bibitem[Boulaguiem et~al., 2022]{boulaguiem2022modeling}
Boulaguiem, Y., Zscheischler, J., Vignotto, E., van~der Wiel, K., and Engelke, S. (2022).
\newblock Modeling and simulating spatial extremes by combining extreme value theory with generative adversarial networks.
\newblock {\em Environmental Data Science}, 1:e5.

\bibitem[Brown and Resnick, 1977]{brown1977extreme}
Brown, B.~M. and Resnick, S.~I. (1977).
\newblock Extreme values of independent stochastic processes.
\newblock {\em Journal of Applied Probability}, 14(4):732--739.

\bibitem[Buishand et~al., 2008]{Buishand2008}
Buishand, T.~A., de~Haan, L., and Zhou, C. (2008).
\newblock {On spatial extremes: With application to a rainfall problem}.
\newblock {\em The Annals of Applied Statistics}, 2(2):624 -- 642.

\bibitem[Cannon, 2010]{cannon2010flexible}
Cannon, A.~J. (2010).
\newblock A flexible nonlinear modelling framework for nonstationary generalized extreme value analysis in hydroclimatology.
\newblock {\em Hydrological Processes: An International Journal}, 24(6):673--685.

\bibitem[Castro-Camilo and Huser, 2020]{castro2020local}
Castro-Camilo, D. and Huser, R. (2020).
\newblock Local likelihood estimation of complex tail dependence structures, applied to {U.S.} precipitation extremes.
\newblock {\em Journal of the American Statistical Association}, 115(531):1037--1054.

\bibitem[Castruccio et~al., 2016]{castruccio2016high}
Castruccio, S., Huser, R., and Genton, M.~G. (2016).
\newblock High-order composite likelihood inference for max-stable distributions and processes.
\newblock {\em Journal of Computational and Graphical Statistics}, 25(4):1212--1229.

\bibitem[Chan et~al., 2018]{chan2018likelihood}
Chan, J., Perrone, V., Spence, J., Jenkins, P., Mathieson, S., and Song, Y. (2018).
\newblock A likelihood-free inference framework for population genetic data using exchangeable neural networks.
\newblock {\em Advances in Neural Information Processing Systems}, 31.

\bibitem[Cisneros et~al., 2024]{cisneros2024deep}
Cisneros, D., Richards, J., Dahal, A., Lombardo, L., and Huser, R. (2024).
\newblock Deep graphical regression for jointly moderate and extreme {A}ustralian wildfires.
\newblock {\em Spatial Statistics}, page 100811.

\bibitem[Coles, 2001]{coles2001introduction}
Coles, S. (2001).
\newblock {\em An Introduction to Statistical Modeling of Extreme Values}, volume 208.
\newblock Springer.

\bibitem[Cremanns and Roos, 2017]{cremanns2017deep}
Cremanns, K. and Roos, D. (2017).
\newblock Deep {G}aussian covariance network.
\newblock {\em arXiv preprint arXiv:1710.06202}.

\bibitem[Davison and Gholamrezaee, 2012]{davison2012geostatistics}
Davison, A.~C. and Gholamrezaee, M.~M. (2012).
\newblock Geostatistics of extremes.
\newblock {\em Proceedings of the Royal Society A: Mathematical, Physical and Engineering Sciences}, 468(2138):581--608.

\bibitem[Davison and Huser, 2015]{davison2015statistics}
Davison, A.~C. and Huser, R. (2015).
\newblock Statistics of extremes.
\newblock {\em Annual Review of Statistics and its Application}, 2:203--235.

\bibitem[Davison et~al., 2013]{davison2013geostatistics}
Davison, A.~C., Huser, R., and Thibaud, E. (2013).
\newblock Geostatistics of dependent and asymptotically independent extremes.
\newblock {\em Mathematical Geosciences}, 45:511--529.

\bibitem[Davison et~al., 2019]{davison2019spatial}
Davison, A.~C., Huser, R., and Thibaud, E. (2019).
\newblock Spatial extremes.
\newblock In Gelfand, A.~E., Fuentes, M., Hoeting, J.~A., and Smith, R.~L., editors, {\em Handbook of Environmental and Ecological Statistics}, pages 711--744. CRC Press.

\bibitem[Davison et~al., 2012]{davison2012}
Davison, A.~C., Padoan, S.~A., and Ribatet, M. (2012).
\newblock Statistical modeling of spatial extremes.
\newblock {\em Statistical Science}, 27(2):161--186.

\bibitem[de~Fondeville and Belzile, 2022]{de2018mvpot}
de~Fondeville, R. and Belzile, L. (2022).
\newblock {mvPot: Multivariate Peaks-over-Threshold Modelling for Spatial Extreme Events}.
\newblock {\em R package version 0.1.5}, 4.

\bibitem[de~Fondeville and Davison, 2018]{de2018high}
de~Fondeville, R. and Davison, A.~C. (2018).
\newblock High-dimensional peaks-over-threshold inference.
\newblock {\em Biometrika}, 105(3):575--592.

\bibitem[de~Fondeville and Davison, 2022]{de2022functional}
de~Fondeville, R. and Davison, A.~C. (2022).
\newblock Functional peaks-over-threshold analysis.
\newblock {\em Journal of the Royal Statistical Society: Series B}, 4(84):1392--1422.

\bibitem[de~Haan, 1984]{de1984spectral}
de~Haan, L. (1984).
\newblock A spectral representation for max-stable processes.
\newblock {\em The Annals of Probability}, 12(4):1194--1204.

\bibitem[de~Haan and Ferreira, 2006]{haan2006extreme}
de~Haan, L. and Ferreira, A. (2006).
\newblock {\em Extreme Value Theory: An Introduction}, volume~3.
\newblock Springer.

\bibitem[Dombry et~al., 2017]{dombry2017bayesian}
Dombry, C., Engelke, S., and Oesting, M. (2017).
\newblock {Bayesian inference for multivariate extreme value distributions}.
\newblock {\em Electronic Journal of Statistics}, 11(2):4813 -- 4844.

\bibitem[Dombry and Ribatet, 2015]{dombry2015functional}
Dombry, C. and Ribatet, M. (2015).
\newblock Functional regular variations, {P}areto processes and peaks over threshold.
\newblock {\em Statistics and Its Interface}, 8(1):9--17.

\bibitem[Engelke et~al., 2011]{engelke2011equivalent}
Engelke, S., Kabluchko, Z., and Schlather, M. (2011).
\newblock An equivalent representation of the {Brown--Resnick} process.
\newblock {\em Statistics \& probability letters}, 81(8):1150--1154.

\bibitem[Ferreira and de~Haan, 2014]{Ferreira2014}
Ferreira, A. and de~Haan, L. (2014).
\newblock {The generalized Pareto process; with a view towards application and simulation}.
\newblock {\em Bernoulli}, 20(4):1717--1737.

\bibitem[Galib et~al., 2022]{galib2022deepextrema}
Galib, A.~H., McDonald, A., Wilson, T., Luo, L., and Tan, P.-N. (2022).
\newblock {DeepExtrema}: A deep learning approach for forecasting block maxima in time series data.
\newblock {\em Proceedings of the Thirty-First International Joint Conference on Artificial Intelligence (IJCAI-22)}, pages 2980--2986.

\bibitem[Gelfand and Schliep, 2016]{gelfand2016spatial}
Gelfand, A.~E. and Schliep, E.~M. (2016).
\newblock Spatial statistics and {G}aussian processes: A beautiful marriage.
\newblock {\em Spatial Statistics}, 18:86--104.

\bibitem[Genton et~al., 2011]{genton2011likelihood}
Genton, M.~G., Ma, Y., and Sang, H. (2011).
\newblock On the likelihood function of {G}aussian max-stable processes.
\newblock {\em Biometrika}, pages 481--488.

\bibitem[Genz and Bretz, 2009]{genz2009computation}
Genz, A. and Bretz, F. (2009).
\newblock {\em Computation of Multivariate Normal and t Probabilities}, volume 195.
\newblock Springer Science \& Business Media.

\bibitem[Gerber and Nychka, 2021]{gerber2021fast}
Gerber, F. and Nychka, D. (2021).
\newblock Fast covariance parameter estimation of spatial {G}aussian process models using neural networks.
\newblock {\em Stat}, 10(1):e382.

\bibitem[Goodfellow et~al., 2016]{goodfellow2016deep}
Goodfellow, I., Bengio, Y., and Courville, A. (2016).
\newblock {\em Deep Learning}.
\newblock MIT press.

\bibitem[Gu et~al., 2018]{gu2018recent}
Gu, J., Wang, Z., Kuen, J., Ma, L., Shahroudy, A., Shuai, B., Liu, T., Wang, X., Wang, G., Cai, J., et~al. (2018).
\newblock Recent advances in convolutional neural networks.
\newblock {\em Pattern Recognition}, 77:354--377.

\bibitem[He et~al., 2015]{he2015spatial}
He, K., Zhang, X., Ren, S., and Sun, J. (2015).
\newblock Spatial pyramid pooling in deep convolutional networks for visual recognition.
\newblock {\em IEEE Transactions on Pattern Analysis and Machine Intelligence}, 37(9):1904--1916.

\bibitem[Healy et~al., 2024]{healy2021inference}
Healy, D., Parnell, A., Thorne, P., and Tawn, J. (2024).
\newblock Inference for extreme spatial temperature events in a changing climate with application to {I}reland.
\newblock {\em Journal of the Royal Statistical Society: Series C}.
\newblock To appear.

\bibitem[Hector and Reich, 2023]{Hector2023}
Hector, E.~C. and Reich, B.~J. (2023).
\newblock Distributed inference for spatial extremes modeling in high dimensions.
\newblock {\em Journal of the American Statistical Association}.
\newblock To appear.

\bibitem[Huser and Davison, 2013]{huser2013composite}
Huser, R. and Davison, A.~C. (2013).
\newblock Composite likelihood estimation for the {Brown--Resnick} process.
\newblock {\em Biometrika}, 100(2):511--518.

\bibitem[Huser and Davison, 2014]{huser2014space}
Huser, R. and Davison, A.~C. (2014).
\newblock Space--time modelling of extreme events.
\newblock {\em Journal of the Royal Statistical Society: Series B}, 76(2):439--461.

\bibitem[Huser et~al., 2016]{huser2016likelihood}
Huser, R., Davison, A.~C., and Genton, M.~G. (2016).
\newblock Likelihood estimators for multivariate extremes.
\newblock {\em Extremes}, 19:79--103.

\bibitem[Huser et~al., 2019]{huser2019full}
Huser, R., Dombry, C., Ribatet, M., and Genton, M.~G. (2019).
\newblock Full likelihood inference for max-stable data.
\newblock {\em Stat}, 8(1):e218.

\bibitem[Huser and Genton, 2016]{huser2016non}
Huser, R. and Genton, M.~G. (2016).
\newblock Non-stationary dependence structures for spatial extremes.
\newblock {\em Journal of Agricultural, Biological, and Environmental Statistics}, 21:470--491.

\bibitem[Huser et~al., 2017]{huser2017bridging}
Huser, R., Opitz, T., and Thibaud, E. (2017).
\newblock Bridging asymptotic independence and dependence in spatial extremes using {G}aussian scale mixtures.
\newblock {\em Spatial Statistics}, 21:166--186.

\bibitem[Huser et~al., 2023]{huser2022vecchia}
Huser, R., Stein, M.~L., and Zhong, P. (2023).
\newblock Vecchia likelihood approximation for accurate and fast inference with intractable spatial max-stable models.
\newblock {\em Journal of Computational and Graphical Statistics}.
\newblock To appear.

\bibitem[Huser and Wadsworth, 2019]{huser2019modeling}
Huser, R. and Wadsworth, J.~L. (2019).
\newblock Modeling spatial processes with unknown extremal dependence class.
\newblock {\em Journal of the American Statistical Association}, 114(525):434--444.

\bibitem[Huser and Wadsworth, 2022]{huser2022advances}
Huser, R. and Wadsworth, J.~L. (2022).
\newblock Advances in statistical modeling of spatial extremes.
\newblock {\em Wiley Interdisciplinary Reviews: Computational Statistics}, 14(1):e1537.

\bibitem[Joe, 1997]{joe1997multivariate}
Joe, H. (1997).
\newblock {\em Multivariate Models and Multivariate Dependence Concepts}.
\newblock CRC press.

\bibitem[Kabluchko et~al., 2009]{kabluchko2009stationary}
Kabluchko, Z., Schlather, M., and de~Haan, L. (2009).
\newblock {Stationary max-stable fields associated to negative definite functions}.
\newblock {\em The Annals of Probability}, 37(5):2042 -- 2065.

\bibitem[Khodeir et~al., 2012]{khodeir2012source}
Khodeir, M., Shamy, M., Alghamdi, M., Zhong, M., Sun, H., Costa, M., Chen, L.-C., and Maciejczyk, P. (2012).
\newblock {Source apportionment and elemental composition of PM2.5 and PM10 in Jeddah City, Saudi Arabia}.
\newblock {\em Atmospheric Pollution Research}, 3(3):331--340.

\bibitem[Klein and Moeschberger, 2003]{klein2003survival}
Klein, J.~P. and Moeschberger, M.~L. (2003).
\newblock {\em Survival Analysis: Techniques for Censored and Truncated Data}, volume 1230.
\newblock Springer.

\bibitem[Krupskii et~al., 2018]{krupskii2018factor}
Krupskii, P., Huser, R., and Genton, M.~G. (2018).
\newblock Factor copula models for replicated spatial data.
\newblock {\em Journal of the American Statistical Association}, 113(521):467--479.

\bibitem[Lehmann and Casella, 1998]{lehmann2006theory}
Lehmann, E.~L. and Casella, G. (1998).
\newblock {\em Theory of Point Estimation}.
\newblock Springer, New York, NY, 2nd edition.

\bibitem[Lelieveld et~al., 2019]{lelieveld2019cardiovascular}
Lelieveld, J., Klingm{\"u}ller, K., Pozzer, A., P{\"o}schl, U., Fnais, M., Daiber, A., and M{\"u}nzel, T. (2019).
\newblock {Cardiovascular disease burden from ambient air pollution in Europe reassessed using novel hazard ratio functions}.
\newblock {\em European Heart Journal}, 40(20):1590--1596.

\bibitem[Lenzi et~al., 2023]{lenzi2021neural}
Lenzi, A., Bessac, J., Rudi, J., and Stein, M.~L. (2023).
\newblock Neural networks for parameter estimation in intractable models.
\newblock {\em Computational Statistics \& Data Analysis}, 185:107762.

\bibitem[Majumder and Reich, 2023]{majumder2022deep}
Majumder, R. and Reich, B.~J. (2023).
\newblock A deep learning synthetic likelihood approximation of a non-stationary spatial model for extreme streamflow forecasting.
\newblock {\em Spatial Statistics}, 55:100755.

\bibitem[Majumder et~al., 2024]{majumder2022modeling}
Majumder, R., Reich, B.~J., and Shaby, B.~A. (2024).
\newblock Modeling extremal streamflow using deep learning approximations and a flexible spatial process.
\newblock {\em The Annals of Applied Statistics}, 18(2):1519--1542.

\bibitem[McCullagh, 2002]{mccullagh2002statistical}
McCullagh, P. (2002).
\newblock What is a statistical model?
\newblock {\em The Annals of Statistics}, 30(5):1225--1310.

\bibitem[McDonald et~al., 2022]{mcdonald2022comet}
McDonald, A., Tan, P.-N., and Luo, L. (2022).
\newblock {COMET} flows: Towards generative modeling of multivariate extremes and tail dependence.
\newblock {\em Proceedings of the Thirty-First International Joint Conference on Artificial Intelligence (IJCAI-22)}, pages 3328--3334.

\bibitem[Munir et~al., 2017]{munir2017analysing}
Munir, S., Habeebullah, T.~M., Mohammed, A.~M., Morsy, E.~A., Rehan, M., Ali, K., et~al. (2017).
\newblock {Analysing PM2.5 and its association with PM10 and meteorology in the arid climate of Makkah, Saudi Arabia}.
\newblock {\em Aerosol and Air Quality Research}, 17(2):453--464.

\bibitem[Opitz, 2013]{opitz2013extremal}
Opitz, T. (2013).
\newblock Extremal t processes: Elliptical domain of attraction and a spectral representation.
\newblock {\em Journal of Multivariate Analysis}, 122:409--413.

\bibitem[Opitz, 2016]{opitz2016modeling}
Opitz, T. (2016).
\newblock Modeling asymptotically independent spatial extremes based on {L}aplace random fields.
\newblock {\em Spatial Statistics}, 16:1--18.

\bibitem[Padoan et~al., 2010]{padoan2010likelihood}
Padoan, S.~A., Ribatet, M., and Sisson, S.~A. (2010).
\newblock Likelihood-based inference for max-stable processes.
\newblock {\em Journal of the American Statistical Association}, 105(489):263--277.

\bibitem[Pagendam et~al., 2023]{Pagendam2023}
Pagendam, D., Janardhanan, S., Dabrowski, J., and MacKinlay, D. (2023).
\newblock A log-additive neural model for spatio-temporal prediction of groundwater levels.
\newblock {\em Spatial Statistics}, 55:100740.

\bibitem[Pasche and Engelke, 2022]{pasche2022neural}
Pasche, O.~C. and Engelke, S. (2022).
\newblock Neural networks for extreme quantile regression with an application to forecasting of flood risk.
\newblock {\em arXiv preprint arXiv:2208.07590}.

\bibitem[Polichetti et~al., 2009]{polichetti2009effects}
Polichetti, G., Cocco, S., Spinali, A., Trimarco, V., and Nunziata, A. (2009).
\newblock {Effects of particulate matter (PM10, PM2.5 and PM1) on the cardiovascular system}.
\newblock {\em Toxicology}, 261(1-2):1--8.

\bibitem[Radev et~al., 2023]{radev2023jana}
Radev, S.~T., Schmitt, M., Pratz, V., Picchini, U., K{\"o}the, U., and B{\"u}rkner, P.-C. (2023).
\newblock {JANA: Jointly amortized neural approximation of complex Bayesian models}.
\newblock In {\em Uncertainty in Artificial Intelligence}, pages 1695--1706. PMLR.

\bibitem[Rai et~al., 2024]{rai2023fast}
Rai, S., Hoffman, A., Lahiri, S., Nychka, D.~W., Sain, S.~R., and Bandyopadhyay, S. (2024).
\newblock Fast parameter estimation of generalized extreme value distribution using neural networks.
\newblock {\em Environmetrics}, 35(3):e2845.

\bibitem[Ribatet, 2022]{spatialextremes}
Ribatet, M. (2022).
\newblock {\em {SpatialExtremes}: modelling spatial extremes}.
\newblock R package version 2.1-0.

\bibitem[Richards and Huser, 2024]{richards2022unifying}
Richards, J. and Huser, R. (2024).
\newblock Regression modelling of spatiotemporal extreme {U.S.} wildfires via partially-interpretable neural networks.
\newblock {\em arXiv preprint arXiv:2208.07581}.

\bibitem[Richards et~al., 2023a]{richards2022insights}
Richards, J., Huser, R., Bevacqua, E., and Zscheischler, J. (2023a).
\newblock Insights into the drivers and spatiotemporal trends of extreme mediterranean wildfires with statistical deep learning.
\newblock {\em Artificial Intelligence for the Earth Systems}, 2(4):e220095.

\bibitem[Richards et~al., 2022]{richards2022modelling}
Richards, J., Tawn, J.~A., and Brown, S. (2022).
\newblock Modelling extremes of spatial aggregates of precipitation using conditional methods.
\newblock {\em The Annals of Applied Statistics}, 16(4):2693--2713.

\bibitem[Richards et~al., 2023b]{richards2023joint}
Richards, J., Tawn, J.~A., and Brown, S. (2023b).
\newblock Joint estimation of extreme spatially aggregated precipitation at different scales through mixture modelling.
\newblock {\em Spatial Statistics}, 53:100725.

\bibitem[Richards and Wadsworth, 2021]{richards2021spatial}
Richards, J. and Wadsworth, J.~L. (2021).
\newblock Spatial deformation for nonstationary extremal dependence.
\newblock {\em Environmetrics}, 32(5):e2671.

\bibitem[Sainsbury-Dale et~al., 2024a]{sainsbury2022fast}
Sainsbury-Dale, M., Zammit-Mangion, A., and Huser, R. (2024a).
\newblock Likelihood-free parameter estimation with neural {B}ayes estimators.
\newblock {\em The American Statistician}, 78(1):1--14.

\bibitem[Sainsbury-Dale et~al., 2024b]{sainsbury2023neural}
Sainsbury-Dale, M., Zammit-Mangion, A., Richards, J., and Huser, R. (2024b).
\newblock Neural {B}ayes estimators for irregular spatial data using graph neural networks.
\newblock {\em arXiv preprint arXiv:2310.02600}.

\bibitem[Sang and Gelfand, 2010]{sang2010continuous}
Sang, H. and Gelfand, A.~E. (2010).
\newblock Continuous spatial process models for spatial extreme values.
\newblock {\em Journal of Agricultural, Biological, and Environmental Statistics}, 15:49--65.

\bibitem[Sang and Genton, 2014]{sang2014tapered}
Sang, H. and Genton, M.~G. (2014).
\newblock Tapered composite likelihood for spatial max-stable models.
\newblock {\em Spatial Statistics}, 8:86--103.

\bibitem[Schlather, 2002]{schlather2002models}
Schlather, M. (2002).
\newblock Models for stationary max-stable random fields.
\newblock {\em Extremes}, 5:33--44.

\bibitem[Schmitt et~al., 2023]{schmitt2023detecting}
Schmitt, M., B{\"u}rkner, P.-C., K{\"o}the, U., and Radev, S.~T. (2023).
\newblock Detecting model misspecification in amortized {B}ayesian inference with neural networks.
\newblock In {\em DAGM German Conference on Pattern Recognition}, pages 541--557. Springer.

\bibitem[Shao et~al., 2024]{shao2022flexible}
Shao, X., Hazra, A., Richards, J., and Huser, R. (2024).
\newblock Flexible modeling of nonstationary extremal dependence using spatially-fused {LASSO} and ridge penalties.
\newblock {\em arXiv preprint arXiv:2210.05792}.

\bibitem[Smith, 1990]{smith1990max}
Smith, R.~L. (1990).
\newblock Max-stable processes and spatial extremes.
\newblock {\em Unpublished manuscript}, 205:1--32.

\bibitem[Stephenson and Tawn, 2005]{stephenson2005exploiting}
Stephenson, A. and Tawn, J. (2005).
\newblock Exploiting occurrence times in likelihood inference for componentwise maxima.
\newblock {\em Biometrika}, 92(1):213--227.

\bibitem[Tariq et~al., 2022]{tariq2022spatial}
Tariq, S., Mariam, A., Mehmood, U., et~al. (2022).
\newblock {Spatial and temporal variations in PM2.5 and associated health risk assessment in Saudi Arabia using remote sensing}.
\newblock {\em Chemosphere}, 308:136296.

\bibitem[Thibaud et~al., 2013]{thibaud2013}
Thibaud, E., Mutzner, R., and Davison, A.~C. (2013).
\newblock Threshold modeling of extreme spatial rainfall.
\newblock {\em Water Resources Research}, 49(8):4633--4644.

\bibitem[Thibaud and Opitz, 2015]{thibaud2015efficient}
Thibaud, E. and Opitz, T. (2015).
\newblock Efficient inference and simulation for elliptical {P}areto processes.
\newblock {\em Biometrika}, 102(4):855--870.

\bibitem[Van~Donkelaar et~al., 2021]{van2021monthly}
Van~Donkelaar, A., Hammer, M.~S., Bindle, L., Brauer, M., Brook, J.~R., Garay, M.~J., Hsu, N.~C., Kalashnikova, O.~V., Kahn, R.~A., Lee, C., et~al. (2021).
\newblock Monthly global estimates of fine particulate matter and their uncertainty.
\newblock {\em Environmental Science \& Technology}, 55(22):15287--15300.

\bibitem[Varin et~al., 2011]{varin2011overview}
Varin, C., Reid, N., and Firth, D. (2011).
\newblock An overview of composite likelihood methods.
\newblock {\em Statistica Sinica}, 21(1):5--42.

\bibitem[Vasiliades et~al., 2015]{vasiliades2015nonstationary}
Vasiliades, L., Galiatsatou, P., and Loukas, A. (2015).
\newblock Nonstationary frequency analysis of annual maximum rainfall using climate covariates.
\newblock {\em Water Resources Management}, 29:339--358.

\bibitem[Wadsworth, 2015]{wadsworth2015occurrence}
Wadsworth, J.~L. (2015).
\newblock On the occurrence times of componentwise maxima and bias in likelihood inference for multivariate max-stable distributions.
\newblock {\em Biometrika}, 102(3):705--711.

\bibitem[Wadsworth, 2018]{spatialADAI}
Wadsworth, J.~L. (2018).
\newblock {SpatialADAI: R package}.
\newblock \url{https://www.lancaster.ac.uk/~wadswojl/SpatialADAI}.
\newblock Accessed: 2023-03-06.

\bibitem[Wadsworth and Tawn, 2012]{wadsworth2012dependence}
Wadsworth, J.~L. and Tawn, J.~A. (2012).
\newblock Dependence modelling for spatial extremes.
\newblock {\em Biometrika}, 99(2):253--272.

\bibitem[Wadsworth and Tawn, 2014]{wadsworth2014efficient}
Wadsworth, J.~L. and Tawn, J.~A. (2014).
\newblock Efficient inference for spatial extreme value processes associated to log-{G}aussian random functions.
\newblock {\em Biometrika}, 101(1):1--15.

\bibitem[Wadsworth and Tawn, 2022]{wadsworth2022higher}
Wadsworth, J.~L. and Tawn, J.~A. (2022).
\newblock Higher-dimensional spatial extremes via single-site conditioning.
\newblock {\em Spatial Statistics}, 51:100677.

\bibitem[Wadsworth et~al., 2017]{wadsworth2017modelling}
Wadsworth, J.~L., Tawn, J.~A., Davison, A.~C., and Elton, D.~M. (2017).
\newblock Modelling across extremal dependence classes.
\newblock {\em Journal of the Royal Statistical Society: Series B}, 79(1):149--175.

\bibitem[Walchessen et~al., 2023]{walchessen2023neural}
Walchessen, J., Lenzi, A., and Kuusela, M. (2023).
\newblock Neural likelihood surfaces for spatial processes with computationally intensive or intractable likelihoods.
\newblock {\em arXiv preprint arXiv:2305.04634}.

\bibitem[Wang et~al., 2023]{Wang2023.01.09.523219}
Wang, Z., Hasenauer, J., and Sch{\"a}lte, Y. (2023).
\newblock Missing data in amortized simulation-based neural posterior estimation.
\newblock {\em bioRxiv preprint bioRxiv:2023.01.09.523219}.

\bibitem[Wilson et~al., 2022]{wilson2022deepgpd}
Wilson, T., Tan, P.-N., and Luo, L. (2022).
\newblock {DeepGPD}: A deep learning approach for modeling geospatio-temporal extreme events.
\newblock {\em Proceedings of the AAAI Conference on Artificial Intelligence}, 36(4):4245--4253.

\bibitem[Wu et~al., 2020]{wu2020comprehensive}
Wu, Z., Pan, S., Chen, F., Long, G., Zhang, C., and Philip, S.~Y. (2020).
\newblock A comprehensive survey on graph neural networks.
\newblock {\em IEEE Transactions on Neural Networks and Learning Systems}, 32(1):4--24.

\bibitem[Xing et~al., 2016]{xing2016impact}
Xing, Y.-F., Xu, Y.-H., Shi, M.-H., and Lian, Y.-X. (2016).
\newblock {The impact of PM2.5 on the human respiratory system}.
\newblock {\em Journal of Thoracic Disease}, 8(1):E69.

\bibitem[Zaheer et~al., 2017]{zaheer2017deep}
Zaheer, M., Kottur, S., Ravanbakhsh, S., Poczos, B., Salakhutdinov, R.~R., and Smola, A.~J. (2017).
\newblock Deep sets.
\newblock {\em Advances in Neural Information Processing Systems}, 30.

\bibitem[Zammit-Mangion and Rougier, 2020]{zammit2020multi}
Zammit-Mangion, A. and Rougier, J. (2020).
\newblock Multi-scale process modelling and distributed computation for spatial data.
\newblock {\em Statistics and Computing}, 30(6):1609--1627.

\bibitem[Zammit-Mangion et~al., 2024]{zammit2024neural}
Zammit-Mangion, A., Sainsbury-Dale, M., and Huser, R. (2024).
\newblock Neural methods for amortised parameter inference.
\newblock {\em arXiv preprint arXiv:2404.12484}.

\bibitem[Zammit-Mangion and Wikle, 2020]{zammit2020deep}
Zammit-Mangion, A. and Wikle, C.~K. (2020).
\newblock Deep integro-difference equation models for spatio-temporal forecasting.
\newblock {\em Spatial Statistics}, 37:100408.

\bibitem[Zhang et~al., 2023]{Zhang.etal:2023}
Zhang, L., Ma, X., Wikle, C.~K., and Huser, R. (2023).
\newblock {Flexible and efficient spatial extremes emulation via variational autoencoders}.
\newblock {arXiv preprint arXiv:2307.11390}.

\bibitem[Zhang et~al., 2022]{zhang2022hierarchical}
Zhang, L., Shaby, B.~A., and Wadsworth, J.~L. (2022).
\newblock Hierarchical transformed scale mixtures for flexible modeling of spatial extremes on datasets with many locations.
\newblock {\em Journal of the American Statistical Association}, 117(539):1357--1369.

\end{thebibliography}
\endgroup
\baselineskip 10pt

\end{document}